\documentclass[fleqn]{2023SCGE}
\usepackage{color}
\usepackage{float}
\usepackage{natbib}
\usepackage{graphicx}
\usepackage{tablefootnote}
\usepackage{amsmath}
\usepackage{etoolbox}
\robustify\cite

\usepackage{url}
\usepackage{hyperref}

\def\arcmin{\hbox{$^\prime$}}
\def\arcsec{\hbox{$^{\prime\prime}$}}

\usepackage[commandnameprefix=ifneeded,markup=bfit]{changes} 
\makeatletter
\AddToHook{cmd/added/before}{\def\Changes@AuthorColor{blue!80!red!80!green}}
\AddToHook{cmd/deleted/before}{\def\Changes@AuthorColor{red}}
\AddToHook{cmd/replaced/before}{\def\Changes@AuthorColor{red}}
\makeatother

\begin{document}

\ensubject{subject}

\ArticleType{Article}
\SpecialTopic{SPECIAL TOPIC: }
\Year{2023}
\Month{January}
\Vol{66}
\No{1}
\DOI{??}
\ArtNo{000000}
\ReceiveDate{January 11, 2023}
\AcceptDate{April 6, 2023}

\title{Scientific Objectives of the Xue-shan-mu-chang 15-meter Submillimeter Telescope}

\author[]{XSMT Project Collaboration Group\footnote{Corresponding authors: Jing~Li, email: lijing@pmo.ac.cn; Yong~Shi, email: shiyong@westlake.edu.cn}}{}
\author[1]{Yiping~Ao}{}%
\author[2]{Jin~Chang}{}%
\author[1]{Zhiwei~Chen}{}
\author[3]{Xiangqun~Cui}{}
\author[4]{\\Kaiyi~Du}{}%
\author[1]{Fujun~Du}{} 
\author[1]{Yan~Gong}{}
\author[5]{Zhanwen~Han}{}
\author[6]{Gregory~Herczeg}{}
\author[6]{Luis~C.~Ho}{}
\author[1]{Jie~Hu}{}
\author[7]{\\Yipeng~Jing}{}
\author[8]{Sihan~Jiao}{}
\author[1]{Binggang~Ju}{}
\author[1]{Jing~Li}{lijing@pmo.ac.cn}
\author[9]{Xiaohu~Li}{}
\author[4]{Xiangdong~Li}{}
\author[4]{Lingrui~Lin}{}
\author[1]{\\Zhenhui~Lin}{}
\author[1]{Daizhong~Liu}{}
\author[1]{Dong~Liu}{}
\author[10]{Guoxi~Liu}{}
\author[1]{Zheng~Lou}{}
\author[1]{Dengrong~Lu}{}
\author[1]{Ruiqing~Mao}{}
\author[1]{\\Wei~Miao}{}
\author[1]{Yuan~Qian}{}
\author[4]{Keping~Qiu}{}
\author[11]{Zhiqiang Shen}{}
\author[12]{Yong~Shi}{shiyong@westlake.edu.cn}
\author[1]{Shengcai~Shi}{}
\author[13]{Chenggang~Shu}{}
\author[1]{\\Jixian~Sun}{}
\author[14]{Xiaohui~Sun}{}
\author[4]{Yichen~Sun}{}
\author[15]{Junzhi~Wang}{}
\author[6]{Ke~Wang}{}
\author[9]{Na~Wang}{}
\author[6]{Ran~Wang}{}
\author[4]{\\Tao~Wang}{}
\author[16,8]{Jingwen~Wu}{}
\author[8]{Xiangping~Wu}{}
\author[1]{Xuefeng~Wu}{}
\author[1]{Di~Xiao}{}
\author[1]{Qijun~Yao}{}
\author[10]{Yong~Yao}{}
\author[1]{\\Wen~Zhang}{}
\author[1]{Xuguo~Zhang}{}
\author[4]{Zhiyu~Zhang}{}
\author[10]{Yuanpeng~Zheng}{}

\AuthorMark{XSMT Collaboration}

\AuthorCitation{XSMT Collaboration et al.}

\address[{\rm 1}]{Purple Mountain Observatory, Chinese Academy of Sciences, Nanjing, 210023, PR China}
\address[{\rm 2}]{School of Astronomy and Space Sciences, University of Science and Technology of China, Hefei, 230026, PR China}
\address[{\rm 3}]{Nanjing Institute of Astronomical Optics \& Technology, Chinese Academy of Sciences, Nanjing, 210042, PR China}
\address[{\rm 4}]{School of Astronomy and Space Science, Nanjing University, Nanjing, 210093, PR China}
\address[{\rm 5}]{Yunnan Observatories, Chinese Academy of Sciences, Kunming, 650216, PR China}
\address[{\rm 6}]{Kavli Institute for Astronomy and Astrophysics, Peking University, Beijing, 100871, PR China}
\address[{\rm 7}]{Department of Astronomy, School of Physics and Astronomy, Shanghai Jiao Tong University, Shanghai, 200240, PR China}
\address[{\rm 8}]{National Astronomical Observatories, Chinese Academy of Sciences, Beijing, 100012, PR China}
\address[{\rm 9}]{Xinjiang Astronomical Observatory, Chinese Academy of Sciences, Urumqi, 830011, PR China}
\address[{\rm 10}]{\hbox{ }The 54th Research Institute of China Electronics Technology Group Corporation, Shijiazhuang, 050081, PR China}
\address[{\rm 11}]{\hbox{ } Shanghai Astronomical Observatory, Chinese Academy of Sciences, Shanghai, 200030, PR China}
\address[{\rm 12}]{\hbox{ } Department of Astronomy, Westlake University, Hangzhou, 310030, PR China}
\address[{\rm 13}]{\hbox{ } Shanghai Key Lab for Astrophysics, Shanghai Normal University, Shanghai, 200234, PR China}
\address[{\rm 14}]{\hbox{ } School of Physics and Astronomy, Yunnan University, Kunming, 650500, PR China}
\address[{\rm 15}]{\hbox{ } Guangxi Key Laboratory for Relativistic Astrophysics, Department of Physics, Guangxi University, Nanning, 530004, PR China}
\address[{\rm 16}]{\hbox{ } University of Chinese Academy of Sciences, Beijing, 100049, PR China}


\abstract{Submillimeter astronomy is poised to revolutionize our understanding of the Universe by revealing cosmic phenomena hidden from optical and near-infrared observations, particularly those associated with interstellar dust, molecular gas, and star formation. The Xue-shan-mu-chang 15-meter submillimeter telescope (XSMT-15m), to be constructed at a premier high-altitude site (4813~m) in Qinghai, China, marks a major milestone for Chinese astronomy, establishing the China mainland’s first independently developed, world-class submillimeter facility. Equipped with state-of-the-art instruments, XSMT-15m will address a diverse range of frontier scientific questions spanning extragalactic astronomy, Galactic structure, time-domain astrophysics, and astrochemistry. In synergy with current and forthcoming observatories, XSMT-15m will illuminate the formation and evolution of galaxies, unravel the physical and chemical processes shaping the interstellar medium, and explore transient phenomena in the submillimeter regime. These capabilities will advance our understanding across extragalactic astronomy, Galactic ecology, astrochemistry, and time-domain astrophysics, inaugurating a new era for submillimeter research in China and the northern hemisphere.}



\maketitle


\tableofcontents

\begin{multicols}{2}

\section{Introduction}\label{section1}
Submillimeter astronomy is a cutting-edge field in astronomy that can unveil astronomical objects and cosmic phenomena invisible in the optical and near-infrared, particularly those associated with  interstellar dust, molecular gas, and star formation. Therefore, submillimeter astronomy plays a crucial role in understanding the origin and evolution of the universe. The Xue-shan-mu-chang 15-meter SubMillimeter telescope (hereafter XSMT-15m) is a planned facility aiming to become one of the world-class submillimeter telescopes. This project is led by Purple Mountain Observatory, Chinese Academy of Sciences, to establish the first independently developed world-class submillimeter telescope in China mainland. XSMT-15m will be in close synergy with other millimeter and submillimeter facilities (e.g., IRAM 30m, JCMT 15m, LMT 50m, SMA, NOEMA) to probe Galactic, extragalactic, and transient science in the northern hemisphere sky, and will also attempt to join force to achieve the next-generation very long baseline black hole imaging. 

XSMT-15m will be constructed at Xue-shan-mu-chang in Delingha,  Qinghai Province, China. Xue-shan-mu-chang is located at an altitude of 4813~m, with an average winter Precipitable Water Vapor (PWV) of 0.85~mm, making it an ideal location for submillimeter observations. XSMT-15m is a 15-meter Ritchey–Chrétien-type telescope with a main mirror surface accuracy of $\sim$30~$\mu$m. Additionally, it has two Nasmyth platforms which will be equipped with state-of-the-art instruments \citep{research.0586}: (1) a multi-color continuum camera covering 230, 345, and 460~GHz bands, (2) a multi-band heterodyne receiver compatible with the next-generation Event Horizon Telescope (ngEHT), and (3) a multi-beam 460~GHz heterodyne receiver. The continuum camera will use Kinetic Inductance Detectors (KIDs) providing a total of $>$10,000 pixels distributed over three bands. A polarimeter will be installed in front of the instrument to enable polarization capability. 
The multi-band heterodyne receiver front-end uses a dual-polarization design, enabling simultaneous dual-polarization observations. The heterodyne receiver will cover four frequency ranges: 85--115 GHz, 211--275 GHz, 275--373 GHz, and 440--510 GHz, with an intermediate frequency coverage of 4--8 GHz and a sideband rejection ratio of over 15~dB. Depending on the observing frequency, the angular resolution will range from 8$^{\prime\prime}$ to 60$^{\prime\prime}$ (see Table~\ref{tab1}). 

These state-of-the-art instruments provide the telescope with distinctive advantages and new capabilities, facilitating advances in a broad range of scientific fields including:
\textbf{extragalactic (EG) astronomy}, 
\textbf{Milky Way (MW) science}, 
\textbf{time-domain (TD) astronomy}, 
and \textbf{astrochemistry (AC)}.

In the following chapters, we present a brief description of key science cases in the above fields (Sections~\ref{sec:EG-GMC}--\ref{sec:AC-Galactic-Chemical}). 
We also briefly discuss which of these science cases are most suitable for first-light observations, and a potential data policy and pipeline development strategy in Section~\ref{sec:data-policy}. 
In Appendix~\ref{appendix:A} and \ref{appendix:B}, we further describe our science cases in the context of Astro2020 and compare with other major current/future submillimeter facilities in the world. 

\begin{table*}[t]
\footnotesize
\caption{Overview of the XSMT-15m submillimeter telescope. Sensitivity look-up table. For line RMS calculation, we use a bandwidth of 1~km~s$^{-1}$ and central frequencies of 86, 230, 345, and 460~GHz. For the continuum RMS calculation, we assumed a confusion limit of 0.28, 0.40, 0.70~mJy at 260, 350 and 460~GHz. Note that continuum RMS estimates are very preliminary at the current stage and mapping overheads are not considered.}
\label{tab1}
\centering
\setlength{\tabcolsep}{12pt} 
\begin{tabular}{|l|c|c|c|c|}
\hline
Antenna Diameter [m]                   & \multicolumn{4}{c|}{15} \\
\hline
Field of View [arcmin$^2$]             & \multicolumn{4}{c|}{$\sim 10^{\prime} \times 10^{\prime}$} \\
\hline
Surface Accuracy [$\mu$m]              & \multicolumn{4}{c|}{$\lesssim 30$} \\
\hline
Target Elevation [$^{\circ}$]          & \multicolumn{4}{c|}{5--90 (45 for RMS calculation)} \\
\hline
$T_\mathrm{receiver}$ [K]              & \multicolumn{4}{c|}{$\lesssim 10 \, h \nu / (k_{B} T)$} \\
\hline
PWV for RMS Calculation [mm]           & 1.5 & 1.5 & 1 & 0.5 \\
\hline
Continuum Band / Frequency [GHz]       & --- & 230--300 & 330--370 & 455--471 \\
\hline
Continuum Bandwidth [GHz]              & --- & 50 & 40 & 16 \\
\hline
Continuum RMS ($t_{\mathrm{on}}=1$h) [mJy~beam$^{-1}$]      & --- & 0.30 & 0.58 & 2.04 \\
\hline
Continuum $t_{\mathrm{on}}$ (RMS=0.2~mJy~beam$^{-1}$) [h]   & --- & 2.0 & 8.3 & 104 \\
\hline
Continuum $t_{\mathrm{on}}$ (confusion limit) [h]   & --- & 1.2 & 2.0 & 8.5 \\
\hline
Line Spectral Window [GHz]             & 85--115 & 196--281 & 268--375 & 378--508 \\
\hline
Line RMS ($t_{\mathrm{on}}=1$h) [mK]   & 5.5 & 3.9 & 5.4 & 18 \\
\hline
\end{tabular}
\end{table*}





\section{EG: GMC Census in Local Group Galaxies}
\label{sec:EG-GMC}

\textbf{Plain summary for non-experts:} \textit{Giant Molecular clouds (GMCs) are the birthplaces of most stars in the Milky Way and the building block of star formation in other galaxies. Understanding the properties of GMCs and their dependence on environments is therefore key to understanding star formation on galactic scales. A recent JCMT-SCUBA2 ultra-deep 850~$\mu$m survey in Andromeda (M31) revealed the first complete view of GMCs distribution within the inner 10 kpc disk of a spiral galaxy, including GMCs both on spiral arms and in inter-arm regions. The inter-arm GMCs are less massive, colder, and less turbulent than their on-arm counterparts, suggesting that GMCs may follow distinct evolutionary pathways of formation and destruction depending on environment. The KID camera of XSMT-15m is expected to have better performance than JCMT-SCUBA2 in continuum mapping surveys. We propose a confusion-limited 850 and 650~$\mu$m survey using XSMT-15m to conduct a GMC census in Local Group galaxies, including M31 outskirt area, M33, and several dwarf galaxies in the Local Group. The SCUBA2-M31 survey has proved that a 15m submillimeter telescope can detect GMCs down to a few $10^{4} M_{\odot}$ at the distance of M31. The proposed project will deliver the first unbiased, large-scale census of GMCs in the local universe, providing key constraints on their global properties and on how environment shapes their evolution.}

\subsection{Motivation}

Most stars, including almost all massive stars, form within Giant Molecular Clouds (GMCs) in the Milky Way (e.g., \citealt{Lada2003}). GMCs are also a basic structure that can be recognized in other galaxies, therefore are a bridge to connect Galactic and extragalactic study on star formation and structure formation. However, our understanding of how GMCs form, evolve, and destruct is still poor. Fundamental questions, like whether GMC are stable structures that survive over a galactic rotation period (e.g., \citealt{Scoville2004}), or short-lived transient structures rapidly destroyed by stellar feedback (e.g., \citealt{Elmegreen2000}), remain unsolved. A recent study (\citealt{Dessauges2019}) shows that GMCs can be much more massive and denser at high redshifts, where conditions are more extreme than in the Milky Way disk. This result challenges the long-standing assumption of universal GMC formation, instead favoring a scenario in which GMC formation is sensitive to their birth environment. Similar conclusions have been drawn from investigations of molecular cloud properties in metal-poor dwarf galaxies (e.g., \citealt{Shi2014}), and in the outer regions of spiral galaxies.

\begin{figure*}[!htbp]
\centering
    \includegraphics[width=0.45\textwidth]{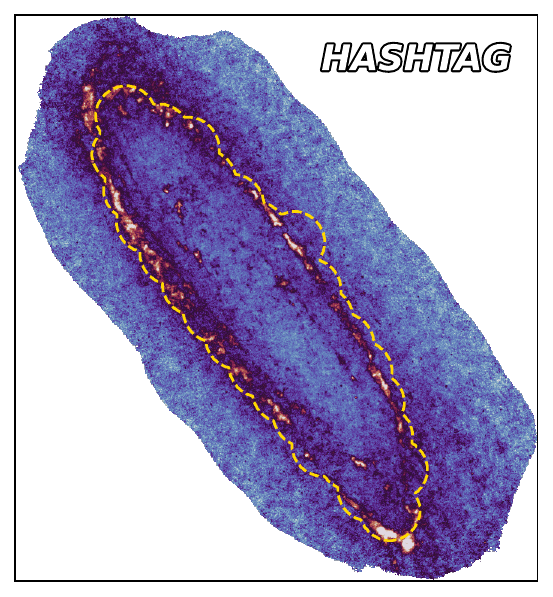}
    \includegraphics[width=0.3\textwidth]{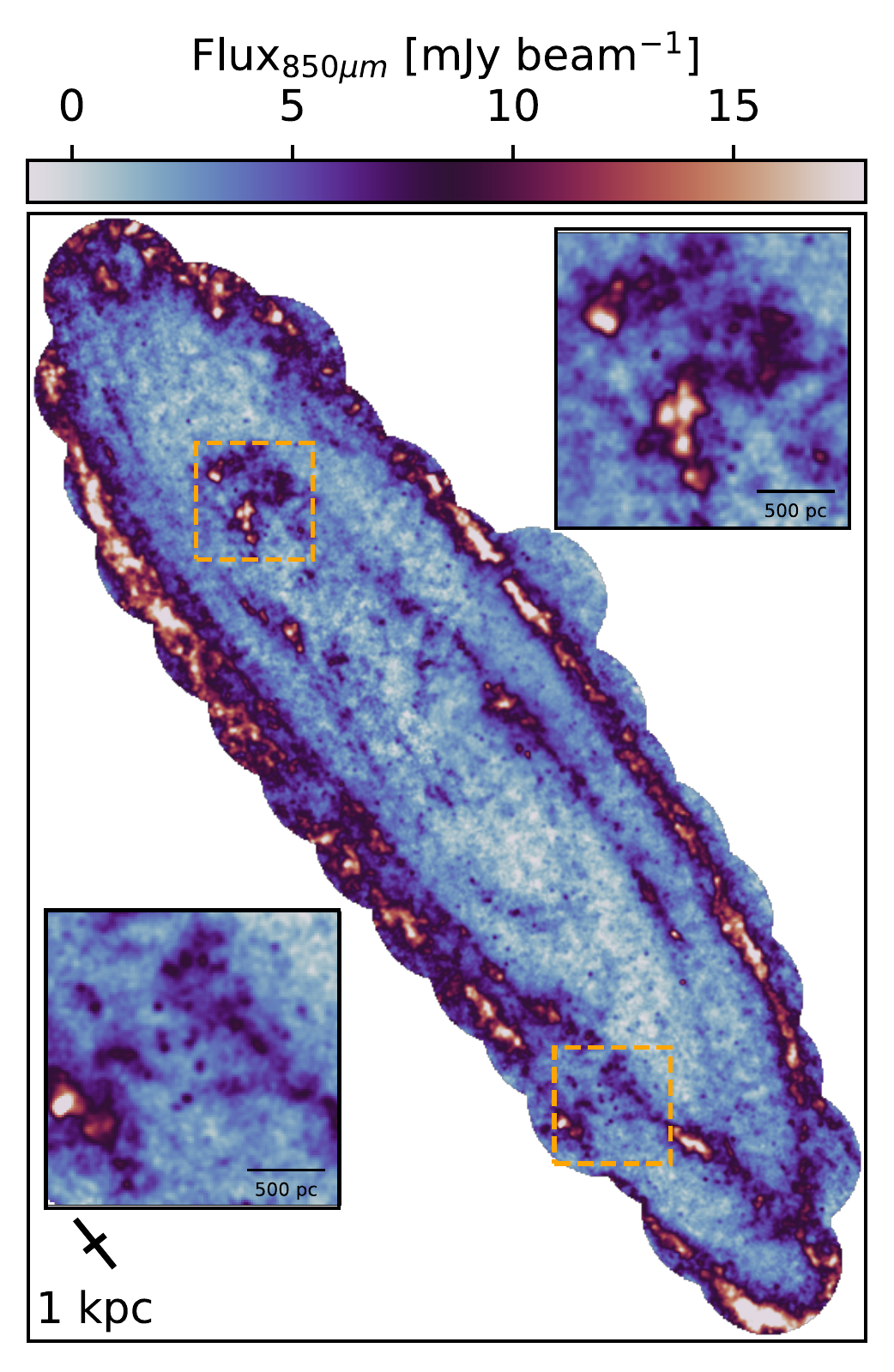}
\caption{{(Left:) SCUBA2- HASHTAG 850~$\mu$m survey on M31 (\citealt{Smith2021}).
The orange contour highlights the field of view of the ultra-deep SCUBA2 850~$\mu$m survey.
(Right:) Jiao et al.(submitted) ultra-deep SCUBA2 850~$\mu$m survey, which identified more than 500~GMCs, including about one third inter-arm GMCs.
Large-scale missing flux was recovered by combining SCUBA-2 850~$\mu$m data with deconvolved Planck 353 GHz observations using the J-comb algorithm \citep{jiao2022SCPMA}.
}\label{fig:GMC-Census}}
\end{figure*}

In the Milky Way, most GMCs are found along spiral arms. Due to the edge-on perspective and the ambiguity of kinematic distances, it is very difficult to obtain a panoramic view of the molecular gas distribution, and especially hard to identify and study inter-arm clouds. Recently, Jiao et al. (submitted)  carried out ultra-deep 850~$\mu$m continuum mapping observations towards M31 using JCMT-SCUBA2 (see Figure~\ref{fig:GMC-Census}). Achieving an RMS noise of $<$1.5~mJy~beam$^{-1}$ that is close to the confusion limit at 850~$\mu$m, they successfully detected more than 500 GMCs within the inner 10 kpc of the stellar disk, including about 190 inter-arm GMCs that had never been discovered before. These inter-arm GMCs are slightly less massive, colder, and more quiescent than on-arm GMCs, as confirmed by a series of follow-up observations. This finding suggests different evolutionary pathways. These detections on on-arm and inter-arm GMCs provide a complete census of GMCs within the 10 kpc stellar disk of M31, reaching masses as low as $2 \times 10^{4} M_{\odot}$, which approaches the traditional lower mass limit of $\sim 10^{4} M_{\odot}$ for GMCs. 
This represents the first complete GMC census of the major disk of a spiral galaxy down to such a low mass threshold, making the statistical studies of GMC properties across a galaxy possible.

Three key factors make this GMC census possible: 1) Cold dust emission at long wavelengths (850~$\mu$m) is the most sensitive tracer of GMCs in galaxies, particularly for colder, less massive GMCs. 2) The distance of M31 matches the 15 m aperture of JCMT, allowing GMCs to be marginally or nearly resolved at 850~$\mu$m (beam size $\sim$50 pc), while most GMC complexes in PHANGS-ALMA (\citealt{Leroy2021}) can not be resolved into individual GMCs. 3) The confusion-limited deep survey enables the detection of cold, inter-arm GMCs. For comparison, most of these inter-arm GMCs were missed in the larger but 3 times shallower SCUBA2 survey of M31, the HASHTAG survey (\citealt{Smith2021}).

Based on the ultra-deep JCMT-M31 survey, we obtained a complete census of GMCs within the inner 10 kpc of the stellar disk of the spiral galaxy M31. Obvious property differences are found between on-arm and inter-arms GMCs. The on-arm GMCs exhibit physical properties similar to those in the Milky Way, whereas inter-arm GMCs may either form by drifting out of the arms or form locally, and seem to be strongly influenced by their surrounding environment. These inter-arm GMCs also show diversity with different levels of gas content and dust-to-gas ratios. These clouds are therefore very good probes to explore how different environment influences the formation and evolution of GMCs.

Following a strategy similar to the JCMT survey, XSMT-15m can be used to cover more galaxies in the Local Group, whose distances are less than or comparable to M31. Using the KID camera in 850 and 1100~$\mu$m confusion-limited deep surveys, we will be able to obtain the complete census of GMCs across these systems. Given that the current JCMT survey covers only the inner 10~kpc of M31, yet outer-disk GMCs may exhibit very different properties due to low metallicities and distinct environments. We therefore propose to extend coverage with XSMT-15m to the outer part of M31 from 10 to 15~kpc. This survey will, for the first time, provide a comprehensive view of GMCs across diverse environments and galaxy types in the Local Group. Such a census is uniquely feasible with a large-aperture single-dish telescope of sufficient sensitivity. While interferometers such as ALMA are not suited for wide-area surveys, they will serve as ideal follow-up facilities to resolve the internal structures of the detected GMCs. 

Based on our experience in M31, detecting GMCs through cold dust continuum emission is far more efficient than gas mapping surveys such as CO. For example, CO detections of cold GMCs require large facilities like the IRAM 30m and integration times of tens of minutes per cloud. In contrast, an XSMT-15m continuum survey will provide the most efficient means of identifying these cold GMCs. Once cataloged through dust emission, high-priority GMCs can then be followed up with spectral line observations (e.g., CO), building an unprecedented GMC database for the Local Universe. CO (1--0) observations with IRAM 30m are already underway toward a large sample of inter-arm clouds in M31, and we plan to propose additional CO measurements for GMCs newly identified in other galaxies. Further follow-up observations will include high-resolution HI imaging with the VLA combined with the FAST–M31 deep HI survey, as well as mid-infrared studies with JWST for selected high-priority targets.

\subsection{Objectives}
The key science goal for this project is to establish the first comprehensive GMC database for the local Universe, enabling us to address fundamental questions: What types of GMCs exist in nearby galaxies? How are they distributed within spirals and other galaxy types? How does the large-scale environment influence their formation and evolution?

Our specific objectives include:
\begin{itemize}
    \item[1.] Conduct a GMC census in Local Group galaxies by identifying GMCs with mass greater than 3$\times 10^{4} M_{\odot}$ in spiral galaxies and a higher mass limit for dwarf galaxies, creating the first GMC database for the local universe.

    \item[2.] Characterize GMC properties and investigate how they vary with galactic environment (e.g., spiral arms, inter-arm regions, galaxy outskirts, and low-metallicity systems).
\end{itemize}





\section{EG: Dust in Nearby Galaxies}

\textbf{Plain summary for non-experts:} \textit{Dust plays a crucial role in the lifecycle of gas in the interstellar medium (ISM) and in the evolution of galaxies. Dust continuum emission dominates the spectral energy distribution (SED) of galaxies from the infrared to the submillimeter. High–spatial-resolution, wide-wavelength-coverage dust SEDs are essential for a thorough understanding of dust properties across different environments on small scales. XSMT-15m, with its KID camera operating at three frequencies, will add critical data points to complement Herschel and Spitzer infrared observations, enabling more robust dust SED modeling and allowing us to investigate variations in dust properties across galaxies.}

\subsection{Motivation}

Although dust accounts for $\lesssim$1\% of the interstellar medium (ISM), it plays a significant role in obscuring starlight and re-emitting energy in the infrared band \citep{Galliano2018}. Dust is crucial in various processes within galaxies: heating the gas in photodissociation regions (PDRs) through the photoelectric effect \citep{Draine1978}, cooling gas via infrared emission, providing surfaces for H$_2$ formation, and shielding molecules from dissociating radiation \citep{Gould1963}. A thorough understanding of dust properties is essential for studying the life cycle of gas in the ISM and the evolution of galaxies.

Our understanding of dust properties primarily comes from studies of the MW. Infrared observations within the MW have revealed radial variations in dust properties, and these studies have laid the foundation for the development of dust models \citep{Draine2003}. However, these models are limited by the relatively narrow range of environmental conditions encountered within our galaxy. An increasing number of nearby galaxy studies provide a wider range of environmental conditions, including extreme cases such as extremely metal-poor environments, which place valuable constraints on fundamental dust models \citep{Stevens2005,Smith2012,Draine2014,Tabatabaei2014,Hunt2015,vanderGiessen2024}. 

Radial variations of dust properties (e.g., dust temperature, $T_{\rm dust}$) have also been observed in extragalactic systems. Many of these studies use a modified blackbody (MBB) model to fit the infrared SEDs, using a fixed emissivity index ($\beta$). However, even in studies exploring the variation of $\beta$ \citep{Galametz2012,Hunt2015}, the fitting of dust SED is limited by the wavelength coverage ($\lesssim 500\ \mu m$), the angular resolution, and the inherent mixing of dust components under different physical conditions. Longer wavelengths are crucial for modeling the properties of cold gas. The JINGLE legacy survey, conducted with JCMT, extended the infrared SED of 193 galaxies in the local Universe to longer wavelengths with 850 $\mu m$ continuum measurements \citep{Saintonge2018}. However, only one measurement between $500\ \mu m \lesssim \lambda \lesssim 1000\ \mu m$ is insufficient to robustly determine the dust emissivity $\beta$ or precisely constrain the dust masses and temperatures. The large IRAM/NIKA2 program IMEGIN \citep{Ejlali2024,Katsioli2024,Ejlali2025}, which performs 1.2 and 2~mm continuum measurements of nearby galaxies, faces contamination from free-free emission, complicating efforts to isolate the dust signal.

XSMT-15m will be able to cover 230 GHz (1.3 mm), 345 GHz (870 $\mu m$), and 460 GHz (650 $\mu m$) bands simultaneously, which will supplement the infrared SED with two spatially-resolved measurements for nearby galaxies at $500\ \mu m \lesssim \lambda \lesssim 1000\ \mu m$. 

\subsection{Objectives}

\begin{figure*}[!htbp]
    \centering
    \includegraphics[width=0.8\textwidth]{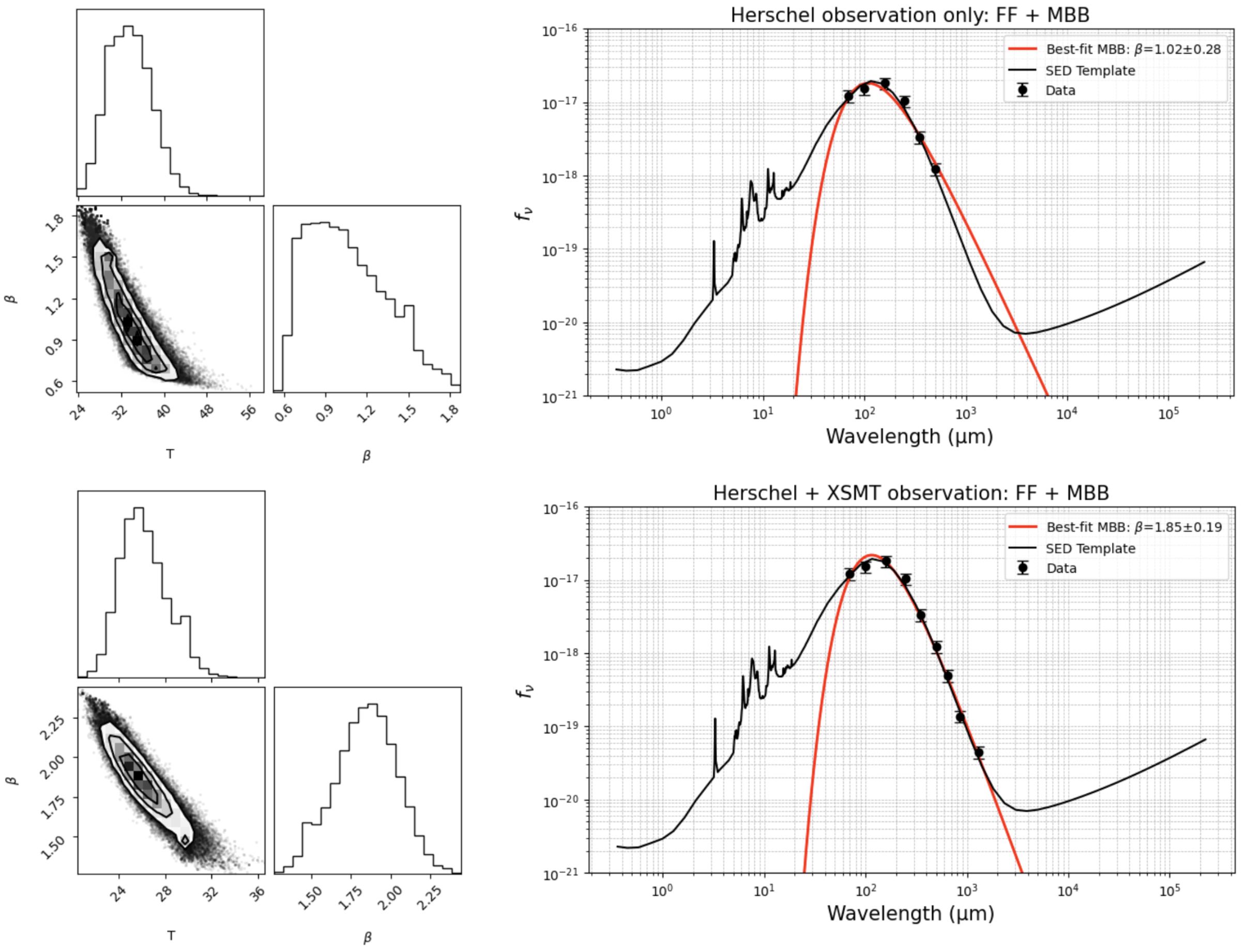}
    \caption{Fitting of the dust SED with a single MBB plus free–free emission model, shown for the Herschel-only dataset and for the combined Herschel+XSMT-15m dataset.}\label{fig:dust_sed_fitting}
\end{figure*}

Many nearby galaxies observed in surveys such as KINGFISH and JINGLE already have extensive multi-band coverage, from the ultraviolet to the radio, including spatially-resolved gas measurements. These galaxies provide ideal samples for detailed studies of dust properties. With the wavelength coverage of the XSMT-15m continuum camera, we aim to achieve the following scientific objectives:

\textbf{1. Constrain dust properties and investigate their variations in nearby galaxies.} 
Dust in the ISM is a complex mixture of grains with varying physical conditions. The common assumption of a fixed $\beta$ when fitting dust SEDs introduces biases in the inferred dust properties \citep{Galliano2018}. Even in studies exploring variations of $\beta$ based on dust SED limited to $\lambda \lesssim 500\ \mu m$, the derived emissivity index is degenerate with temperature mixing. At $100\ \mu m \lesssim \lambda \lesssim 1000\ \mu m$, only large amounts of very cold dust with $T\lesssim 10\ K$ could bias the estimates. Mid-IR emission of galaxies is strongly contaminated by stellar light, while Herschel's far-IR bands (350 $\mu m$, 500 $\mu m$) suffer from poor angular resolution, preventing spatially-resolved dust studies in nearby galaxies. 
The advent of single-dish telescopes with large FoVs, such as NIKA2 at IRAM-30m, has made spatially-resolved continuum mapping feasible for extended targets. However, continuum emission at 1.3 and 2~mm is often contaminated by free-free emission, synchrotron radiation, and submillimeter excess. Therefore, the measurements of dust emission at $500\ \mu m \lesssim \lambda \lesssim 1000\ \mu m$ are crucial for providing more accurate constraints on the intrinsic emissivity, mass, and temperature of dust. In Figure~\ref{fig:dust_sed_fitting}, we present the model fitting of the dust SED. Using the SED templates from \citet{Dale2014}, we generate mock observational data for the Herschel and XSMT-15m bands, assuming an uncertainty of 0.8\,dex for all bands. We fit the dust SED with a single MBB model plus a free-free emission component, considering two cases: Herschel-only observations and combined Herschel+XSMT-15m observations. The best-fitting results are shown in the figure. With the inclusion of XSMT-15m continuum mapping at 650\,$\mu$m, 850\,$\mu$m, and 1.3\,mm, the model fitting significantly improves, and the best-fit $\beta$ parameter exhibits a reduced uncertainty. XSMT-15m will thus enable spatially-resolved measurements at 650\,$\mu$m, 870\,$\mu$m, and 1.3\,mm, allowing robust determinations of the dust emissivity index $\beta$, and placing precise constraints on dust mass and temperature. This capability will yield valuable insights into the variation of dust properties across a wide range of environments in nearby galaxies.

\begin{figure*}[!htbp]
    \centering
    \includegraphics[width=0.8\textwidth]{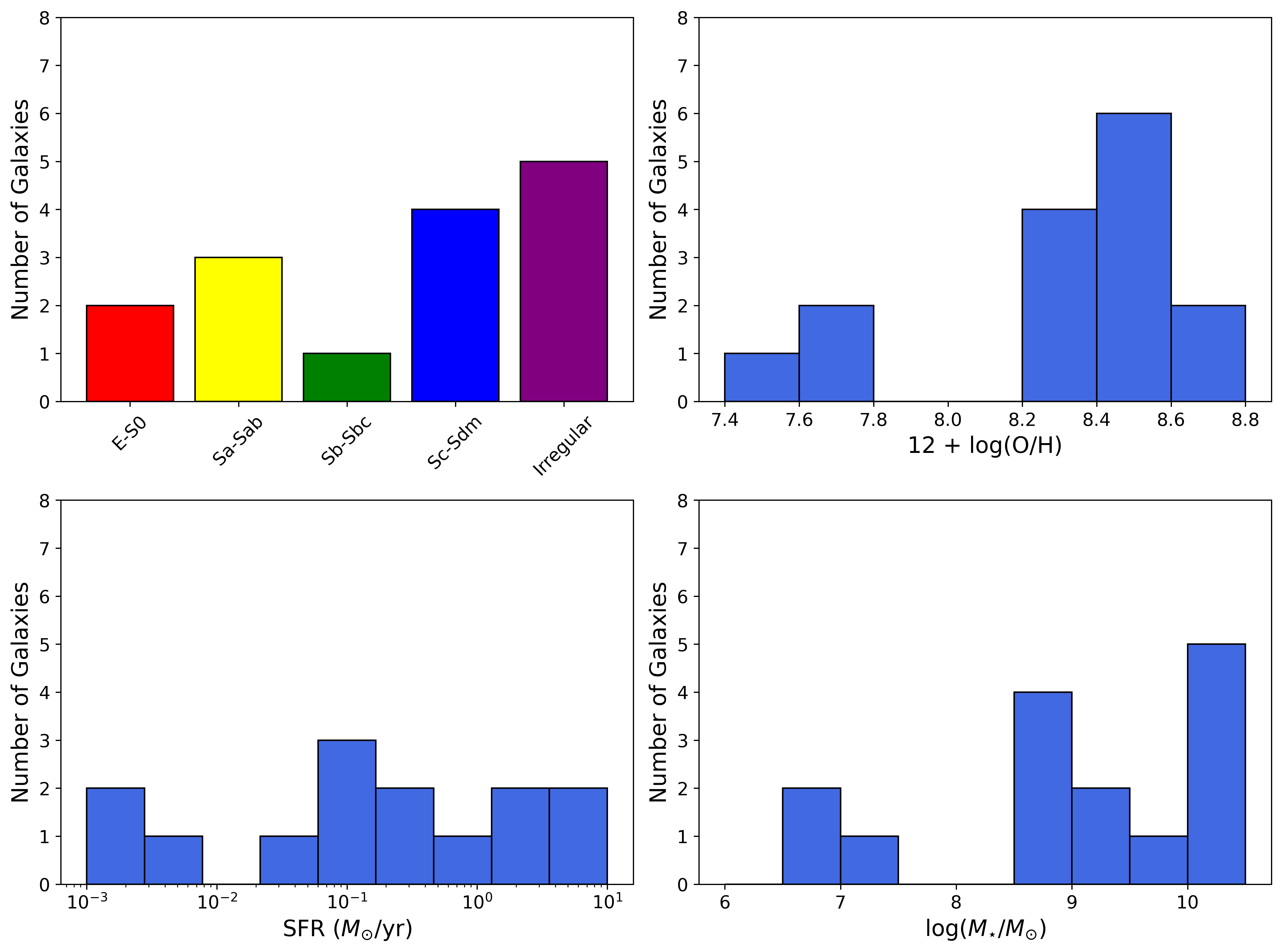}
    \caption{Distributions of morphological type, metallicity, star formation rate (SFR), and stellar mass for the selected sample.}
    \label{fig:kingfish_sample}
\end{figure*}

\textbf{2. Disentangle spatially-resolved galaxy SEDs from dust, free-free, and synchrotron components.}
Dust emission dominates the infrared SED of galaxies, while the contributions from free-free and synchrotron emission become more significant at longer wavelengths ($\lambda > 1\ \rm mm$). With the full Herschel and Spitzer datasets, combined with spatially-resolved observations from XSMT-15m at $650\ \mu m$ and $870\ \mu m$, robust models of dust emission can be established. The contributions from free-free and synchrotron emission can be constrained using XSMT-15m's continuum measurements at 1.3 mm (230 GHz). Thus, it will be feasible to separate spatially-resolved galaxy SEDs into dust, free–free, and synchrotron components.

\textbf{3. Investigate the evolution of the dust-to-gas mass ratio within galaxies.}
Dust continuum emission has been proposed as an alternative tracer of gas mass in galaxies \citep{Eales2012,Magdis2012,Scoville2016}, providing a substitute for the CO ($J=1-0$) measurements. If dust properties are universal, or if their variations with redshift and environment are well understood, this method could become a powerful tool for systematic surveys of gas masses in high-redshift galaxies. XSMT-15m continuum mapping of dust emission will provide more accurate constraints on dust masses across the entire disks of various types of nearby galaxies. When combined with gas observations, this will allow us to determine the dust-to-gas mass ratios in different environments, shedding light on their variation within galaxies and across galaxy populations. 

\textbf{4. Study the submillimeter excess in the SED of galaxies.}
Previous studies have identified an excess of emission at submillimeter wavelengths in the SED of low-metallicity galaxies \citep{Galliano2003,Galliano2005,Galametz2009,Galametz2011,Galliano2011,Dale2012}. The origin of this submillimeter excess remains unclear, posing a challenge to our understanding of dust properties in low-metallicity environments. Several possible explanations for the excess have been proposed: (i) very cold dust in dense clumps \citep{Galliano2003,Galliano2005,Galametz2009}; (ii) temperature-dependent submillimeter emissivity \citep{Meny2007}; (iii) spinning dust emission \citep{Bot2010,Planck2011}. To better understand the origin of this submillimeter excess, spatially-resolved observations with broader wavelength coverage (especially $\lambda \gtrsim 500\ \mu m$) are needed to explore dust properties across different environments on small scales. The three XSMT-15m bands are affected differently by the submillimeter excess, allowing the instrument to distinguish or rule out its possible origins.

\vspace{10mm}

\begin{table*}[!htbp]
\centering
\caption{Approximate mapping information of NGC~2976}\label{tab:NGC2976}
\begin{tabular}{c|c|c|c}
    \hline
    Source & \multicolumn{3}{c}{NGC~2976} \\
    \hline
    RA, DEC   & \multicolumn{3}{c}{146.813810, 67.916679} \\
    $\Delta_{x}\times \Delta_{y}$ [arcmin$^2$] & \multicolumn{3}{c}{17.6$\times$5.9} \\
    pwv [mm] & \multicolumn{3}{c}{$\lesssim1mm$} \\
    Elevation [deg] & \multicolumn{3}{c}{50} \\
    \hline
    band & 230GHz (1.3mm) & 345GHz (850$\mu m$) & 460 (650$\mu u$) \\
    \hline
    Estimated $\Sigma$(IR) (MJy/sr)  & 0.30 & 0.97 & 2.25\\
    $t_{beam}$ (hr) & 5 & 2 & 3 \\
    Achieved S/N & 5.9 & 9.2 & 8.3\\
    \hline
    Total observing time [hours] & \multicolumn{3}{c}{20.9} \\
    \hline
\end{tabular}
\end{table*}

Our preliminary sample is selected from the KINGFISH sample \citep{Kennicutt2011,Hunt2015}, which possesses IR data ($\lesssim$ 500 $\mu m$) from space-based observatories such as Spitzer and Herschel. We select galaxies located in the northern celestial hemisphere, with disk size larger than 2~$arcmin^2$ and final mapping area smaller than 120~$arcmin^2$. This leaves us a sample of 15 nearby galaxies. The distributions of morphology type, metallicity, SFR, and stellar mass are shown in Figure~\ref{fig:kingfish_sample}. Table~\ref{tab:NGC2976} lists the approximate mapping information for NGC~2976. The total observing time for this sample is about 300 hr, including overheads, pointing adjustments, map coverage, the fraction of good pixels, and mapping speed.





\section{EG: The Circumgalactic Medium of Nearby Galaxies }

\textbf{Plain summary for non-experts:} \textit{Galaxies are dynamic systems that continuously exchange gas with their surroundings through an extended region known as the circumgalactic medium (CGM). This vast halo regulates galaxy growth, star formation, and long-term evolution. Once thought to consist primarily of hot gas, the CGM is now known to contain substantial amounts of cold gas as well, although this component remains poorly characterized. This project aims to map the distribution of cold molecular gas around nearby galaxies with XSMT-15m. The results will provide new insights into the processes by which galaxies recycle baryons and will help to address one of the outstanding questions in cosmology: the origin and location of the Universe’s missing baryons.}

\subsection{Motivation}

Galaxies in modern astrophysical research are not just basic gravitationally bound systems, but complex ecosystems of physical processes. They are embedded in dark matter halos and host supermassive black holes at their centers. Spanning tens of kiloparsecs (kpc), galaxies are dominated by stars and ISM in their disks, while dark matter and the circumgalactic medium (CGM) prevail beyond the disk. The CGM serves as a dynamic interface, mediating the exchange of material between galaxies and the surrounding dark matter halos and large-scale structures. Its extent can reach hundreds of kpc. For example, the Milky Way's CGM spans nearly 200~kpc in diameter. 

Over billions of years, galaxies accrete primordial gas from the cosmic web. This gas enters the halo and disk, gradually losing angular momentum before settling into central regions. Along the way, it forms dense molecular clouds and star-forming regions under the influence of self-gravity, turbulence, and shocks. These processes trigger feedback from star formation. When some gas penetrates the disk and approaches the supermassive black hole, strong accretion occurs, igniting active galactic nuclei (AGN). AGN release intense radiation, powerful jets, and massive outflows, which are known as AGN feedback. The interplay between star formation and AGN feedback regulates the galactic ecosystem, with material either expelled into the CGM or cycling back into the disk like a fountain.

\begin{figure*}[!htbp]
    \centering
    \includegraphics[width=0.95\textwidth]{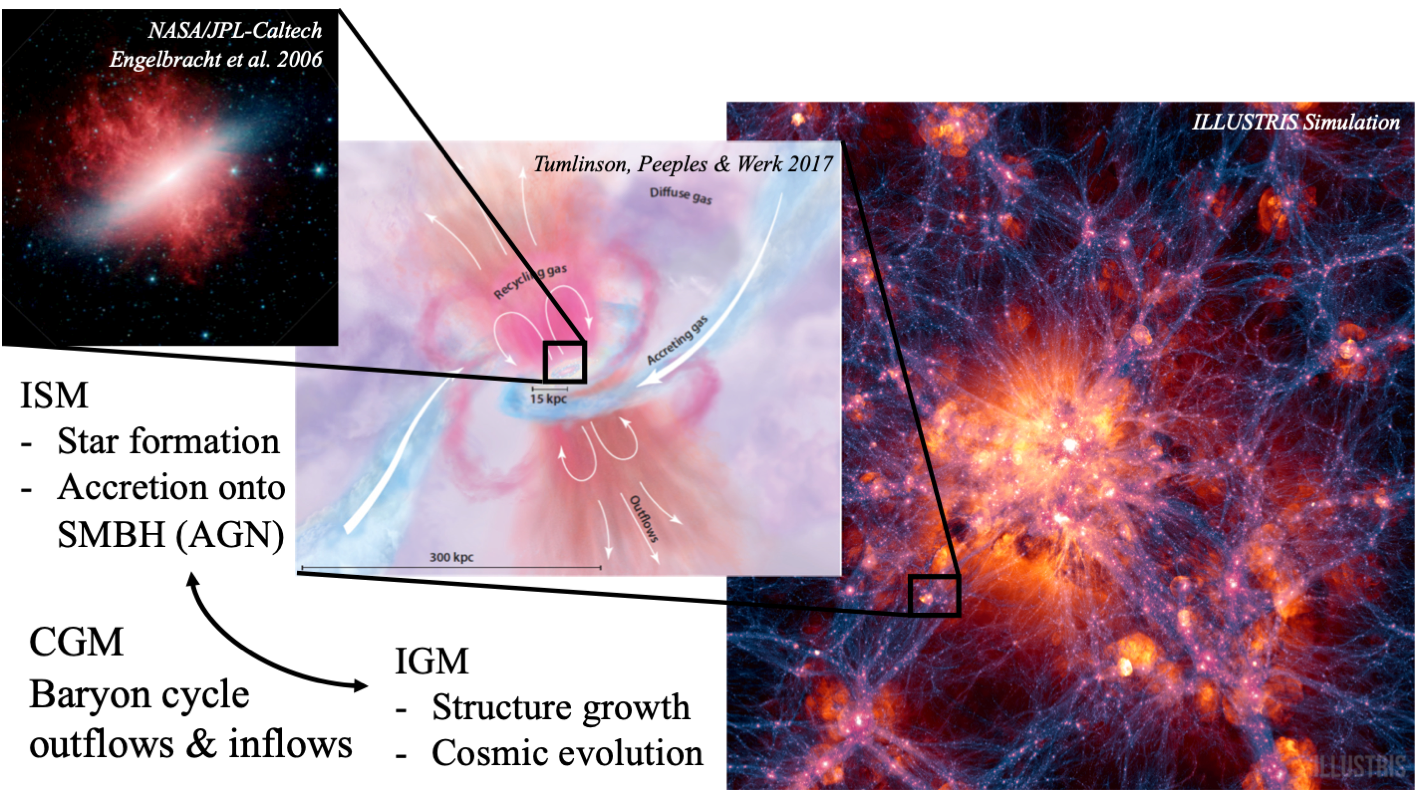}
    \caption{Gas cycling driving the formation and evolution of galaxies, with the CGM connected to the ISM and the Intergalactic Medium (IGM), as well as the larger cosmic network structure. Its fundamental properties are directly related to the process of baryon cycling. Adapted from \citet{Cicone2019} with permission from the corresponding author.}
    \label{fig:CGM}
\end{figure*}

Therefore, the CGM serves as the hub connecting galaxies to their growth environment and is a key component of the galactic ecosystem (Figure~\ref{fig:CGM}). Its physical properties, spatial distribution, and dynamical state impose strong constraints on galaxy formation and evolution theories, while also providing crucial evidence for solving the long-standing problem of the missing baryons in the Universe \citep{Tumlinson2017}.

Our current understanding of the CGM is still very limited. The traditional view holds that CGM is dominated by diffuse, hot gas (with temperatures as high as 10$^6$~K). This ``hot'' gas itself is difficult to probe directly and can only be detected through absorption lines against background quasars or high-ionization emission lines. The former method has limited spatial coverage, while the latter method is only sensitive at lower redshifts. Future millimeter/submillimeter observations of the CGM will help us better understand this hot gas using the Sunyaev-Zel'dovich (SZ) effect, in which hot electrons imprint a characteristic scattering signature on background photons. 

Recent breakthroughs have revealed the CGM’s multiphase nature: in addition to hot gas, it also contains significant reservoirs of cold neutral and molecular gas with $T < 10^4$ K \citep{Cicone2019}. Yet, the properties and distribution of these cold phases remain poorly constrained. Single-dish millimeter/submillimeter telescopes are the only means to obtain complete information about them. The primary goal of this project is to test galaxy formation and evolution models from a gas-circulation perspective by characterizing the cold CGM around galaxies at low to intermediate-redshift ($z \sim 0.2$), thereby unveiling the coupled evolution of galaxies and their CGM.

While the C$^+$ line can effectively detect warm neutral/atomic gas in the CGM \citep{Cicone2015}, CO lines come from the cold molecular phase \citep{Emonts2016}. Therefore, high-resolution CO spectroscopy in the millimeter/submillimeter wavelengths is a primary method for directly detecting cold CGM gas. However, interferometric observations are limited by their baseline lengths and can only recover structures below a certain scale (typically $\lesssim$10\arcsec). The size of the CGM can reach up to $\sim$100 kpc, which corresponds to angular scales of $\sim$110\arcmin\ for nearby galaxies at a distance of 3 Mpc, and $\sim$32\arcsec\ for galaxies at a redshift of $\sim$0.2, and even as small as $\sim$13\arcsec\ for galaxies at a redshift of 2. Thus, for low to intermediate redshift galaxies, interferometric observations will filter out most of the extended CGM emission, leaving single-dish telescopes as the only viable option. Moreover, single-dish telescopes can be equipped with advanced multi-beam receivers, which compensate for their smaller collecting area and enable efficient mapping of large-scale structures. 

\subsection{Objectives}

Key science goals include:

\begin{itemize}
    \item Characterize the ``multiphase" nature of the CGM using CO spectroscopy, linking the cold molecular component to the broader baryon cycle and testing predictions of galaxy formation models.
    \item Map the extended CGM on large scales in representative nearby galaxies to quantify its mass, spatial distribution, and kinematics, thereby constraining gas accretion and feedback processes in galaxy evolution. 
\end{itemize}

\section{EG: Thermal and Kinematic Sunyaev-Zel'dovich Effects of Galaxy Clusters}

\textbf{Plain summary for non-experts:} \textit{As the largest gravitationally-bound systems in the Universe, galaxy clusters serve as powerful probes of cosmology and laboratories for studies of galaxy evolution in dense environments. They contain a large amount of hot gas, which can be probed through the SZ effect -- the distortion of the Cosmic Microwave Background (CMB). XSMT-15m with a multi-color KID camera in three frequencies is unique in probing the SZ effect, especially the kinematic SZ effect.}

\subsection{Motivation}

Galaxy clusters are the largest self-gravitationally bound systems in the Universe. They contain large quantities of gas, member galaxies, and dark matter. In the hierarchical model of cosmic structure formation, galaxy clusters form and evolve through the merging of galaxy groups and clusters and continuous accretion of surrounding gas \citep{Kravtsov2012}. Galaxy clusters offer powerful probes of the large-scale structure, velocity field of the Universe, and cosmological parameters \citep{Carlstrom2002,Maccio2007,Allen2011,Ludlow2012,Planck2016}. However, these cosmological constraints with galaxy clusters are affected by physical processes, such as non-thermal pressure, shocks, AGN feedbacks, and the clusters' dynamical and virialization states that can cause departures from thermal equilibrium and bias our measurements.

Hot gas, the most abundant baryonic component in galaxy clusters, reflects the clusters' accretion and merging processes. While the hot gas in the intracluster medium (ICM) can be observed through X-ray emission, the SZ effects provide a unique way to probe the hot gas because it is independent of redshift. The SZ effect consists of the thermal-SZ (tSZ) effect \citep{Sunyaev1972}, which is the spectral distortion of the CMB caused by the scattering of the CMB photons off the high-energy electrons, and the kinematic-SZ (kSZ) effect \citep{Sunyaev1980}, arising from the CMB Doppler shift induced by the Doppler effect of the cluster bulk velocity on the scattered CMB photons. The tSZ effect is proportional to the line-of-sight (LOS) integral of the electron pressure \citep{Hasselfield2013,Planck2016}, while the kSZ effect is related to the integrated LOS electron density and the gas velocity with respect to the CMB reference frame. Currently, there's a scarcity of high-resolution and high-sensitivity tSZ observations in merging galaxy clusters. Moreover, only one galaxy cluster has spatial-resolved kSZ observation with a signal-to-noise ratio (SNR) greater than 3 (MACS~J0717.5+3745, \citealt{Adam2017}). This limits our understanding of how galaxy clusters form and evolve. Besides, these kSZ detections of individual clusters are usually based on two-band observations, while there are at least three free parameters even under the optimal conditions, thus additional assumptions with X-ray observations are needed. 

XSMT-15m will be powerful to obtain spatially-resolved tSZ and kSZ maps of galaxy clusters at intermediate redshifts, as the continuum camera on XSMT-15m will be able to cover 230 GHz, 345 GHz, and 460 GHz, which are located around the flux peaks of tSZ and kSZ effects. There are at least three parameters in the analytical expression of the SZ effect ($\frac{\Delta T_{\rm tSZ}}{T_{\rm CMB}}=f(x)y=f(x)\int n_e\frac{k_B T_e}{m_e c^2} \sigma_Tdl$ and $\frac{\Delta T_{\rm kSZ}}{T_{\rm CMB}}=-\tau_e \frac{v_{\rm pec}}{c}$). Thus, at least three measurements are required to gain the $kT$, $\tau$ and $v_{\rm pec}$ of galaxy clusters via the SZ effect. Compared to SZ surveys conducted by other telescopes (e.g., MUSTANG2 at GBT, NIKA2 at IRAM 30M), XSMT-15m offers more frequency bands and a larger FoV, which will be very efficient for mapping the SZ effect. Another advantage of XSMT-15m is that its location possesses good weather conditions with a median PWV of $\sim$0.85~mm in winter. Observations in the high-frequency bands are highly dependent on PWV levels. A few SZ effect surveys and large programs are now being conducted via NIKA2 (150 GHz and 260 GHz), and XSMT-15m is very important in supplementing the observations of the SZ effect at 345 GHz and 460 GHz.

\subsection{Objectives}


Tens of massive galaxy clusters at intermediate redshift (0.5$\lesssim$z$\lesssim$1.0) are well studied with optical and X-ray observations. These clusters show a wide range of dynamical states, from dynamically relaxed clusters (such as MACSJ~1423.8+2404 at z$\sim$0.545), to merging clusters that show clear substructures (such as MS~1054.4-0321 at z$\sim$0.831 and CL~J0152.7-1357 at z$\sim$0.833), as shown in X-ray images in Figure~\ref{fig:MACSJ1424_MS1054_CLJ153}. These targets, especially merging clusters, are valuable sources to study the assembly of galaxy clusters. XSMT-15m will be able to provide large-FoV and spatially-resolved tSZ and kSZ effect maps of these targets. This will enable at least the following scientific objectives:

\begin{figure*}[!htbp]
    \centering
    \includegraphics[width=0.95\textwidth]{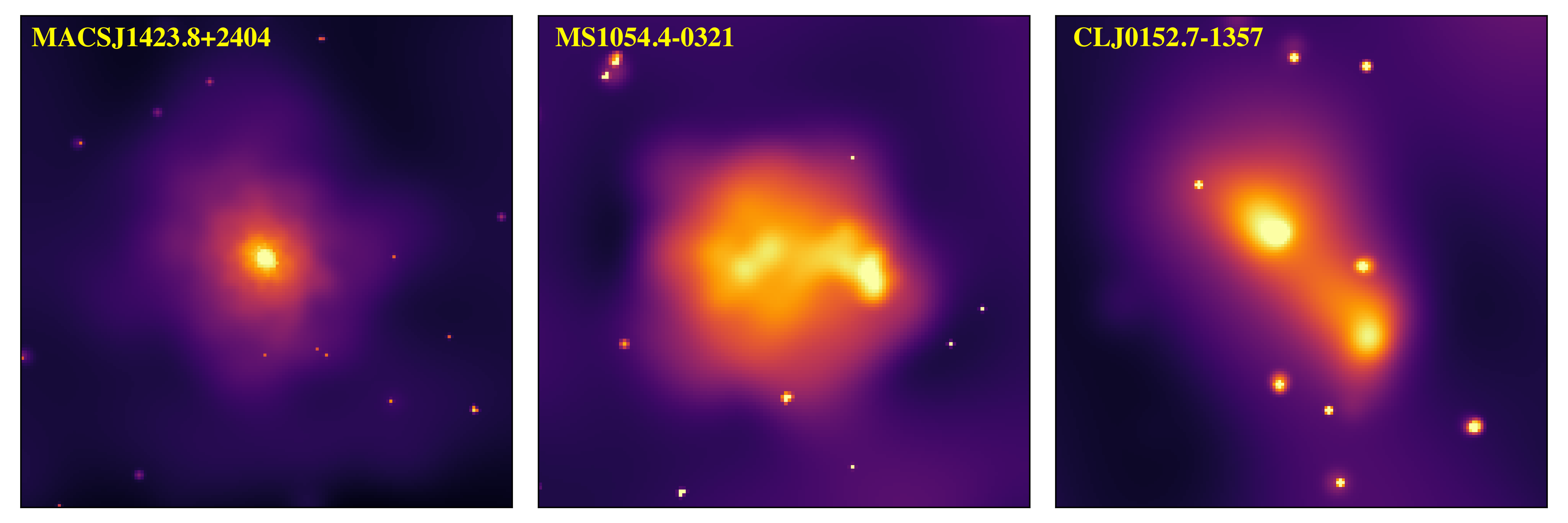}
    \caption{Chandra X-ray images of MACSJ~1423.8+2404 (left panel), MS~1054.4-0321 (middle panel) and CL~J0152.7-1357 (right panel) in 0.5-7.0 keV band.}
    \label{fig:MACSJ1424_MS1054_CLJ153}
\end{figure*}

\textbf{1. Extending the spatially-resolved non-thermal pressure analysis of the galaxy clusters to z$>$0.7 and investigating related physical processes.} 
Accretion, merging, and baryonic feedback processes lead to departures from spherical symmetry in the mass distribution of galaxy clusters, which in turn introduces bias in the estimation of key physical properties such as mass, centroid, and peculiar velocity \citep{Lee2018,Yuan2020}. As a result, studying the pressure profiles of hot gas is crucial for utilizing galaxy clusters as cosmological probes. Quantifying the non-thermal pressure fraction across a broad range of cluster masses and redshifts can significantly improve the precision of these cosmological constraints. However, the current non-thermal pressure analysis of galaxy clusters—based on X-ray, SZ effect, and lensing data—is limited to fewer than twenty clusters and restricted to redshifts lower than 0.7, with spatial resolutions for SZ observations typically above 1$^{\prime}$ \citep{Sayers2021}. XSMT-15m will extend non-thermal pressure analysis to redshifts greater than 0.8 with higher resolution, enabling a more comprehensive SZ effect survey for clusters across a wider range of masses and redshifts. This will be crucial for calibrating scaling relations, such as the $Y_{\rm SZ}-M$ relation and the concentration-mass relation, which are key to improving cosmological constraints.

\textbf{2. Obtaining spatially-resolved kSZ maps of galaxy cluster under merging at intermediate redshifts and studying the LOS gas motions of substructures.}
The kSZ signal has been detected in only a few galaxy clusters \citep{Sayers2013, Adam2017}, or through statistical measurements \citep{Hand2012, Planck2016a}. To date, MACS~J0717.5+3745 remains the only galaxy cluster with spatially-resolved kSZ measurements with S/N$>$3, limiting our understanding of gas motion in merging clusters \citep{Sayers2013, Adam2017}. XSMT-15m will significantly expand the sample of spatially-resolved kSZ measurements for individual galaxy clusters. Studies of merging clusters have shown that the velocities of subclusters can reach a few thousand km~s$^{-1}$ \citep{Springel2007, Bradac2008, Adam2017}, making systems like MS1054.4-0321 and CLJ0152.7-1357 promising targets for investigating the merger process and internal gas velocity distribution through spatially-resolved mapping of the kSZ effect. Expanding the kSZ detection sample will enhance our understanding of merger dynamics at the substructure level.

\textbf{3. Measuring the bulk motion velocities of galaxy clusters and constraining the structure growth of the Universe:}
The peculiar motion of galaxies and clusters serves as a direct and less-biased tracer of the Universe's structural growth \citep{Peebles1980, Koda2014, Shi2024}. The kSZ effect is a promising tool for measuring the peculiar velocities of galaxy clusters. With its band coverage and high mapping speed, XSMT-15m will allow for systematically capturing the kSZ information of large samples of galaxy clusters. In studies of the kSZ effect in disturbed clusters, both through observations \citep{Adam2017} and cosmological simulations \citep{Du2025}, we can select a suitable sample of galaxy clusters based on their dynamical states and internal baryonic feedback.

\begin{table*}[hbt]
\centering
\caption{Preliminary sample}\label{tab:SZE_sample}
\resizebox{\textwidth}{!}{
\begin{tabular}{lccccccccccc}
\hline
Target & RA & Dec & z & $M_{500}$ & $R_{500}$ & Map Area & $t_{\rm beam}$ & $t_{\rm total}$ & $(v/\sigma)_{reg}$ & $(v/\sigma)_{app}$ \\
 & (deg) & (deg) &  & ($10^{14} M_\odot$) & (arcmin) & (arcmin$^2$) & (hr) & (hr) &  &  \\
\hline
MACSJ1149.5+2223 & 177.39622 & 22.40304  & 0.536 & 18.7 & 4.00 & 8.0$\times$19.7 & 6.0 & 32.0 & 3.1 & 3.3 \\
MACSJ1423.8+2404 & 215.949   & 24.0785   & 0.545 & 6.9  & 2.90 & 5.8$\times$17.5 & 7.5 & 30.0 & 3.0 & 3.4 \\
MACSJ0744.8+3927 & 116.21863 & 39.45759  & 0.698 & 12.5 & 2.95 & 5.9$\times$17.6 & 7.0 & 29.3 & 3.1 & 3.5 \\
MS1054.4-0321    & 164.25125 & -3.62472  & 0.831 & 14.2 & 2.70 & 5.4$\times$17.1 & 7.6 & 30.0 & 3.0 & 3.4 \\
CLJ0152.7-1357   & 28.16625  & -13.97417 & 0.833 & 10.0 & 2.10 & 4.2$\times$16.0 & 9.5 & 32.4 & 3.1 & 3.7 \\
\hline
\end{tabular}
}
\end{table*}

\begin{figure}[H]
    \centering
    \includegraphics[width=0.49\textwidth]{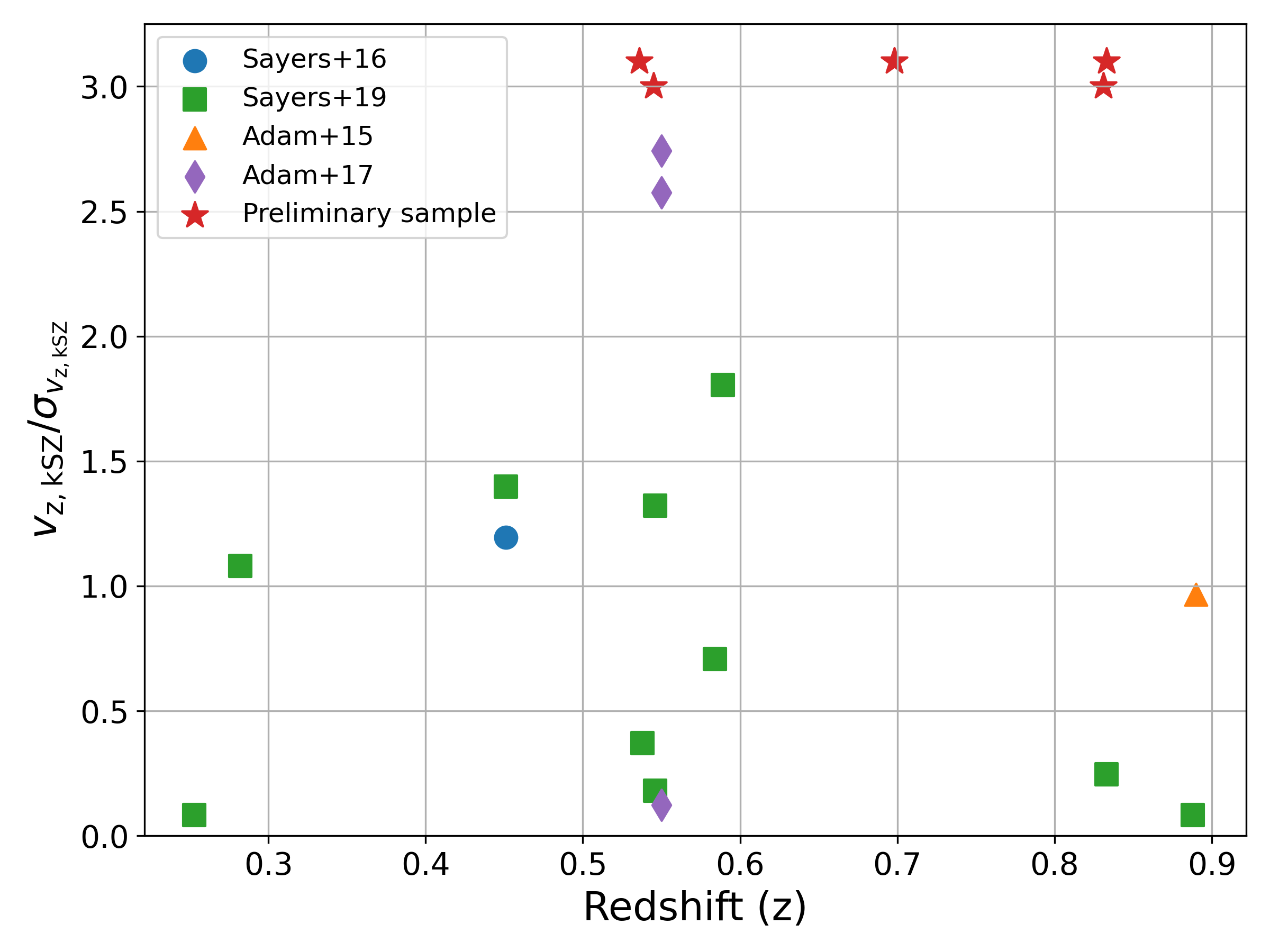}
    \caption{This figure shows the ratio of LOS peculiar velocities ($v_{\rm z,kSZ}$) and uncertainties ($\sigma_{v_{\rm z,kSZ}}$) based on the kSZ effect from literature. With the proposed observation time listed in Table~\ref{tab:SZE_sample}, the reached $v_{\rm z,kSZ}/\sigma_{v_{\rm z,kSZ}}$ of our proposed sample are shown by red stars.}
    \label{fig:ksz_v_sigma_redshift}
\end{figure}

Our preliminary sample is listed in Table~\ref{tab:SZE_sample}. This sample contains galaxy clusters with different redshifts and in different dynamic states. The map area is roughly designed based on the observation guidelines of NIKA2. $t_{\rm beam}$ is the integration time per beam, which determines the sensitivities. Since XSMT-15m will only observe two bands simultaneously, and the high-frequency observation could be severely affected by the atmospheric absorption, we roughly set the 0.5 $t_{\rm beam}$ for 230~GHz, 0.5 $t_{\rm beam}$ for 345~GHz, and $t_{\rm beam}$ for 460~GHz, respectively. $t_{\rm total}$ is approximately estimated based on the NIKA2 observation guidelines, accounting for overheads, map area, the fraction of good pixels, mapping speed, etc. Only few galaxy clusters are reported with LOS peculiar velocities obtained via the kSZ effect, shown in Figure~\ref{fig:ksz_v_sigma_redshift}. \cite{Sayers2019} reported $v_{\rm z, kSZ}$ of ten massive galaxy clusters, which show an ensemble-mean $\langle v_{\rm z} \rangle$ of 430$\pm$210~km~s$^{-1}$. We assume both the LOS approaching or regressing bulk motion (relative to CMB) of galaxy clusters $\gtrsim$400~km~s$^{-1}$. With the proposed observation time, $v_{\rm z,kSZ}/\sigma_{v_{\rm z,kSZ}}$ of our sample, with resolution smoothed to 30$^{\prime\prime}$, are shown by the red stars in Figure~\ref{fig:ksz_v_sigma_redshift}.

\vspace{5cm}

\vspace{2cm}

\section{EG: Galaxy Evolution across the Cosmic Epoch}


\textbf{Plain summary for non-experts:} \textit{From the first galaxies that formed after the Big Bang to the large spirals and ellipticals we see now, galaxies have grown and changed over billions of years. By studying galaxies at different distances, astronomers look back in time to trace how they formed stars, merged, and built larger structures. XSMT-15m will observe northern deep fields in synergy with JWST, Euclid, and CSST to understand galaxy evolution and early Universe --- especially, by revealing the obscured early galaxies, identifying infant proto-clusters, and tracing their formation histories through cosmic time.}

\subsection{Motivation}

\begin{figure*}[htb]
    \centering
    \includegraphics[width=0.95\textwidth]{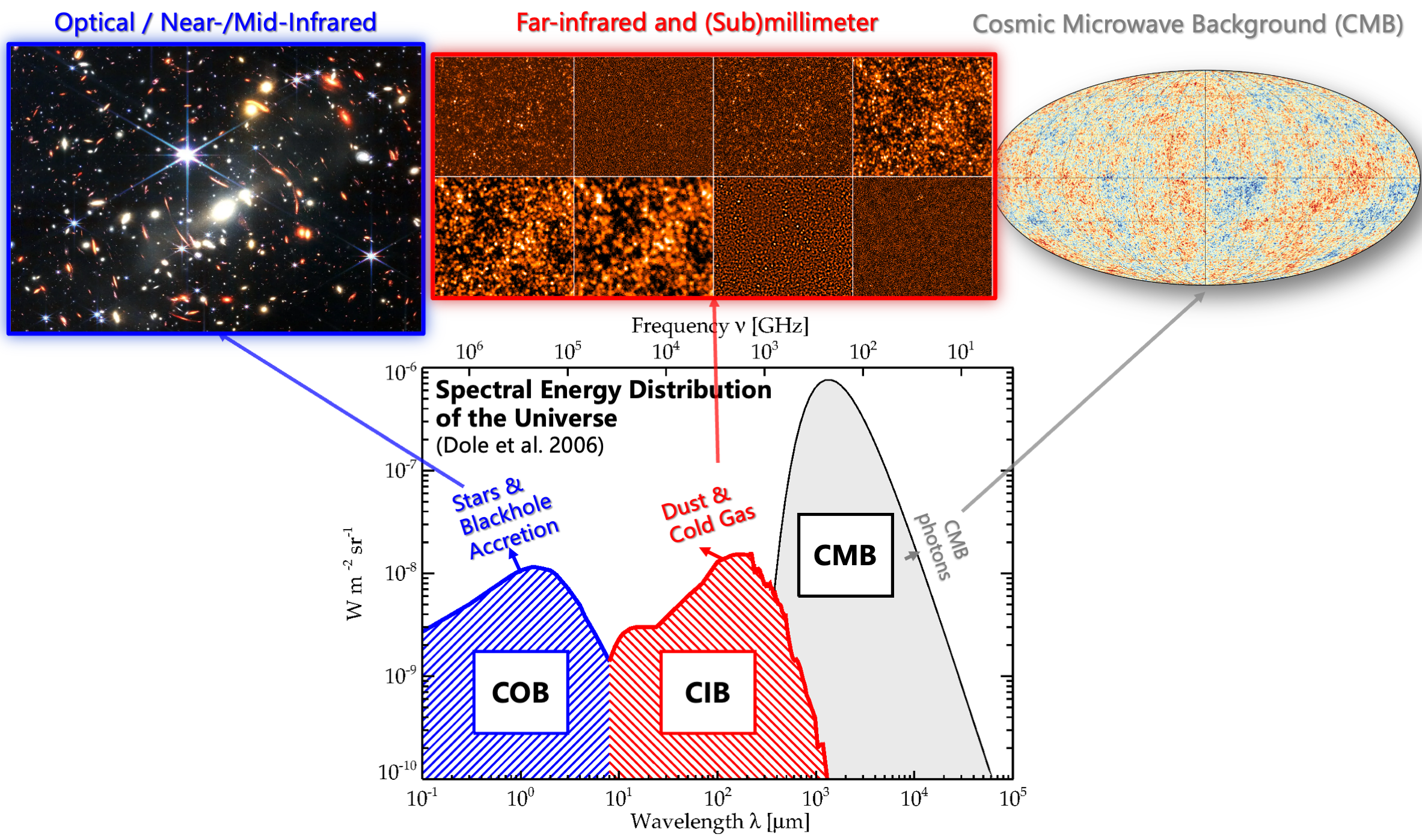}
    \caption{Deep fields and cosmic background radiation. Top panels are: \textit{JWST} deep field at optical and near-/mid-infrared wavelengths (Image Credit: NASA, ESA, CSA, and STScI), \textit{Herschel} deep field at far-infrared and submm wavelengths (based on the public data from \citealt{Elbaz2011}, \citealt{Roseboom2010}, \citealt{Simpson2019} and \citealt{Schinnerer2007}), and the \textit{Planck} cosmic microwave background map (based on public data from \citealt{Planck2014a}), respectively. Bottom panel shows the total energy distribution of the cosmic optical background (COB), cosmic infrared background (CIB) and cosmic microwave background (CMB), adapted from \cite{Dole2006} with permission from A\&A.}
    \label{fig:Universe-COB-CIB-CMB}
\end{figure*}


Galaxies and (proto-)clusters are building blocks of our Universe. 
Deep-field observations play a crucial role in the study of early galaxies and proto-clusters. By targeting certain areas of the sky with advanced telescopes, one can uncover hidden populations of distant galaxies, including submillimeter galaxies (SMGs; \citealt{Smail1997, Hughes1998, Blain2002}), dusty star-forming galaxies (DSFGs, \citealt{Casey2014}), extremely-red galaxies (HIEROs, \citealt{Huang2011, Wang2016a}), heavily-obscured quasars (\citealt{Eisenhardt2012, Wu2012, Tsai2015}), optically-dark galaxies (\citealt{Franco2018, Wang2019, Barrufet2023}), as well as the precursors of lower-redshift galaxy clusters, which are known as proto-clusters (\citealt{Miley2004, Capak2011, Wang2016b, Sillassen2024, Zhou2024}).

Figure~\ref{fig:Universe-COB-CIB-CMB} shows a schematic diagram of optical and near-/mid-IR deep field, far-IR and (sub)millimeter (submm) deep field, and the CMB, corresponding to the three peaks of the total energy distribution of the Universe: the cosmic optical background (COB), the cosmic IR background (CIB), and the CMB. 
COB mainly consists of emission from stars and black hole accretion disks, whereas CIB mainly comes from dust and cold gas emission in galaxies. 
Deep-field surveys covering optical to submillimeter wavelengths are an indispensable approach to studying COB, CIB, and the underlying physics related to galaxy formation and evolution.

Far-IR/submillimeter deep-field surveys conducted by \textit{Spitzer}, \textit{Herschel}, and ground-based submillimeter telescopes like the JCMT and the South Pole Telescope (SPT) have revealed the peak of cosmic star formation rate density (CSFRD; \citealt{Madau2014}) around the cosmic noon ($z\sim 1$--$3$) and a population of intriguing SMGs, DSFGs, and proto-clusters (\citealt{Walter2012, Riechers2013, Vieira2013, Casey2014, Miller2018, Spilker2023}) in the first two billion years of the Universe ($z \gtrsim 3$). Despite these advances, there are still several major questions regarding our understanding of the dusty Universe in its early epoch.

\begin{figure}[H]
  \centering
  \includegraphics[width=0.49\textwidth]{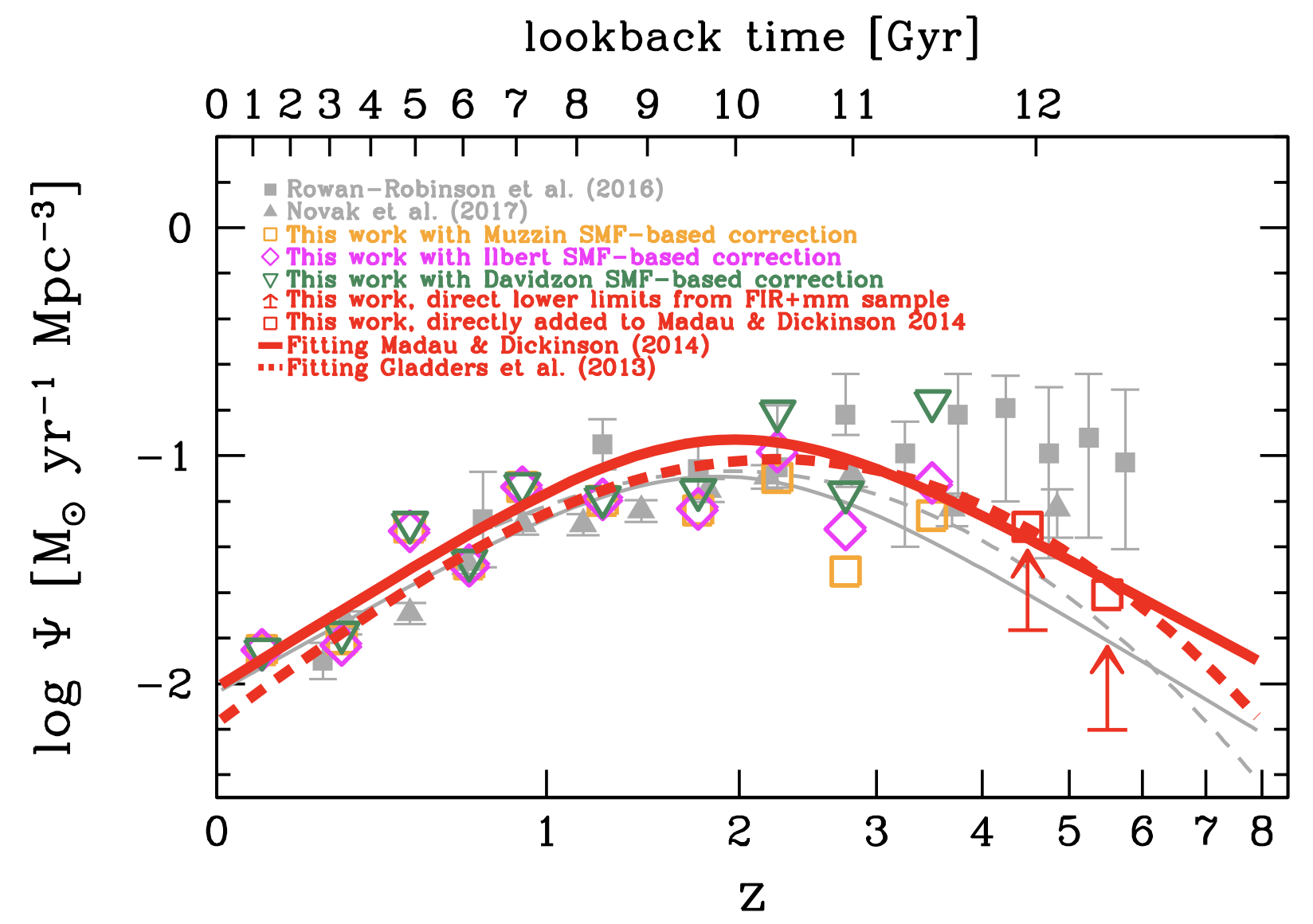}
  \caption{The cosmic star formation rate densities based on FIR and submm data. Figure is taken from \citet{Liu2018}. The $z > 3$ obscured star formation history is still very poorly constrained.}
  \label{fig:sfh}
\end{figure}

Firstly, the CSFRD is still very poorly constrained in the early cosmic time. While near-/mid-IR surveys with \textit{JWST} and \textit{Euclid} have probed galaxies up to photo-$z\sim 20$ \citep{Carniani2024}, far-IR/submm surveys with \textit{Herschel} and JCMT have mostly detected dusty galaxies at $z < 6$ with only a small fraction spectroscopically confirmed. As shown in Figure~\ref{fig:sfh}, the CSFRD has a huge uncertainty at $z > 3$ \citep{Rowan2016, Liu2018}. This affects our understanding of how many stars formed at each cosmic time and which galaxy populations contribute mostly to the cosmic star formation. Future progress requires enlarging the submillimeter survey areas and conducting more spectroscopic scans.

Secondly, the hunting of dusty galaxies in the earliest cosmic time is still limited to not very high-$z$, with only very few confirmed SMGs (SFR~$\gtrsim 1000 \ \mathrm{M_{\odot}\,yr^{-1}}$) at $z \sim 6$ (HFLS3, \citealt{Riechers2013}; SPT0311-58 \citealt{Strandet2017}; and G09 83808, \citealt{Zavala2018}). From \textit{JWST} and $z>6$ quasar studies, we can expect a very rapid star formation even in the epoch of cosmic dawn, but the most vigorously star-forming, dust-obscured systems are still elusive.

\begin{figure*}[htb]
    \centering
    \includegraphics[width=0.9\textwidth, trim=0 10mm 0 0]{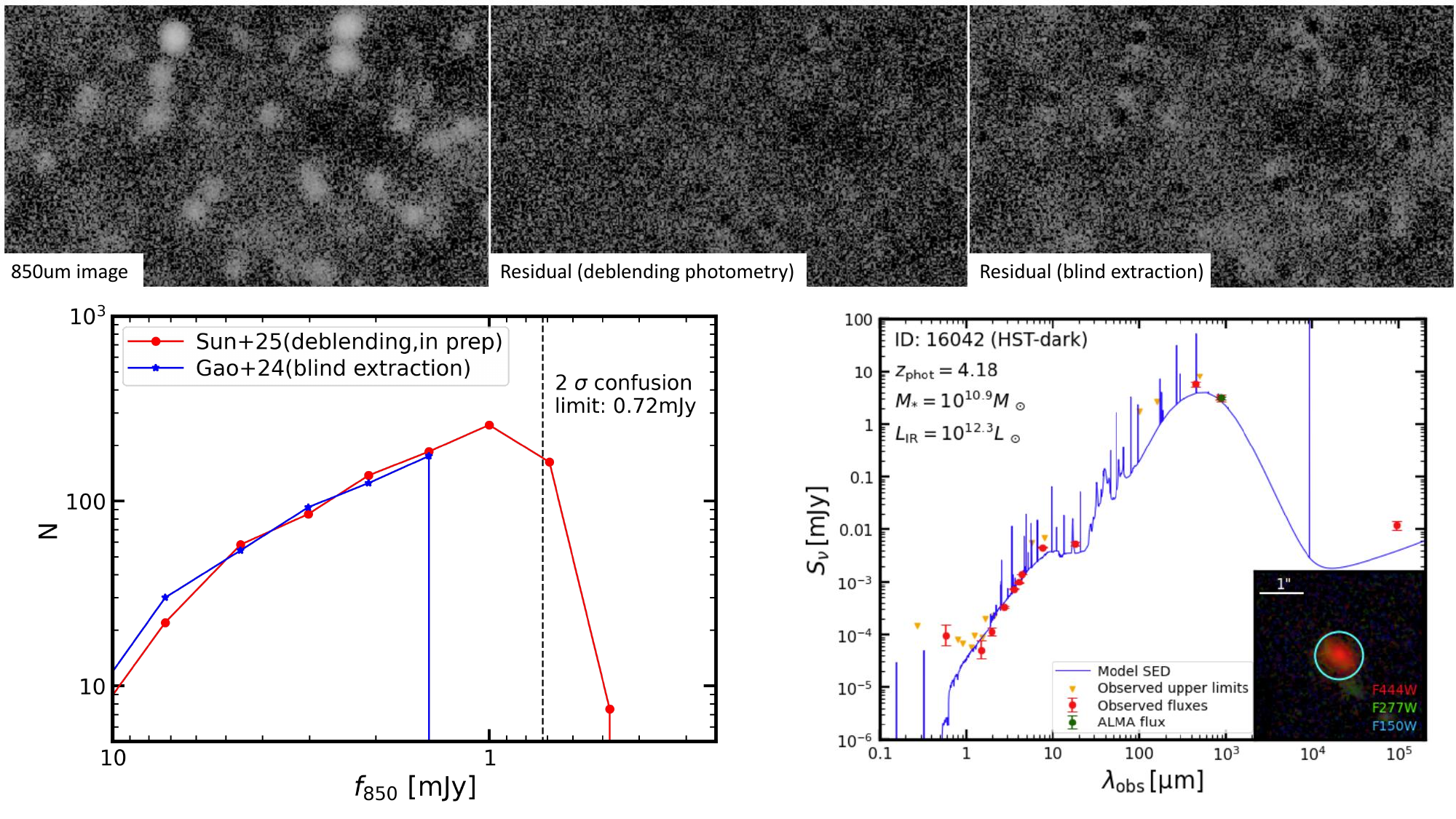}
    \caption{Comparison of source extraction results between traditional blind extraction and prior-based deblending methods. Top panels are the SCUBA2/850$\mu$m image and its residual maps based on the two extraction methods. The bottom left panels shows the comparison of the number counts from the two source extraction methods. The bottom right panel shows an example of the best-fit SED of an infrared-luminous galaxy at $z = 4.18$ with an 850$\mu$m detection based on prior-based extraction, which was missed from blind extractions.}
    \label{fig:figure_850combined}
\end{figure*}

Thirdly, we are entering a golden era of wide multi-wavelength surveys of galaxies and (proto-)clusters. New observations from JWST, Euclid, and the upcoming CSST and SKA will be essential for submillimeter deep fields, as they can provide far more complete prior catalogs for IR-bright sources compared to previous surveys, greatly improving submillimeter source extraction much below the confusion limit. Figure~\ref{fig:figure_850combined} shows an example of source extraction results with JWST priors using the SCUBA2/850$\mu$m map in the CANDELS-COSMOS region from the STUDIES survey (Sun, Wang, et al. 2025, in prep). Such synergies between JWST/Euclid/Roman/CSST and submillimeter deep-field surveys will be central in the next decade to unveil the most intensive star formation approaching the cosmic dawn.

\subsection{Objectives}

We aim to provide essential submillimeter deep-field coverage in synergy with JWST/Euclid\footnote{Euclid deep fields: \url{https://www.euclid-ec.org/science/overview/}.}/CSST galaxy surveys in the northern hemisphere. The survey will adopt a multi-tiered strategy, as outlined in Table~\ref{tab:summ deep fields}, so as to cover different areas and depths.

\begin{table*}[htb]
    \centering
    \normalsize
    \caption{Preliminary survey tier plan.
    \label{tab:summ deep fields}
    }
    \begin{tabularx}{0.92\textwidth}{>{\hsize=.32\hsize}X | >{\hsize=1.2\hsize}X | >{\hsize=.55\hsize}X | >{\hsize=.53\hsize}X | >{\hsize=.45\hsize}X}
        \hline
        \hline
        Tier & Fields & Synergy & Depth ($1\sigma$)$^a$ & Integration (on source) \\
        \hline
        Pointed & 
            $N \sim 30$ proto-clusters, SMGs and AGNs (1 Fov $\sim 100$~arcmin$^2$) &
            JWST, Euclid, CSST & 
            $\sim 0.25$~mJy~beam$^{-1}$ & 
            $\sim 200$ hours \\
        Deep & 
            GOODS-North or EGS or others (0.5~deg$^2$), and COSMOS (2~deg$^2$) & 
            JWST, Euclid, CSST & 
            $\sim 0.4$~mJy~beam$^{-1}$ & 
            $\sim 200$ hours \\
        Wide & 
            Euclid Deep Field North (EDF-North, 23~deg$^2$)$^b$ & 
            JWST, Euclid, CSST & 
            $\sim 1.2$~mJy~beam$^{-1}$ & 
            $\sim 200$ hours \\
        \hline
        \hline
    \end{tabularx}
    \begin{tablenotes}
    \item $^a$ The confusion limit for a 15m telescope at 870~$\mu$m is $\sigma_{\mathrm{confusion,\,870{\mu}m}} \sim 0.4$~mJy~beam$^{-1}$ \citep{Simpson2019}. 
    \item $^b$ The EDF-North field (center RA Dec 17:58:55.9 +66:01:03.7) includes the \textit{JWST} North Ecliptic Pole (NEP) field. 
    \end{tablenotes}
\end{table*}

Key science questions to be addressed:

\textbf{i) Measuring the cosmic star formation history approaching the cosmic dawn.}

Since there will be no more large far-IR space telescopes like \textit{Herschel} in the near future, submillimeter deep fields will be unique to probe the cosmic star formation history. Euclid and future CSST deep fields will extend to tens of square degrees, whereas ground-based submillimeter deep-field surveys were mostly too small or too shallow (the SPT survey covers $\sim2500$~deg$^2$ with a $1\sigma$ depth of $\sim 5$~mJy at 1.4~mm; the S2COSMOS survey covers $\sim 1.6$~deg$^2$ with $1\sigma \sim 1.2$~mJy at 870~$\mu$m; the eS2COSMOS survey covers $\sim 2$~deg$^2$ with $1\sigma \sim 0.9$~mJy at 870~$\mu$m). 

Compared to existing submillimeter bolometer detectors, such as SCUBA-2 on the JCMT, XSMT-15m's KID continuum camera will be much more efficient in surveying the dusty Universe because it has a large FoV of $\sim$100~arcmin$^2$ and can simultaneously observe two of the three continuum bands. One continuum band will be a new band at 460~GHz which is different from the SCUBA-2 450~$\mu$m (666~GHz) and can not only improve the SED constraints but also serve as a ``medium-band'' ($\Delta \nu \lesssim 15$~GHz) for very bright $z \sim 3.1$ [CII]\,\footnote{Assuming a typical [CII] line flux of 10~Jy~km~s$^{-1}$ at $z=3.1$ for an ultra-luminous galaxy with $L_{\mathrm{IR}}=10^{13}\;\mathrm{L_{\odot}}$, and a bandwidth of 10~GHz which corresponds to 9700~km~s$^{-1}$ at 464~GHz, then the averaged line flux within the 460~GHz band is $\sim 1.1$~mJy.} or low-$z$ CO line emitters. 

With wide- and deep-tier XSMT-15m submillimeter deep-field observations, we expect to detect fainter DSFGs with SFRs $\gtrsim 100\,\mathrm{M_{\odot}\;yr^{-1}}$ over $\gtrsim 2$~deg$^2$ area and brighter SMGs with SFRs over $\gtrsim 500\,\mathrm{M_{\odot}\,yr^{-1}}$ over $\gtrsim 20$~deg$^2$, spanning a wide redshift range. These samples will provide significantly improved statistics on the cosmic SFR density at $z>3$. When  combined with the multi-wavelength data from JWST, Euclid, and CSST, they will achieve the science goal of constraining the cosmic star formation history approaching the cosmic dawn. 

\textbf{ii) Measuring the evolution of gas and dust content over cosmic time.}
The XSMT-15m submillimeter deep-field survey will allow deblending of individual main-sequence star-forming galaxies and measuring their dust and gas properties across a wide range of cosmic time. This includes the following science goals: 

a) Using the photometric data to constrain the SED shape, hence to determine the dust temperature, emissivity, and dust mass \citep{Bethermin2015}. 

b) Combining the derived SFRs of individual galaxies and their stellar masses from JWST/Euclid/CSST to constrain the dichotomy of star-formation main sequence and starburst, and to trace the evolution of the main sequence's scatter and `bending' at $z>3$ \citep{Schreiber2015}. 

c) By adopting reasonable dust-to-gas mass ratios, submillimeter luminosity can yield rough estimates of cold ISM masses \citep{Scoville2014, Scoville2016, Liudz2019b}. The evolution of cold ISM at $z>3$ is still much debated and will be critical for understanding early galaxy assembly.

\textbf{iii) Constraining feedback from AGNs at high redshift.} The feedback from AGNs is proposed to play an important role in regulating the evolution of their galaxies, which could heat up and expel the gas from central regions, thereby suppressing star formation in the host galaxies (\citealt{Cicone2014, Fiore2017, Izumi2023, Vayner2024}). 
A key research is constraining the cold gas and dust in high-$z$ AGNs. 
Recent studies of molecular gas fractions in X-ray selected AGN samples at $z=1$ to 3 show evidence of gas depletion in quasar host galaxies (\citealt{Circosta2021, Bertola2024}). Therefore, how AGN feedback affects cold gas content and star formation in their host galaxies is still a key question for understanding the co-evolution of galaxy and supermassive black hole (SMBH). The XSMT-15m submillimeter deep-field survey will also help to understand AGNs in those deep fields, especially the wide-tier field, which is probably large enough to include a sample of bright high-$z$ AGNs.

\textbf{iv) Constraining proto-cluster formation and evolution.}
Proto-clusters are high-$z$ overdense objects that are progenitors of the densest galaxy clusters in the local Universe. Submillimeter galaxies are thought to be the signpost of high-$z$ overdense regions, and many of them are found to be associated with proto-clusters (\citealt{Riechers2014, Wang2016b, Sillassen2024, Zhou2024}). The NOEMA Forming Cluster Survey (NICE) large program (PIs: Wang \& Daddi) has developed a method to efficiently select high-$z$ proto-clusters in various deep fields including COSMOS, Lockman Hole, etc., probing massive dark matter halos at $2 < z < 4$. Very massive proto-clusters are still rare to find in existing small-area surveys. A deeper and wider submillimeter survey with XSMT-15m will increase the finding of the most distant and massive proto-clusters, shedding light on the most massive dark matter halos in the early cosmic epoch. 



\section{EG: Black Hole Imaging}

\textbf{Plain summary for non-experts:} \textit{The Next Generation Event Horizon Telescope (ngEHT) is an ambitious international project that will link the most powerful (sub)millimeter facilities worldwide into an Earth-sized virtual telescope, enabling direct imaging of photon orbits near the event horizon of supermassive black holes. If integrated into the ngEHT array,  XSMT-15m, with its strategic location in China, would provide unique baselines and thereby make a significant contribution to future findings on the nature of black holes.}

\subsection{Motivation}
The Event Horizon Telescope (EHT) achieved a historic milestone by capturing the first image of a black hole, combining a global network of millimeter telescopes into an Earth-sized telescope using very long baseline interferometry (VLBI; \citealt{EHT-Collaboration-2019}). This image resolves the shadow of the SMBH at the center of the giant elliptical galaxy M87, revealing the expected ring-like morphology predicted by General relativity (GR). The ring arises because photons orbiting the SMBH are bent and distorted by its strongly curved space-time. Most of the photons detected by the EHT originate from synchrotron radiation in the accretion disk surrounding M87$^{*}$, although the hot dusty torus predicted by theory may also contribute to some degree.

The angular resolution of VLBI observations currently ranges from $\sim$20~$\mu$as to sub-arcsecond scales, depending on the longest baseline (comparable to the Earth’s diameter) and the highest observing frequency (up to $\sim$230~GHz). To achieve an even higher resolution, the next-generation EHT (ngEHT) has been proposed to extend the current EHT into the submillimeter regime. By incorporating additional telescopes operating at $\gtrsim$300~GHz, the ngEHT will enable black hole imaging at higher angular resolution and even time-resolved movies (e.g., photon rings, structural and brightness variations). This will allow tests of the nested photon rings predicted by general relativity, the variability of accretion disks in GR magnetohydrodynamic (GRMHD) models, the precession of relativistic jets, and magnetic-field structures. Such observations will provide new insights into the fundamental physics of black holes, among the most profound mysteries in astrophysics.

Table~\ref{tab:EHT-stations} presents the geographical positions of the current EHT stations along with the planned location of XSMT-15m. The addition of XSMT-15m would provide a strategically valuable site that fills a gap in the existing network. If incorporated into the ngEHT, XSMT-15m would deliver unique baselines that enhance the $uv$ coverage (see Figure~\ref{fig:uv-track} for the mock $uv$ tracks), thereby improving image fidelity and dynamic range. 


\begin{table*}[hbtp]
    \centering
    \caption{\textbf{EHT stations}
    \label{tab:EHT-stations}}
    \begin{tabular}{c|c|c|c}
    \hline\hline
\textbf{Telescope} & \textbf{X} & \textbf{Y} & \textbf{Z} \\
    \hline
ALMA (phased)   & 2225061.873     & $-$5440061.953  & $-$2481682.084 \\
APEX            & 2225039.530     & $-$5441197.629  & $-$2479303.360 \\
JCMT            & $-$5464584.676  & $-$2493001.170  & 2150653.982  \\
SMA (phased)    & $-$5464588.447  & $-$2492884.038  & 2150756.452  \\
LMT             & $-$768715.632   & $-$5988507.072  & 2063354.852  \\
IRAM 30-meter   & 5088967.900     & $-$301681.600   & 3825015.800  \\
SMT             & $-$1828796.200  & $-$5054406.800  & 3427865.200  \\
SPT             & 792.600         & $-$802.600      & $-$6359569.200 \\
GLT             & 541547.0        & $-$1387978.6    & 6180982.0    \\
NOEMA (phased)  & 4524000.4       & 468042.1        & 4460309.8    \\
KP 12M          & $-$1995954.4    & $-$5037389.4    & 3357044.3    \\
\textbf{XMST 15-meter}   & $-$577785.471           & 5004576.366          & 3906373.570         \\
    \hline\hline
    \end{tabular}
    \begin{tablenotes}
    \item Notes. EHT station coordinates from \url{https://eventhorizontelescope.org/for-astronomers/proposals} (Mar. 2025) and XSMT-15m's coordinate appended in the last row. Antenna positions are in meters in an Earth-centered, right-handed coordinate system. The origin is the Earth's center, X is to the Greenwich meridian, Z is in the direction of the north pole, and Y is oriented to give a right-handed coordinate system with X and Z.
    \end{tablenotes}
\end{table*}

\subsection{Objectives}

XSMT-15m is capable of observations up to $\sim$800~GHz. However, to align with the plausible ngEHT receiver bands, a multi-band heterodyne receiver operating at 86, 230, and 345~GHz simultaneously will be developed and installed on XSMT-15m as a first-generation instrument. This will enable at least the following scientific objectives:

\begin{enumerate}
\itemsep0em
    \item Resolving the different $n$-number photon rings of SMBHs as predicted by the GR. This would require an angular resolution of $\lesssim$10$\mu$as, that is, an observing frequency of $\gtrsim$460GHz. 
    \item Measuring the polarized submillimeter emission around SMBH. 
    \item Constraining the energy distribution of the synchrotron emission of the accretion disk and the possible outer dust emission. 
\end{enumerate}

\begin{figure}[H]
    \centering
    \includegraphics[width=0.5\textwidth]{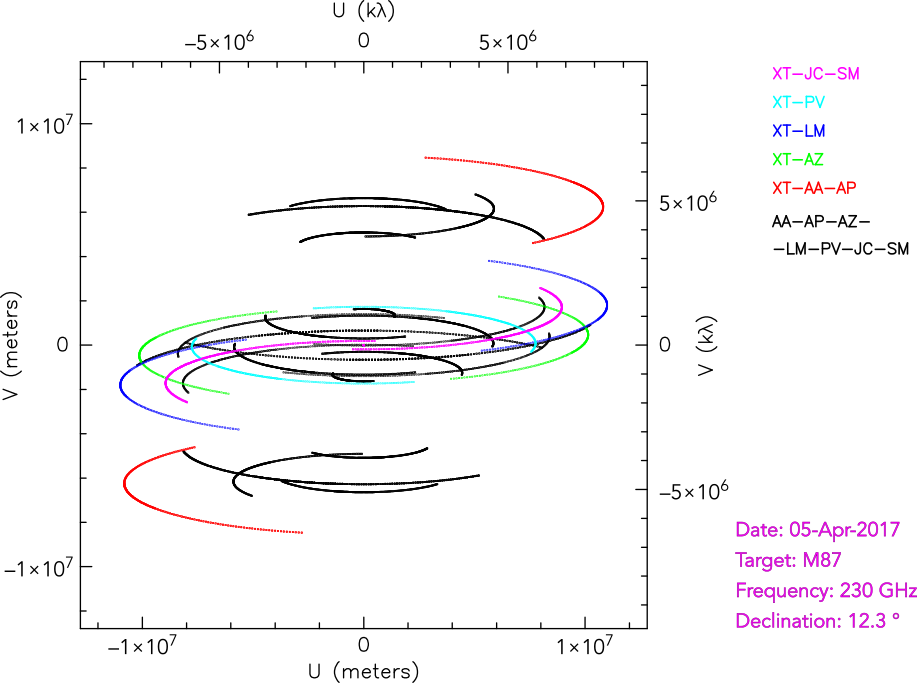}
    \caption{Simulated $uv$ track for a mock EHT observation with XSMT 15m (XT). The observing target and date are set to the same as the EHT's observation for M87 SMBH on Apr. 05, 2017, but showing $uv$ tracks starting from $-$4 to $+$4 hours around the transit. Black points show the original baselines among EHT stations, and colored points show the new baselines enabled by connecting XSMT 15m to the JCMT (JC), SMA (SM), IRAM 30m (PV), The Large Millimeter Telescope (LM), The Submillimeter Telescope at the Arizona Radio Observatory (AZ), ALMA (AA), and APEX (AP). 
    \label{fig:uv-track}
    }
\end{figure}

To assess the potential contribution of XSMT-15m to the $uv$ coverage, we used the sites listed in Table~\ref{tab:EHT-stations} and performed baseline simulations with the GILDAS \texttt{UV\_track} task. Figure~\ref{fig:uv-track} shows the simulated baselines for an 8~h tracking run (for illustration) on the date of the actual EHT observations of M87 at 230~GHz \citep{EHT-Collaboration-2019}. The original EHT baselines, i.e., without XSMT-15m, are shown in black, while those including XSMT-15m appear in color. In particular, the XSMT–ALMA (XT–AA) and XSMT–APEX (XT–AP) baselines, shown in red, are the most extended, exceeding even the longest baselines of the original EHT.

The likelihood of XSMT-15m joining the ngEHT is high, given that only a few submillimeter facilities worldwide exceed 10~m in diameter. Moreover, higher-frequency test observations may be feasible between XSMT-15m and ALMA and/or APEX, as these sites support Band~8 observations and already possess (or soon will possess) Band~8 heterodyne receivers.

\section{MW: Probing Multi-Scale Magnetic Fields in Molecular Clouds through Dust Polarization}

\textbf{Plain summary for non-experts:} \textit{Stars are born within molecular clouds, where magnetic fields are thought to play an important role in shaping their structures and influencing how stars are born. However, what exact role they play remains poorly constrained, largely due to the observational challenges of measuring magnetic fields across a wide range of spatial scales and density regimes. Polarized emission from dust grains offers a powerful means of tracing magnetic field orientations. Although significant progress has been made through existing observations, there remains a critical need for wide-field polarization mapping at high angular resolution to fully characterize magnetic field structures across the multiple scales of molecular clouds. The upcoming XSMT-15m telescope will address this limitation by delivering sensitive, wide-field, high-resolution ($\sim$15\arcsec) polarization maps of dust emission. These observations will enable a comprehensive, multi-scale characterization of magnetic field structures across diverse molecular cloud environments, providing critical insight into the interplay between magnetic fields, gravity, and turbulence in the formation of filaments, dense cores, and ultimately stars, thereby advancing our understanding of star formation across the Milky Way.}

\subsection{Motivation}

Magnetic fields play a fundamental role in shaping the structure and dynamics of molecular clouds, and are therefore critical for understanding star formation processes \citep[e.g.,][]{2012ARA&A..50...29C,2017ARA&A..55..111H,2022FrASS...9.3556L}. Theories of star formation diverge significantly in the importance that they assign to magnetic fields: strong-field models argue that magnetic fields regulate cloud evolution and core collapse via ambipolar diffusion \citep[e.g.,][]{2004ApJ...616..283T}, while weak-field models emphasize turbulence as the dominant force \citep[e.g.,][]{2004RvMP...76..125M}. More recent models or simulations incorporate both magnetic fields and turbulence as key agents \citep[e.g.,][]{2019FrASS...6....5H}. Crucially, distinguishing between these competing scenarios relies on accurate measurements of magnetic fields, which are essential for revealing the physical mechanisms that govern star formation in molecular clouds.

Despite their crucial role in influencing star formation, magnetic fields remain challenging to accurately measure from an observational perspective. Various methods have been developed to study magnetic fields \citep[e.g.,][]{2012ARA&A..50...29C}, including Zeeman-effect observations \citep[e.g.,][]{2019FrASS...6...66C}, polarization of starlight from background stars \citep[e.g.,][]{2000AJ....119..923H}, polarized dust emission \citep[e.g.,][]{2017ApJ...846..122P}, the Goldreich–Kylafis (GK) effect observed in spectral lines \citep[e.g.,][]{2011Natur.479..499L}, velocity gradient \citep[e.g.,][]{2019NatAs...3..776H}, and intensity gradient techniques \citep[e.g.,][]{2012ApJ...747...79K}. Among these observational techniques, linearly polarized emission from dust grains offers distinct advantages, as it provides a direct tracer of magnetic field orientations independent of gas dynamics, efficiently probes extensive spatial scales, and enables detailed mapping of magnetic field morphologies throughout entire molecular clouds.

Thanks to the Planck satellite, all-sky measurements of polarized dust emission are now available at angular resolutions of $\gtrsim$5\arcmin, providing critical insight into magnetic field structures on large scales \citep[e.g.,][]{planck2015,soler2019}. On the other end of the scale, interferometric observations have significantly advanced our understanding of dust polarization on sub-arcsecond to a few arcsecond scales, enabling detailed studies of magnetic fields in dense cores and filaments within molecular clouds \citep[e.g.,][]{2014ApJ...792..116Z,2019FrASS...6....3H,2023ApJ...945..160L}. However, growing evidence indicates that magnetic fields are dynamically important across a broad range of spatial scales \citep[e.g.,][]{2015Natur.520..518L}, highlighting the importance of a coherent, multi-scale characterization of magnetic field structures. Intermediate-scale observations, such as those obtained with JCMT’s POL-2 \citep[e.g.,][]{2017ApJ...846..122P} and SOFIA’s HAWC+ \citep[e.g.,][]{2020NatAs...4.1195P}, are beginning to fill this gap. However, systematic multi-scale studies of polarized dust emission across entire molecular clouds remain scarce, largely due to the observational expense of obtaining high-resolution polarization maps over wide fields. The upcoming XSMT-15m telescope can help to overcome this limitation by enabling sensitive dust polarization measurements across large spatial scales with an angular resolution of $\lesssim$15\arcsec. The multi-color KID continuum camera on  XSMT-15m will operate at 230, 345, and 460~GHz. This capability will offer a valuable complement to current instruments such as NIKA2 on IRAM-30m \citep{2017A&A...599A..34R,2024PASP..136k5001R} and POL-2 on JCMT \citep{2016SPIE.9914E..03F}, enhancing our ability to probe dust properties and magnetic field structures in the submillimeter regime.

\subsection{Objectives}
\begin{figure*}[!htbp]
\includegraphics[height=.33\linewidth]{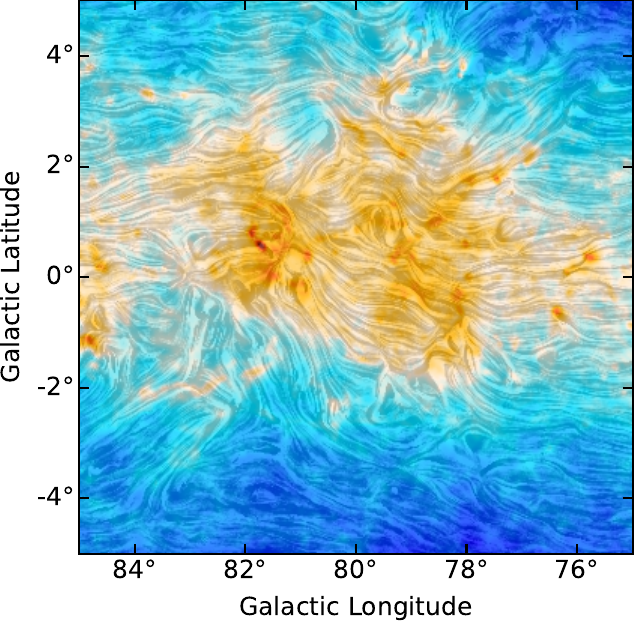}
\includegraphics[height=.33\linewidth]{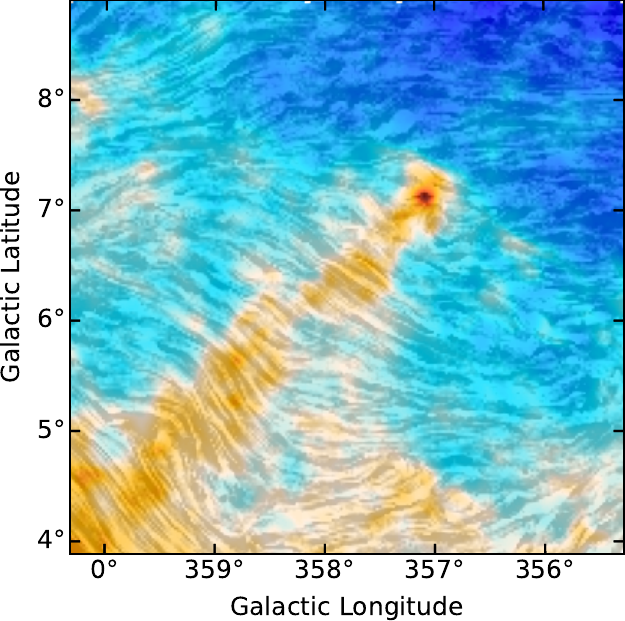}
\includegraphics[height=.33\linewidth]{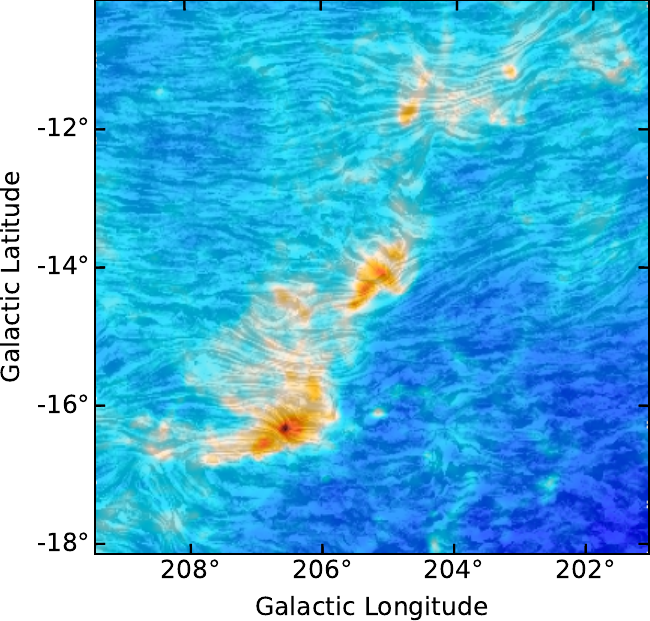}
\caption{Plane-of-the-sky magnetic field distribution of Cygnus X (left), Orion B (middle), and the Pipe Nebula (right) measured by the \textit{Planck} 353~GHz dust continuum observations. The overlaid pattern, produced using the line integral convolution (LIC) method \citep{Cabral93}, indicates the orientation of magnetic field lines.}
\label{fig:pol}
\end{figure*}

As part of our effort to probe magnetic fields across multiple spatial scales in molecular clouds, we have selected dense molecular clouds in Cygnus X, Orion B, and the Pipe Nebula as our primary targets (see Figure~\ref{fig:pol}). These regions, spanning different levels of star formation activity, provide an excellent opportunity to explore the variation of magnetic fields across distinct environments. Cygnus X, one of the closest and most active high-mass star-forming regions in the Milky Way \citep[e.g.,][]{2008hsf1.book...36R,2014AJ....148...11K}, offers insights into how magnetic fields influence the formation of high-mass stars and their feedback processes. In contrast, Orion B, a prototypical region for studying low- to high-mass star formation \citep[e.g.,][]{2013ApJ...766L..17S}, allows us to examine magnetic fields in both active and less active phases of star formation. The Pipe Nebula, a low-mass star-forming region with very low star-formation activity \citep[e.g.,][]{1999PASJ...51..871O}, serves as a reference for studying the early stages of star formation in a more quiescent environment. Therefore, these sources offer exceptional opportunities for investigating the scale-dependent roles of magnetic fields in diverse star formation environments. 

Multi-scale dust polarization measurements offer a comprehensive and coherent view of magnetic field structures across the hierarchical architecture of molecular clouds. Low-angular-resolution magnetic field data are already available for the selected regions from Planck measurements \citep[e.g.,][see Figure~\ref{fig:pol} for instance]{planck2015}. However, high-angular-resolution magnetic field data are currently limited to a few small fields within these regions \citep[e.g.,][]{2009ApJS..182..143M,2022ApJ...941..122C}. To investigate the role of magnetic fields in the multi-scale physics of molecular clouds, we therefore propose to carry out large-scale polarimetric mapping observations toward the selected regions using the KID continuum camera on the XSMT-15m telescope. The main scientific objectives are summarized below:
\begin{itemize}
    
    \item \textbf{To comprehensively map and infer the three-dimensional magnetic field structure in molecular clouds through synergy with multi-wavelength observations.} Planck’s all-sky maps provide essential large-scale context \citep[e.g.,][]{planck2015}, while optical polarization traces magnetic fields in the diffuse, low-density ISM \citep[e.g.,][]{2000AJ....119..923H}. By bridging these with the XSMT-15 m observations, we will connect the magnetic field morphology from the Galactic scale down to the dense filaments and cores, tracing field evolution across a wide range of density regimes. With the advent of the SKA, both Zeeman splitting measurements and Faraday rotation measure  techniques will be significantly enhanced, enabling determinations of the magnetic field component along the line of sight toward a greatly increased number of positions across the sky. In synergy with our proposed dust polarization observations, this will allow us to infer three-dimensional magnetic field structure at a higher resolution than previously achieved \citep[e.g.,][]{2022FrASS...9.0027T}. \\

    \item \textbf{To quantitatively assess the interplay among magnetic fields, turbulence, and gravity across multiple spatial scales in molecular clouds.} Observations of NGC~6334 reveal that this relative orientation remains unchanged from 10 to 0.1~pc \citep{2015Natur.520..518L}, implying that magnetic fields remain dynamically significant over a broad range of scales. In contrast, transitions in the relative orientation on small scales in Serpens South \citep{2020NatAs...4.1195P} are interpreted as signatures of magnetic supercriticality, where gravity begins to dominate. By systematically investigating these multi-scale transitions, our goal is to constrain the scale-dependent roles of magnetic fields, turbulence, and gravity in regulating the formation and evolution of filaments, dense cores, and ultimately stars within molecular clouds. \\
    
    \item \textbf{To investigate the efficiency of dust grain alignment as a function of physical conditions within molecular clouds.} The linear polarization degree serves as a key diagnostic of the alignment efficiency of dust grains with respect to the local magnetic field \citep[e.g.,][]{2025arXiv250116079T}. Observational studies increasingly support the radiative torque mechanism as the primary driver of dust grain alignment in molecular clouds \citep[e.g.,][]{2014A&A...569L...1A}. Nevertheless, robust interpretation remains challenging due to projection effects associated with the complex three-dimensional magnetic field configuration, which can bias the two-dimensional polarization measurements. These challenges can be alleviated through statistical analyses, forward modeling, and careful comparison with ancillary data. The polarimetric mapping capabilities of XSMT-15m will allow us to systematically study the spatial variations in linear polarization degree across a range of physical environments and spatial scales. Moreover, synergy with polarization measurements from other instruments (i.e., NIKA2, HAWC+, POL-2, etc.) enables a more comprehensive comparison of the wavelength dependence of the linear polarization degree with theoretical model predictions \citep[e.g.,][]{2025arXiv250116079T}. By quantifying how alignment efficiency changes with local conditions, we aim to place critical constraints on dust grain alignment mechanisms within molecular clouds.

\end{itemize}

The complete dataset will serve as an invaluable resource for the entire astronomical community, enabling a wide range of studies beyond the immediate scope of this project and providing a lasting foundation for future advances in the field.

\section{MW: Dynamic States of Molecular Clouds in the Galactic Outer Disk}

\textbf{Plain summary for non-experts:} \textit{Dynamic properties of molecular clouds determine the initial conditions for star formation. The outer disk of the Milky Way provides the nearest laboratory for the dynamics of molecular clouds under low-metallicity conditions, offering critical insights into star formation in other metal-poor environments, such as local dwarf galaxies and galaxy populations in the early cosmic epoch. The heterodyne receiver on XSMT-15m enables a comprehensive census of the dynamic states of a large sample of Galactic outer disk molecular clouds on parsec scales, targeting their molecular gas tracers at 1.3 and 0.6~mm wavelengths.}

\subsection{Motivation}

The dynamic properties of molecular clouds are governed by the interplay of self-gravity, turbulence, external pressure, and magnetic fields. These processes determine how a cloud will collapse under the effect of self-gravity and then form stars \citep{McKee2007}. Past studies of molecular clouds in the Milky Way \citep{Larson1981} and external galaxies
\citep{Bolatto2008} have found power-law correlations between the cloud physical parameters (size, velocity dispersion, mass). These correlations are often attributed to the virial equilibrium between the turbulent kinetic energy ($E_{\rm k}$) and the self-gravitational energy ($E_{\rm g}$) of molecular clouds, which is quantified as the virial parameter ($\alpha_{\rm vir}=2E_{\rm k}/|E_{\rm g}|$) equal to one. Such virial equilibrium is often used as the initial condition for star formation in numerical simulations \citep{McKee2003}.

\begin{figure*}[!htbp]
 \centering
\includegraphics[width=0.56\linewidth]{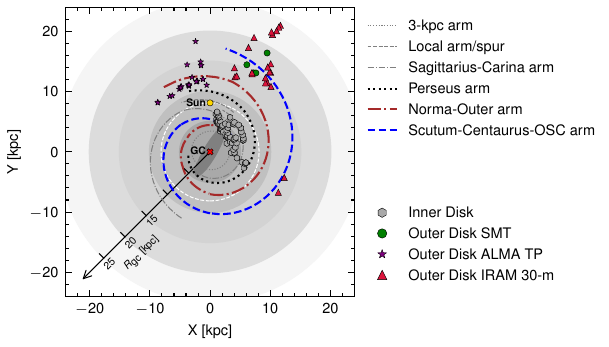}
\includegraphics[width=0.43\linewidth]{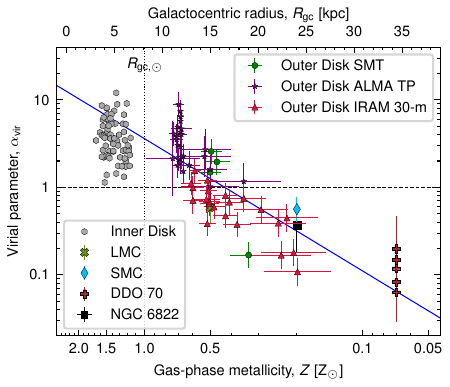}
\caption{Adapted from \citet{Lin2025}. {\it Left:} Galactic face-on view of the distribution of molecular cloud samples. {\it Right:} Variation of cloud virial parameter with the gas phase metallicity.}
\label{fig: GMPMC}
\end{figure*}

However, previous studies on the dynamic states of molecular clouds are often based on optically thick low-$J$ $^{12}$CO transitions. This method not only introduces large uncertainties to molecular gas mass measurements
\citep{Bolatto2013} but also overestimates velocity dispersions due to
opacity broadening \citep{Hacar2016}. In contrast, optically thin
rotational transitions of the rarer $^{13}$CO molecules \citep{Wilson1994} are better H$_{2}$ column density tracers than $^{12}$CO lines \citep{Heyer2009}. Nevertheless, the faint $^{13}$CO lines limit previous $^{13}$CO cloud dynamics studies only within a Galactocentric radius ($R_{\rm gc}$) of less than 12 kpc, leaving the dynamic properties of molecular clouds in the far outer disk of our Galaxy poorly explored.

The outer disk of the Milky Way, where the Galactocentric radius is beyond the Solar circle \citep{Knapen2017}, is characterized by very different physical conditions than those in the inner Galaxy, such as low gas-phase metallicity \citep{Mendez-Delgado2022}, low mid-plane pressure \citep{Wolfire2003}, and low levels of turbulence injection \citep{Miville-Deschenes2017}. These environments  not only reflect the early formation stage of the Galactic thin disk \citep{Chiappini2009}, but also resemble the general conditions of low-metallicity, gas-rich galaxies \citep{Shi2014}.

\citet{Lin2025} perform a survey of a small sample of Galactic outer disk molecular clouds (Figure~\ref{fig: GMPMC}), targeting the $^{13}$CO $J=1-0$ or $J=2-1$ lines. With a physical resolution of $\approx$1~pc, they find that molecular clouds in the inner Galaxy are mostly in a supervirial state ($\alpha_{\rm vir}>1$) while molecular clouds in the outer Galaxy are systematically in a subvirial state ($\alpha_{\rm vir}<1$). Using the Galactic radial gradient of gas-phase metallicity \citep{Mendez-Delgado2022}, \citet{Lin2025} estimate the gas-phase metallicity of the Galactic molecular clouds based on their $R_{\rm gc}$. By including molecular clouds in nearby metal-poor dwarf galaxies, a robust trend of $\alpha_{\rm vir}$ as a function of the gas-phase metallicity is found (Figure~\ref{fig: GMPMC}, \citealt{Lin2025}). This trend is followed by molecular clouds in both the Milky Way and nearby metal-poor dwarf galaxies, showing systematic superviral states in metal-rich conditions and subvirial states in metal-poor conditions. The subvirial states systematically found in metal-poor conditions indicate another cloud-supporting mechanism being prevalent at low metallicity. {\it A direct inference from this trend is that the initial condition of star formation varies with the gas-phase metallicity.} Despite the lack of direct magnetic field measurements, \citet{Lin2025} showed that field strengths comparable to those typically measured in the Milky Way could provide the support needed to explain the subvirial states in low-metallicity environments.

However, \citet{Lin2025} is limited in two aspects: a). The sample size is small. Among the 30,000 Galactic outer disk clouds discovered with $^{12}$CO in the northern sky \citep{Sun2024}, thanks to the Milky Way Imaging Scroll Painting (MWISP) survey \citep{2019ApJS..240....9S}, \citet{Lin2025} only covers a very small sample for the strongest $^{13}$CO emitters. A large variety of target properties, such as those associated with H{\sc ii} regions and those with relatively weak lines, are not yet studied. Clouds with very large Galactocentric distances ($\sim 20-25$ kpc) are still very limited. b). The $^{13}$CO emission is strongly limited in its sensitivity and chemical environments. As the $^{12}$C/$^{13}$C ratio increases with the Galactocentric distance, $^{13}$CO may only arise from dense molecular cores. Therefore, another optically-thin H$_2$ gas tracer, [C{\sc i}] 1--0, would provide a new window. XSMT-15m provides the opportunity to conduct a comprehensive survey of the Galactic outer disk clouds identified by the PMO 13.7-m telescope \citep{Sun2015, Sun2024}, targeting the $^{13}$CO (2--1) line and [C{\sc i}] (1--0) transitions.

\subsection{Objectives}

With the capabilities of XSMT-15m, we are able to map $^{13}$CO (2--1) and [C{\sc i}] (1--0) emission of the brightest clouds selected from the sample of  \citet{Sun2024}, achieving a physical resolution of $\lesssim 1-2$ pc for the most distant objects. With these data, we will measure the cloud physical parameters following the procedures in \citet{Lin2025}. We will then conduct the following investigations.

\textbf{$\bullet$ Relations among the cloud physical parameters.} We will
explore the correlations among the cloud radius, the velocity dispersion, and the molecular gas mass, which are commonly known as the Larson's relations
\citep{Larson1981}. We will divide the whole sample into different $R_{\rm gc}$ bins, allowing us to examine how these relations evolve as a function of $R_{\rm gc}$ (i.e., gas phase metallicity). 

\textbf{$\bullet$ Examine the $\alpha_{\rm vir}-R_{\rm gc}$ trend.} We will
calculate the cloud virial parameter and test the $\alpha_{\rm vir}-R_{\rm gc}$ trend reported by \citet{Lin2025}. In that study, most of the outer disk clouds are in the second Galactic quadrant and there are only six clouds at $R_{\rm gc}>$20 kpc. With a larger sample, we will extend the analysis to the largest Galactic radii and fully validate the trend across the entire Galactic plane.

\textbf{$\bullet$ Benchmark [C{\sc i}] (1–0) as a tracer at low metallicity.} The [C{\sc i}] (1--0) line has been proposed as a new molecular gas tracer because its excitation requirements are comparable to those of low-$J$ CO lines. In most cases, [C{\sc i}] (1--0) is optically thin and therefore it has the potential to accurately trace molecular gas even at low metallicities
\citep{Bisbas2025}. We will observe a sub-sample to benchmark the use of [C{\sc i}] (1--0) as a molecular gas tracer.

\textbf{$\bullet$ Calibrate the metallicity-dependence of the CO-to-H$_{\rm 2}$
conversion factor ($X_{\rm CO}$).} With molecular gas masses measured from $^{13}$CO emission, we will calculate the conversion factor between the CO luminosity and molecular gas mass for the outer disk clouds. We will estimate the gas-phase metallicity of the outer disk clouds based on their $R_{\rm gc}$ and then calibrate the dependence of $X_{\rm CO}$ on metallicity. We will also compare our results with the $X_{\rm CO}$ values measured in metal-poor galaxies based on dust emission \citep{Bolatto2013} as well as those from chemical models \citep{Bisbas2025}.


\section{MW: Connecting Dense Gas Properties of the Milky Way with Galaxies}

\textbf{Plain summary for non-experts:} \textit{Stars are born in dense clouds of gas, yet accurately measuring this gas across galaxies remains a challenge. This project will survey star-forming regions within the Milky Way to calibrate how molecular emission relates to actual gas mass. The findings will refine estimates of star-forming material in distant galaxies, offering a clearer picture of how stars take shape across the cosmos.}

\subsection{Motivation}

Dense molecular gas is the key for studying star formation in galaxies. Stars, especially massive stars, form essentially and exclusively in the dense cores of GMCs \citep{2008ASPC..390...52E}. Low-$J$ CO lines trace the total amount of molecular gas content, but are not sensitive to dense cores. In contrast, the transitions of molecules with large dipole moments (e.g., HCN, HCO$^+$, HNC, and CS) have critical densities $n_{\rm crit} > 10^4$~cm$^{-3}$, and are thus better tools to probe the relationship between gas and star formation. 

\begin{figure}[H]
    \centering
   \includegraphics[width=0.4\textwidth]{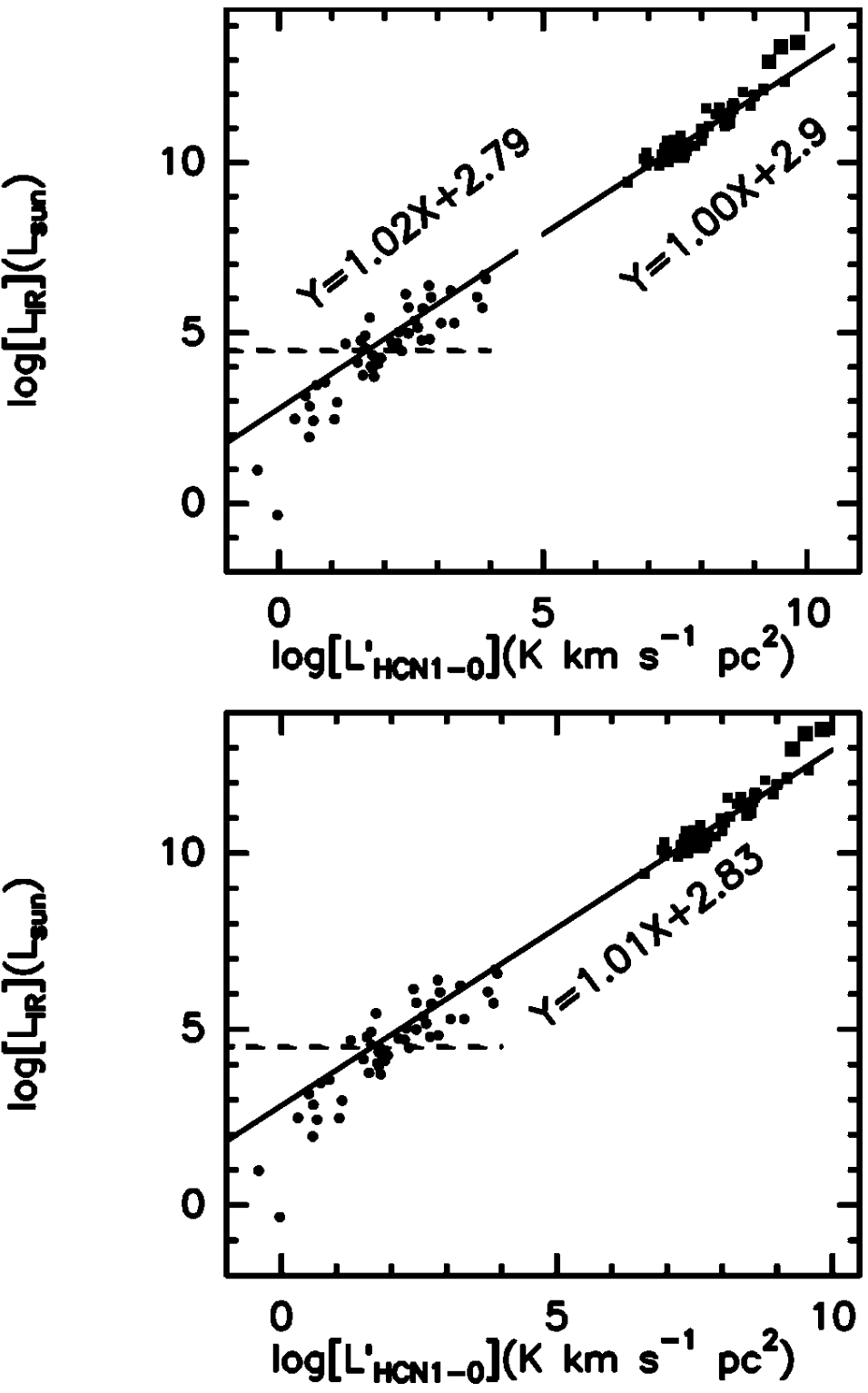}
    \caption{The relation between HCN (1--0) luminosity and infrared luminosity \citep{2005ApJ...635L.173W}.}
    \label{fig:Wu2005}
\end{figure}

With observations of HCN (1--0) toward 65 galaxies, \citet{2004ApJ...606..271G} found a strong linear correlation between HCN (1--0) and infrared luminosities ($N\sim$1). This relation has been extended to Galactic dense cores \citep[e.g.,][see Figure~\ref{fig:Wu2005}]{2005ApJ...635L.173W} and possibly high-$z$ galaxies and quasi-stellar objects as well \citep[e.g.,][]{2007ApJ...660L..93G}. Follow-up CS (5--4) observations in 24 IR-bright galaxies confirmed that this linear correlation is still valid for gas as dense as $n_{\rm H_2} \sim 10^6$~cm$^{-3}$ \citep{2011MNRAS.416L..21W}. Similarly, an HCN (4--3) survey toward 20 nearby star-forming galaxies also showed such a linear correlation \citep{2014ApJ...784L..31Z}. Multiple dense gas tracers, including HCN (1--0), HNC (1--0), HCO$^+$ (1--0), and CS (3--2), observed in $\sim$70 local galaxies  also show the consistent linear relation with infrared luminosity as the tracer of star formation rate for all dense gas tracers \citep{2021MNRAS.503.4508L}. These results indicate that, despite scatter in individual galaxies, standard dense gas tracers provide a robust first-order approximation of the dense gas–star formation relation.

However, dense molecular gas masses, as the most important parameter for studying star formation in galaxies, remain difficult to determine. The standard approach is to convert line luminosities to gas masses using a so-called conversion factor: CO lines for the total molecular gas, and high–dipole-moment molecules such as HCN, HNC, HCO$^+$, and CS for dense gas. Yet these lines are typically optically thick, introducing large uncertainties in the derived gas masses. These uncertainties arise not only from excitation conditions and molecular abundances, but also from assumptions about line opacities. Moreover, galaxies with different star formation types, including normal star-forming disks, circumnuclear starbursts, spiral-arm starbursts, merger-induced overlap starbursts, and systems with low star formation activity, may exhibit different dense-gas-mass-to-line-luminosity ratios for these tracers. Therefore, mapping observations of these molecular lines toward massive star-forming cores in the Milky Way, spanning a range of evolutionary stages, environments, and metallicities, are essential to calibrate the conversion factors from line luminosity to dense gas mass under different physical conditions.

\subsection{Objectives}

To determine the conversion factor from line luminosity to dense gas mass under different physical conditions, we aim to perform a calibration in the Milky Way. Key parameters for dense molecular cores include dense-gas masses, line luminosities, line opacities, excitation conditions, evolutionary stages, and possibly metallicities. A reasonably large sample (up to several hundred dense molecular cores spanning different evolutionary stages) will be needed for this study. Even though a sample of massive star-forming cores has been mapped with JCMT as part of the large program MAJORS (Massive, Active, JCMT-Observed Regions of Star Formation), the observations are only done with the optically thick dense gas tracers (HCN and HCO$^+$ 3--2), and thus lack the important information that  optically thin tracers can provide.

For each dense core, the measured parameters will include typical line widths  and spatial distribution of transitions of dense gas tracers. Mapping observations of these tracers, together with their optically thin isotopologues, will enable estimates of the average opacity and/or the opacity distribution in each source. Line luminosities will be derived from velocity- and spatially-integrated fluxes combined with source distances, determined either from maser parallaxes or from kinematic methods based on the Galactic rotation curve. Dense-gas masses can be estimated using several independent approaches: (i) virial masses from optically thin line widths and physical sizes, (ii) gas masses from submillimeter dust continuum emission, and (iii) masses inferred from molecular column densities using optically thin tracers. By comparing line luminosities with masses derived from these methods for each source, we aim to calibrate the conversion between line luminosity and dense-gas mass in galaxies under different physical conditions.

\section{MW: Supernova Remnants}

\textbf{Plain summary for non-experts:} \textit{Where does cosmic dust come from? One source is the stellar winds of low- to intermediate-mass stars that have evolved to the asymptotic giant branch (AGB). However, this process is too slow to explain the large amounts of dust observed in the Galaxy. Core-collapse supernovae, the explosive deaths of massive stars, are therefore considered another major source. Polarization observations of young supernova remnants (SNRs), such as Cas~A and the Crab Nebula, have confirmed the presence of cold dust. But what about other young SNRs? Here, we propose to observe some of the youngest SNRs in the Galaxy to search for polarized dust emission. These observations will provide critical insights into the role of supernovae in dust production.}

\subsection{Motivation}

Core-collapse supernovae~(SNe) are one of the primary dust factories that account for the large reservoirs of dust observed in the early universe~\citep[e.g.,][]{sarangi+18}. However, it remains unclear how much dust an SN can produce and what the dust components are. Observations of young SN remnants (SNRs) are essential to answer these questions.

Previous observations have primarily focused on two young SNRs: Cas~A and the Crab Nebula. The SCUBA observations of Cas~A ($\sim$350 yr) with the JCMT revealed, for the first time, a cold dust component of $\sim$~2--4~$M_\odot$ and $\sim$~18~K~\citep{dunne+03}. Polarization observations of Cas~A with SCUPOL on the JCMT detected dust polarization fractions up to $\sim$30\% at 850~$\mu$m~\citep{dunne+09}, while SOFIA HAWC+ observations detected dust polarization fractions up to 20\% at 154~$\mu$m towards a smaller region of Cas~A~\citep{Rho+23}. The Herschel observations of another young SNR, the Crab Nebula ($\sim$1000~yr), also found a cool dust component of $\sim$~0.24~$M_\odot$ and $\sim$~28~K or $\sim$~0.11~$M_\odot$ and $\sim$~34~K for silicates and carbonaceous grains, respectively~\citep{gomez+12b}. Although \citet{deLooze+19} obtained a smaller mass for the cool dust component of the Crab Nebula, they found a high condensation efficiency, suggesting that supernovae could provide substantial contributions to the dust budget. Recently, polarized emission at 89 and 154~$\mu$m has been detected, indicating a cold dust component with a mass of about 0.1~$M_\odot$ for silicate grains~\citep{chastenet+22}.

Polarization observations of both Cas~A and the Crab Nebula not only confirm the existence of cold dust in young SNRs, but also differentiate carbonaceous and silicate grains. The next question is: what about other young SNRs?

\subsection{Objectives}

\begin{figure*}[!htbp]
    \centering
    \includegraphics[width=0.46\textwidth]{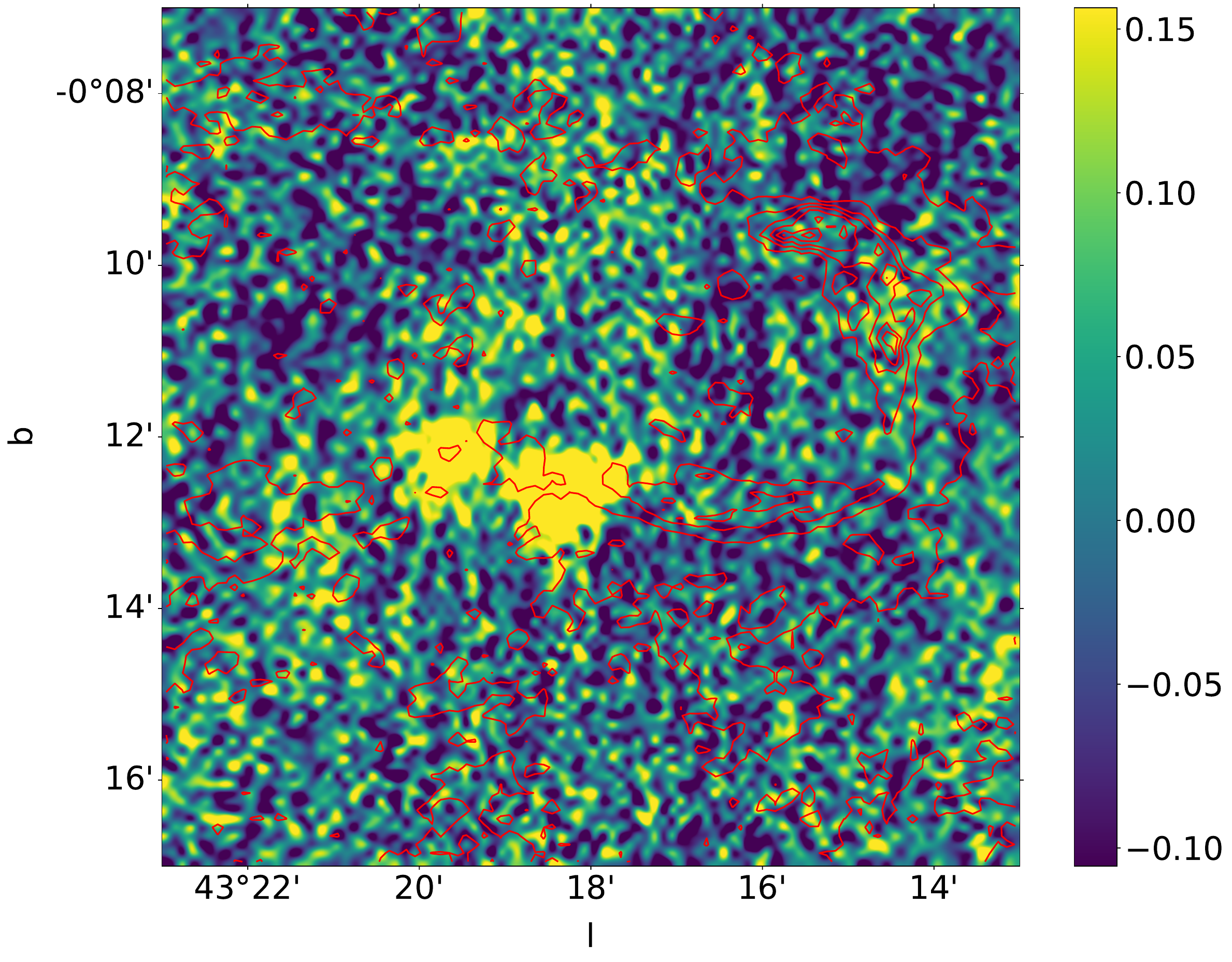}
    \includegraphics[width=0.46\textwidth]{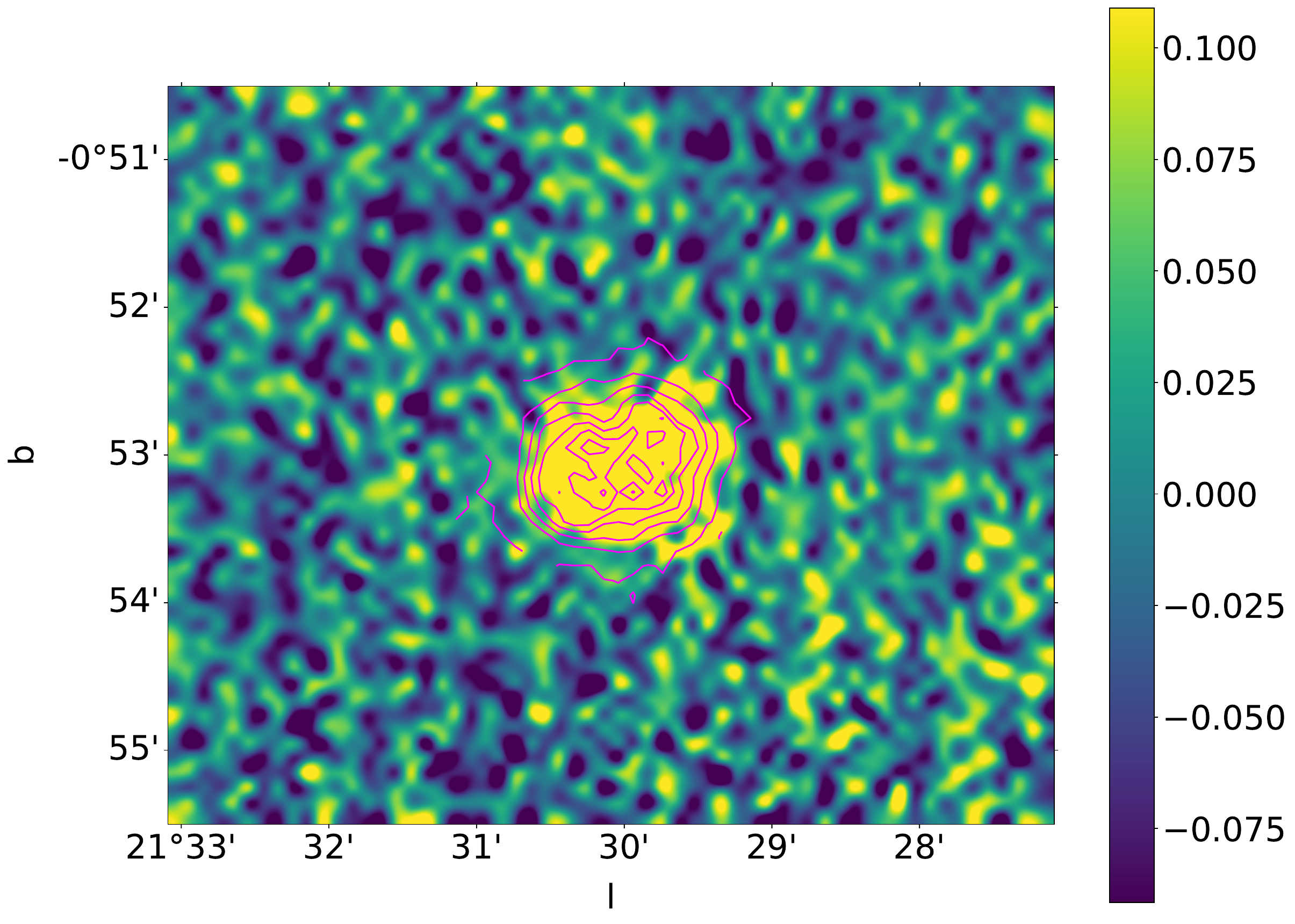}
    \caption{Supernova remnants: W49B and G21.5$-$0.9. Images from LABOCA on the APEX telescope are overlaid with the radio contours from VLA.}
    \label{fig:supernova remnants}
\end{figure*}

We propose to conduct a systematic observation of a sample of young SNRs~(Table~\ref{tab:snr}) with XSMT-15m. The sample is built as follows.
\begin{itemize}
\item[(1)] Core-collapse SNRs younger than $\sim$2~kyr selected from the catalog of \citet{chawner+20} with far infrared emission detected in the Hi-GAL survey~\citep{Molinari+10}. The resulting sample includes G11.2$-$0.3, G15.9+0.2, G21.5$-$0.9, G27.4+0.0 G29.7$-$0.3, G34.7$-$0.4, G43.3$-$0.2, and G54.1$+$0.3. Several of these SNRs are aslo associated with pulsar wind nebulae (PWNe).

\item[(2)] Cas~A, for which a complete polarization map at 353~GHz is still missed. 

\item[(3)] SNR 3C~58 ($\sim$800~yr) showing strong emission at 353~GHz~\citep{planck+16}. 

\item[(4)] The Crab Nebula, included to verify the polarization capabilities.
\end{itemize}

In addition to the core-collapsed SNRs described above, we propose to observe the Type Ia SNRs Kepler and Tycho. Dust formation in Type Ia explosions is expected to be challenging, as radioactive heating enhances grain destruction. Nevertheless, the model by \citet{nozawa+11} predicts that up to $\sim$0.2 $M_\odot$ of dust could still be produced. Observations of Kepler by \citet{morgan+03} claimed a dust mass of $\sim$1 $M_\odot$, whereas \citet{gomez+12a} found no evidence for cold dust. Polarization measurements at 353~GHz, which are highly sensitive to cold dust, will therefore provide stringent constraints on dust formation in Type Ia SNRs.

\begin{table*}[!htbp]
    \centering
    \caption{\textbf{Young SNRs to be observed.}
    \label{tab:snr}}
    \begin{tabular}{cccc}
    \hline\hline
\textbf{SNR} & \textbf{Type} & \textbf{Age (kyr)} & Reference \\
    \hline
G11.2$-$0.3     &  PWN           & 1.4-2.4 & \citet{Borkowski+16} \\
G15.9+0.2       &  SNR           & $\leq$2.4 & \citet{Reynolds+06, Klochkov+16}\\
G21.5$-$0.9     & PWN    & 0.87 & \citet{Bietenholz+11,Bocchino+05}\\
G27.4+0.0~(Kes 73) & SNR & 0.8-2.1 &\citet{Priestley+22}\\
G29.7$-$0.3~(Kes 75) & PWN & $<$0.84 & \citet{Morton+07}\\
G34.7$-$0.4    & PWN     & 2 & \citet{Wolszczan+91}\\
G43.3$-$0.2~(W49B)    & SNR & 1-4 & \citet{Zhu+14}\\
G54.1+0.3 & PWN & 2 & \citet{Bocchino+10}\\
G111.7$-$2.1 (Cas A) & SNR & 0.35 & \citet{Rho+23}\\
G130.7+3.1 (3C 58) & PWN &  0.8 & \citet{Bietenholz+13}\\
G184.6$-$5.8 (Crab Nebula) & PWN &0.97& \citet{Green+04}\\
G120.1+1.4 (Tycho)$^a$ &&0.4&\citet{gomez+12a}\\
G4.5+6.8 (Kepler)$^a$ & &0.4& \citet{gomez+12a}\\
    \hline\hline
    \end{tabular}
    \begin{tablenotes}
    \item $^a$: Type Ia SNRs.
    \end{tablenotes}
\end{table*}

With the polarization observations of the young SNR sample, we aim to address the following questions.

\begin{itemize}
    \item How does dust form and evolve during the lifetime of an SNR? Our sample spans ages from a few hundred to several thousand years, corresponding to different stages of reverse-shock evolution, when dust grains are processed and destroyed. As reviewed by \citet{micelotta+18}, the surviving dust mass remains roughly constant during the first few hundred years, but diverges at later times. Thus, measuring dust masses in SNRs of different ages provides critical constraints on models of dust processing in SNRs.
    
    \item What are the compositions of dust in SNRs? Traditionally, dust is modeled with two components: carbonaceous and silicate, with the latter contributing more to polarization because of the alignment with the magnetic field. Recently, \citet{hensley+23} proposed a one-component model with uniform compositions of carbon-rich and silicon-rich grains, called astrodust. On the basis of this model, a uniform fractional polarization from far-infrared to submillimeter can be reproduced. By observing a large sample of SNRs, we can directly compare the stardust produced in supernovae with that found in the ISM.
    
    \item What is the magnetic field traced by dust? The magnetic field in SNRs can be studied with radio polarization observations of the synchrotron emission. However, Faraday rotation by the magnetized medium inside and in front of the SNRs modulates polarized emission, which makes the measurement of the magnetic field challenging. In contrast, polarization observations at 850~$\mu$m experience virtually no Faraday rotation and thus trace the intrinsic magnetic field. 
    
\end{itemize}

For all these SNRs, there are low-frequency ($\lesssim$10~GHz) polarization observations that are dominated by synchrotron emission. Figure~\ref{fig:supernova remnants} illustrates this for W49B and G21.5$-$0.9. By extrapolating the observations to 345~GHz, we can obtain and subtract the contribution of synchrotron emission to isolate the dust polarization. The line emission by molecules, such as CO and HCN, may contribute to the total emission but will not affect polarization. This approach ensures that the detected submillimeter polarization can be robustly attributed to dust.


\section{TD: Protostellar Variability}

\textbf{Plain summary for non-experts:} \textit{Stars are born hidden within dense clouds, making it hard to observe their early growth. By monitoring how their brightness changes in submillimeter light, astronomers can track how baby stars gain mass—often in sudden bursts rather than steadily. China’s upcoming XSMT-15m telescope will enable the most sensitive and extensive survey yet, watching hundreds of young stars over a decade to uncover the true nature of stellar birth and solve the long-standing ``Luminosity Problem".}

\subsection{Motivation}

Stars form via the gravitational collapse of molecular cloud cores. The initial phase of the growth of a protostar occurs steadily, driven by the gravitational infall of material in the surrounding envelope ($\sim$ 1000--10000 AU). A protoplanetary disk ($\sim$ 0.1--100 AU) forms early in this process (e.g., \citealt{Jorgensen2008,Maud2019}) and channels most of the accreting material from the envelope to the protostar via a loss of angular momentum, likely due to viscous interactions and MHD instabilities.

Accretion from the disk onto the star is expected to be variable due to instabilities in both the inner and outer disk (see review by \citealt{Armitage2016}). Observationally, variability is a fundamental
property of accretion at the late (Class II) stages in the evolution of young stellar objects, as seen in both small routine bursts \citep{Cody2017} and in rare, huge bursts of $10^2$--$10^4$ times enhancement in accretion rate (FUor and EXor outbursts; see reviews by \citealt{Audard2014} and \citealt{Fischer2023}). The accretion variability that is thought to be prevalent during the main phases of stellar growth has far-reaching implications for many of the most important aspects of star formation, including reconciling the decades-old discrepancy between theoretical and observed luminosities of young stars known as the ``Luminosity Problem'' (e.g., \citealt{Kenyon1990,Dunham2010}), the physical and chemical structure of the circumstellar envelopes and the disk (e.g., \citealt{LeeJE2007, Harsono2015, Hsieh2019}), and the location of the stellar birth region that sets the initial conditions for subsequent stellar evolution (e.g., \citealt{Baraffe2009, Kunitomo2017}).

In this era of time-domain astrophysics, many wide-field and all-sky monitoring optical surveys, such as ASAS-SN, ZTF, and WFST, are revealing variability in optically bright objects, including some young stellar objects. However, during the dominant stages of stellar growth, the protostar is deeply embedded in its natal envelope and is not visible at short wavelengths. For protostars that are detected in the infrared, any variability may be caused by accretion, extinction, or even both (e.g., \citealt{Hillenbrand2013}), and is therefore challenging to interpret (e.g., \citealt{Yoon2022}). In contrast, any changes in the submillimeter are caused by a change in protostellar luminosity, and therefore the accretion rate (\citealt{Johnstone2013}). The envelope acts as a thermometer, with dust continuum emission that changes with the stellar luminosity (e.g., \citealt{Baek2020}).

Long-term monitoring of nearby star-forming regions with XSMT-15m can provide a scientific legacy by testing models for stellar assembly with variability measurements for far more objects than currently possible. The first dedicated submillimeter monitoring program, the JCMT Transient Survey (\citealt{Herczeg2017}), established that protostellar variability in the submillimeter is common (e.g., \citealt{Mairs2024, ChenZhiwei2025}). Compared to JCMT/SCUBA-2, the improvement in sensitivity of XSMT-15m will facilitate a larger program, with sufficient statistics to test models of protostellar variability by including fainter objects and increasing the chances for catching a rare, large outburst. Spectral indices from the 3-band monitoring would allow us to correct for interstellar heating of envelopes and better interpret variability, including from AGN and coronal flares.

\begin{figure*}[!htbp]
    \centering
    \includegraphics[width=\textwidth]{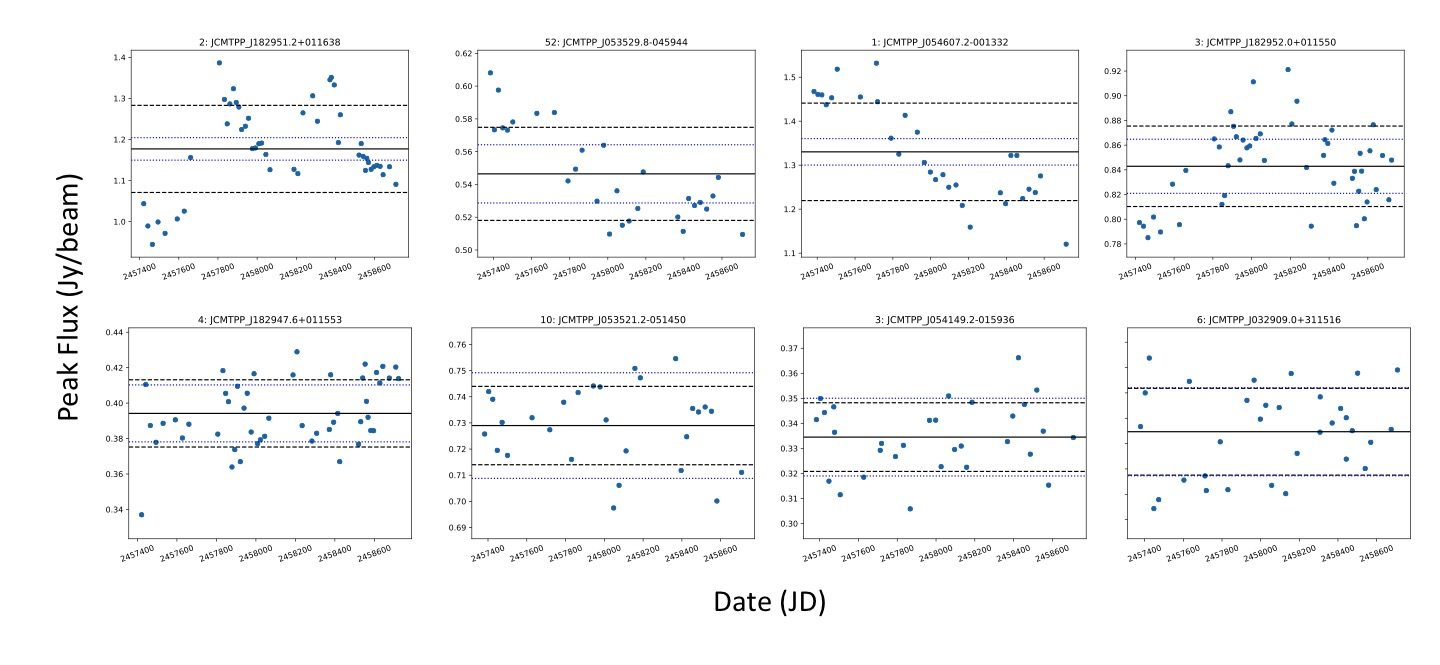}
    \caption{%
    Sub-mm light curves of four variables (top panels) and four non-variables (bottom panels) from the JCMT Transient program \citep{LeeYH2021, Mairs2024}.
    The blue dotted lines indicate the 1-$\sigma$ range expected for the source brightness, and the black dashed lines indicates the measured range for the source. Note that the y-axes (flux ranges) differ for each object.
    }
    \label{fig:protostellar-variability-figure-1}
\end{figure*}

\subsection{Objectives}

Most protostellar luminosities fall far below those expected from energy release by steady accretion over protostellar lifetimes, known as the Luminosity Problem. The resolution of this discrepancy likely requires a time dependence in the accretion rate, either in episodic bursts (e.g., \citealt{Dunham2010}) or in a rapidly decaying accretion rate (\citealt{Fischer2017}). The primary long-term goal of submillimeter monitoring of star-forming regions is to directly test the episodic accretion burst hypothesis by measuring the frequency, size, and duration of large outbursts in protostars.

\begin{figure*}[!htbp]
    \centering
    \includegraphics[width=\textwidth]{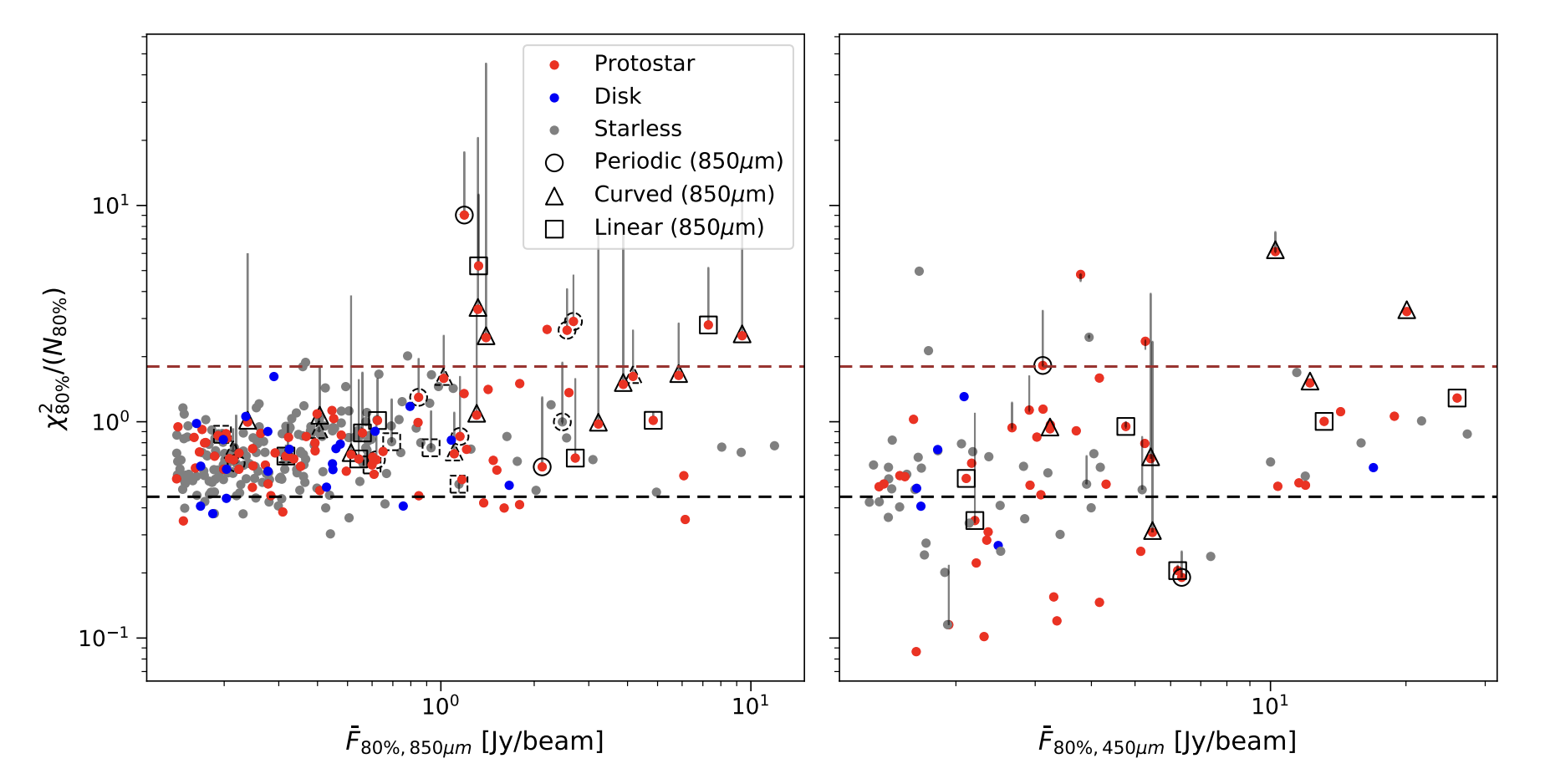}
    \caption{From \citet{Mairs2024}, scatter plot of chi-square-fit value divided by the number of observations for each source at 850~$\mu$m (left) and~450 $\mu$m (right), with variables marked. The robust variables at 850~$\mu$m are also annotated. Most variables are brighter than 1~Jy~beam$^{-1}$ because of sensitivity limits.}
    \label{fig:protostellar-variability-figure-2}
\end{figure*}

The JCMT Transient survey monitored eight nearby low-mass star-forming regions for 8 years (2016--2023) and six intermediate-mass star-forming regions for 3.5 years. Light curves for eight selected objects (four submillimeter variables and four submillimeter non-variables) are shown in Figure~\ref{fig:protostellar-variability-figure-1}. 
In the low-mass star-forming regions, $\sim$50\% of all bright sources show variability, often at the 2–3\%/year level (Figure~\ref{fig:protostellar-variability-figure-2}). Submillimeter variations are proportional to $L^{0.25}$ (e.g., \citealt{Baek2020}), so the corresponding changes in luminosity and accretion rate are much larger, though modest. The detection of variability in faint sources is still uncommon because of limited sensitivity.

Expansion to more distant intermediate mass star-forming regions added several variables (\citealt{Park2024}; \citealt{ChenZhiwei2025}; Zhang et al. in prep). Identification and interpretation of variables improve when submillimeter monitoring is combined with mid-IR monitoring (NEOWISE, \citealt{ContrerasPena2020, ChenZhiwei2025}). These intermediate regions offer many more potential sources, along with greater challenges due to interstellar heating and source confusion.

The roughly $\sim$20 variable sources have revealed fascinating case studies. Cyclical accretion bursts of EC 53 reveal the filling and draining of the inner accretion disk (\citealt{LeeYH2020}), with extensive JWST and ALMA follow-up to assess the implications for disk chemistry (e.g., Lee et al. submitted). The combination of single-dish with interferometric (SMA and ALMA) monitoring allows the mapping of light echoes, similar to reverberation mapping (\citealt{Francis2022}). Other submillimeter variables, including HOPS 373 (\citealt{Yoon2022}) and HOPS 358 (\citealt{Sheehan2025}), provide powerful laboratories for outflows and envelope chemistry.

In total, the JCMT Transient Survey has monitored roughly 250 sources bright enough for variability analysis. The submillimeter monitoring at XSMT-15m will be $\sim$10 times faster than SCUBA-2, opening up new opportunities for transient science. This faster monitoring will allow an increase in the number to $\sim$500 sources in $\sim$20 fields. If a protostellar outburst occurs every 103 years (\citealt{Fischer2019}), then we should detect 5 large bursts in a 10-year survey, sufficient for robust tests of variability in stellar assembly.

A robust submillimeter monitoring program would be enhanced with coordination of complementary modeling and observational programs. Radiative transfer modeling of bursts through envelopes is necessary to interpret submillimeter variability, including placing constraints on envelope properties (e.g., \citealt{Baek2020}). The accretion variability provides a robust test of disk instabilities, predicted in magnetohydrodynamic models (e.g., \citealt{ZhuZhaohuan2010}). Coordinated outflow monitoring could allow us to convert outflow emission into accretion histories, thereby providing a thousand-year record of accretion (e.g., \citealt{Takahashi2024, Dutta2024, ZhouWei2024}). Submillimeter variables have provided powerful case studies for chemical and outflow properties, with follow-up from ALMA and other facilities (e.g., \citealt{LeeJE2019, LeeYH2020, Yoon2022, Sheehan2025}).

Transient science in the submillimeter may also have powerful applications to other fields, including dust-obscured quasars, radio flares from stellar coronae, supernova discovery and feedback, and AGB stars. The JCMT Transient program has already detected one coronal flare, possibly the strongest radio flare ever recorded from a stellar-like coronae \citep{Mairs2019}, and two AGN \citep{Johnstone2022,Park2024}. The stacked observations from a monitoring survey will also provide exquisite images for measuring faint filaments, very low luminosity protostars, and disks.

The goal of this monitoring program is to assess sustained changes in the luminosity of protostars. Following the JCMT Transient Survey, the optimal cadence is roughly one visit per month per field. The envelope acts as a bolometer, with light travel time and scattering smoothing out any variability over $\sim$2 weeks. Monthly monitoring maximizes the sensitivity to weak changes over long timescales and facilitates stacking to assess variability in faint objects.

Integration times should be set to achieve a 5-$\sigma$ sensitivity of $\sim$20~mJy~beam$^{-1}$ at 850~$\mu$m across a 30\arcmin$\times$~30\arcmin\,field, which should be achievable in roughly 15~min. For comparison, SCUBA-2 achieves a 1-$\sigma$ sensitivity of 12~mJy~beam$^{-1}$ across a 30\arcmin\,diameter field in 40~min integration, so XSMT-15m would be a much faster survey. 

With this advantage of the KID camera over SCUBA-2 bolometer arrays, we could reasonably survey $\sim$500 protostars in $\sim$20 regions with 9 epochs per year, or a total of 45 hrs per year (preliminary estimates, not considering overheads). The total program should last 10 years to provide robust and long-term statistics. The northern winter is ideal for monitoring star formation, highlighted by Perseus (300~pc), Orion (400~pc), and Taurus (400~pc) along with more distant intermediate-mass star-forming regions in Cygnus (1~kpc). Evaluation of specific fields will be required for a final selection.

The high-mass star-forming regions, like Cygnus, are crowded and more confused. Nevertheless, each selected region includes enough bright, concentrated point sources to initially employ a similar approach to flux calibration. The confusion due to extended emission in these regions, however, introduces new challenges in tracking source stability that will require some efforts and innovation to overcome if we are to produce the same unprecedented sensitivity levels as we achieved for the low mass regions.

A monitoring program should require a 1\% calibration accuracy for bright sources in flux calibration at 850 $\mu$m (and 5\% at shorter wavelengths), feasible through auto-correlation functions applied to data \citep{Mairs2024}. With a 5$\sigma$ sensitivity of 20~mJy~beam$^{-1}$, that means that for a source brightness of 400~mJy~beam$^{-1}$, the flux calibration uncertainty (1$\sigma$ = 1\% = 4~mJy) and random noise (1$\sigma$ = 4~mJy) would be equal at 400~mJy~beam$^{-1}$. Current detections are heavily biased to targets brighter than 1~Jy~beam$^{-1}$ because of the sensitivity limits of SCUBA-2.

This flux calibration places stringent requirements on the quality of the data reduction, including pointing, beam shape, and contrast. The process of developing a pipeline for monitoring will lead to improvements in understanding the data, with benefits for all programs.

\section{TD: Submillimeter Time-domain Astronomy (High-Energy)}

\textbf{Plain summary for non-experts:} \textit{Submillimeter time-domain astronomy focuses on transient and explosive cosmic phenomena within the submillimeter wavelength range, bridging gaps between radio and infrared observations. This emerging field aims to uncover extreme astrophysical processes, such as relativistic jet dynamics, neutron star mergers, and magnetar-driven bursts, by leveraging the unique sensitivity of submillimeter emission to probe low-opacity environments and high-energy particle acceleration. Key challenges include achieving high-level time resolution, sub-mJy sensitivity, and coordinated multi-messenger observations to revolutionize our understanding of compact objects and cosmic transients.}

\subsection{Motivation}
With the rapid development of cosmic transient phenomena in recent years, time-domain astronomy has emerged as one of the most promising fields for revolutionary breakthroughs in modern astronomy, which was listed as one of the three most critical research areas in the Astro2020 Decadal Survey. Transients are typically associated with compact objects (neutron stars, black holes), including electromagnetic counterparts of gravitational waves (GWs), gamma-ray bursts (GRBs), fast radio bursts (FRBs), supernovae (SNe), tidal disruption events (TDEs) and so on. Studying these phenomena can reveal the formation, evolution, and death of compact objects. The radiation from these explosive events spans nearly the entire electromagnetic spectrum, making multi-wavelength and multi-messenger observations essential for uncovering physics under extreme conditions, such as the equation of state of ultra-dense matter, physical processes in extremely strong magnetic fields, and the acceleration and propagation of ultra-high-energy particles.

In the past two years, multiple domestic ground- and space-based telescopes have been completed (e.g., WFST, Mephisto, EP, and SVOM) and are now focusing on transient phenomena. Internationally, very large survey projects like LSST, Euclid, and WFIRST will also become operational soon. Compared to high-energy, optical, and radio bands, submillimeter time-domain astronomy remains relatively underdeveloped. Current submillimeter telescopes such as JCMT, ALMA, and IRAM have limited engagement in time-domain studies, leaving significant gaps in submillimeter observations of these cosmic transients. Dedicated observational efforts in this regime could yield groundbreaking discoveries from scratch. Furthermore, the novel submillimeter window could fill critical gaps in the spectral energy distribution of these transients, enabling deeper insights into the nature of compact objects and associated extreme physical processes. XSMT-15m is expected to play an important role in this respect.


\subsection{Objectives}
With the advent of the XSMT-15m telescope, we will be able to promptly track sources of interest. Accordingly, we aim to address the following scientific objectives:
\begin{itemize}
    \item{\it Detecting submillimeter counterparts of GW events.} Theoretical predictions have suggested that binary neutron star mergers or neutron star-black hole mergers would produce rich multi-wavelength electromagnetic signals. The milestone event GW170817 showed notable gamma-ray, X-ray, optical and radio signals \citep{Abbott2017}, however, ALMA conducted multiple observations between 1.4 and 44 days post-merger but detected no submillimeter signal \citep{Kim2017}. The collision between merger ejecta with circumstellar material would generate bright submillimeter radiation if the merger remnant is a long-lived massive neutron star \citep{GaoH2013}. Successfully capturing this signal in the submillimeter band would be a landmark achievement in multi-messenger astronomy.
    
    \item{{\it Detecting submillimeter prompt and afterglow emission of GRBs.} Although a general physical framework for GRBs has been established, fundamental questions remain unresolved, including jet composition and magnetic field configuration, prompt emission mechanisms, and others.  High-energy observations alone are insufficient to distinguish between competing models. To date, no GRB prompt emission has been detected in the submillimeter band, and only a handful of well-monitored afterglow events exist in this regime \citep{Laskar2019}. Future progress urgently requires expanding the submillimeter sample of GRBs to bridge the spectral gap between radio and optical wavelengths. This will enable comprehensive characterization of the SED and evolution of accelerated electrons, constrain jet composition and properties of the circum-burst environment (e.g., dust torus radiation), and resolve key physical uncertainties.}

    \item{{\it Detecting submillimeter counterparts of FRBs.} FRBs represents one of the most rapidly advancing frontiers. Multi-wavelength counterpart observations remain the most effective approach to unraveling FRB radiation mechanisms and origins. However, beyond the radio band, only FRB 200428 has been definitively linked to X-ray bursts from the Galactic magnetar SGR 1935+2154 \citep{CHIME2020}, demonstrating that magnetars can produce FRBs. Current FRB observations are limited to frequencies up to 8 GHz, yet studies suggest magnetars can emit submillimeter pulses \citep{Torne2017}. This raises a critical question: can FRB energy spectra extend into the submillimeter regime?  Detecting submillimeter bursts would open an entirely new observational window for FRB research. Furthermore, FRBs may produce multi-wavelength afterglows \citep{Yi2014}, including submillimeter emission.}

    \item{{\it Performing systematic submillimeter surveys of SNe.} Over the past three decades, the field of SN research has undergone rapid growth. Nevertheless, fundamental questions about SN physics remain unresolved, and the increasing discovery of peculiar SNe continues to challenge traditional theoretical models. Theoretical studies predict that interaction-powered SNe could produce strong radio-submillimeter radiation \citep{Bietenholz2021}. However, systematic submillimeter surveys of SNe have yet to be conducted due to the limited sensitivity of small-aperture telescopes and the demanding follow-up requirements of large facilities. This observational gap represents a critical missing piece in SN studies. Future submillimeter observations of SNe could provide critical constraints on explosion physics, circumstellar material distribution, and potential late-time dust torus radiation.}
    
    \item{{\it Follow-up observations of TDEs in the submillimeter band.} TDEs evolve over relatively long timescales—lasting weeks to years—making them observationally accessible. A subset of TDEs produce relativistic jets that accelerate particles and generate multi-wavelength non-thermal radiation, linking TDEs to particle acceleration and galactic nuclear environments \citep{Yuan2016}. Compared to low-frequency radio emission, submillimeter radiation offers a crucial observational window for early-phase non-thermal emission in TDEs due to its lower optical depth, enabling critical studies of TDE physics at extremely early stages. Key scientific questions in TDE research include: statistical properties of quiescent SMBHs and their environments, the incidence and mechanisms of relativistic jet production in TDEs, and the evolution of black hole jets and high-energy particle acceleration mechanisms.}
    
\end{itemize}

\section{AC: Molecules in Diverse Environments}

\textbf{Plain summary for non-experts:} \textit{Molecules --- ranging from the simplest to the more complex --- exist throughout the Universe, from interstellar space to planets. They provide critical insights into how galaxies evolve, stars form, and planets like Earth come to be. With the XSMT-15m telescope, we aim to carry out sensitive observations of these molecules across a wide range of astrophysical environments, thereby uncovering how local physical conditions shape the formation, distribution, and evolution of these molecules over time.}

\subsection{Motivation}

The first interstellar molecules were detected more than 80 years ago \citep{1937ApJ....86..483S,1940PASP...52..187M}, but astrochemistry began to flourish only in the 1960s with detections of NH$_{3}$, H$_{2}$O, and H$_{2}$CO \citep[e.g.,][]{2013ChRv..113.8707H,2022ApJS..259...30M}. Since then, molecular observations have played a crucial role in advancing our understanding of a wide range of astrophysical processes --- from interstellar chemical evolution \citep[e.g.,][]{1998ARA&A..36..317V,2009ARA&A..47..427H}, to the formation of planets \citep[e.g.,][]{2011ApJ...740...14O}. To date, more than 330 molecular species have been identified in various environments, including diffuse and translucent clouds, dense dark clouds, active star-forming regions, and chemically rich circumstellar envelopes surrounding evolved stars \citep[e.g.,][]{websiteOfCDMS,2022ApJS..259...30M}. The chemical richness of the ISM --- spanning from simple diatomic species to complex organic molecules --- encodes the evolutionary history of interstellar gas and links the processes in molecular clouds to the material reservoirs that shape planetary systems and potentially seed the ingredients for life \citep[e.g.,][]{2009ARA&A..47..427H,2012A&ARv..20...56C,2016ARA&A..54..181G}. Therefore, exploring the molecular inventory across various environments is essential for understanding how molecules evolve with the evolution of the ISM and ultimately for bridging the connection to the origin of life on Earth (see Figure~1 in \citealt{2012A&ARv..20...56C}, for instance).

Molecular line surveys are a key tool for probing the physical and chemical conditions of astronomical sources, offering the only means to obtain a complete and unbiased inventory of their molecular content. Despite numerous existing line surveys and the long history of submillimeter telescopes, comprehensive spectral line surveys within the frequency range of 440--504~GHz remain rare. Existing line surveys overlapping this frequency range are summarized in Table~\ref{tab:linesurvey}, highlighting that this part of the spectrum remains relatively unexplored and that many astrophysical environments have yet to be systematically studied. Thus we propose a systematic characterization of the spectral properties of prominent sources across different evolutionary stages in this frequency range. This approach is expected to reveal numerous previously unidentified spectroscopic features and shed new light on the evolution of the molecular universe at submillimeter wavelengths.

\begin{figure*}[!htbp]
    \centering
    \includegraphics[width=0.95\textwidth]{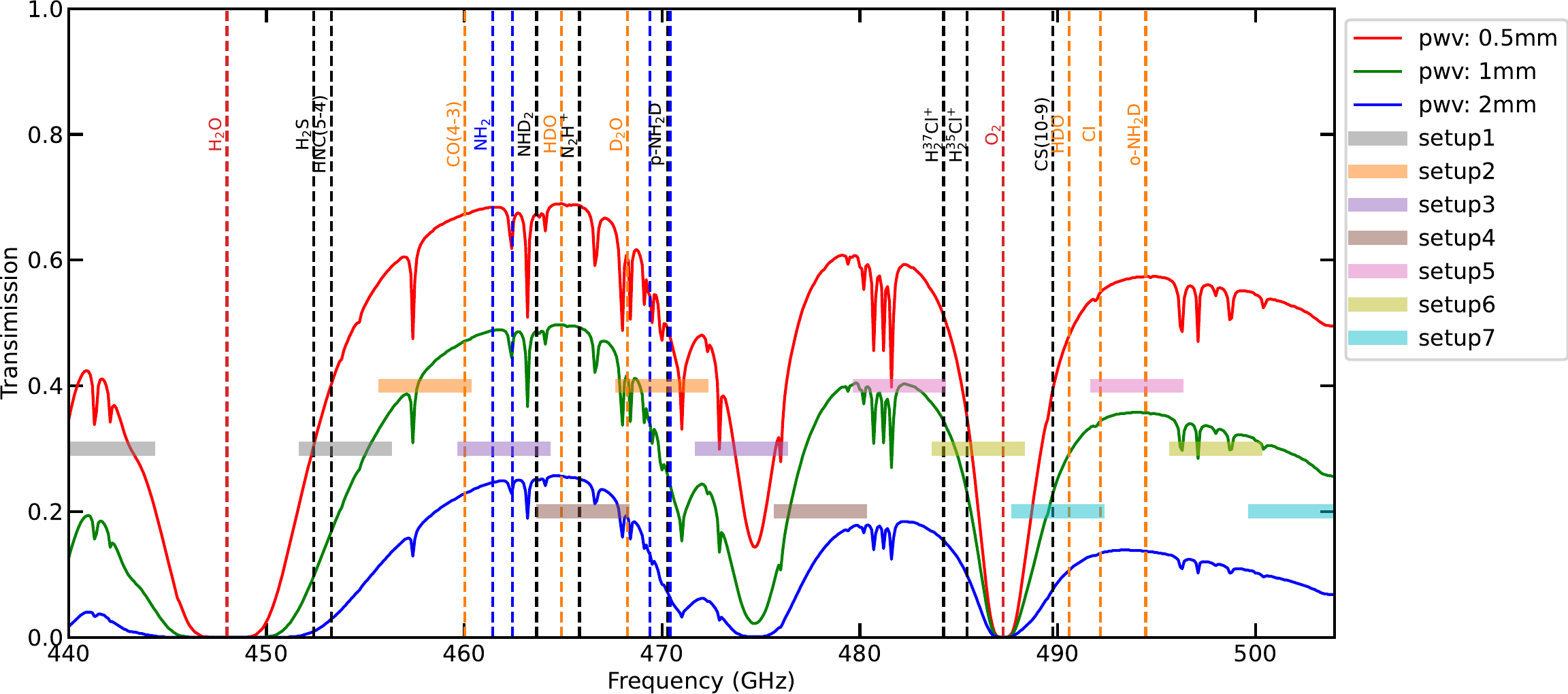}
    \caption{Selected frequency ranges with major target molecular lines indicated. The red, green, and blue solid lines represent the atmospheric transmission at a typical PWV value of 0.5~mm, 1~mm, and 2~mm, respectively. The proposed frequency setups are indicated by the horizontal lines.}
    \label{fig:xsmc-line}
\end{figure*}

Compared to other telescopes, XSMT-15m stands out because of its exceptional access to high-frequency bands in the northern sky, enabled by the extremely low PWV at its site. Figure~\ref{fig:xsmc-line} presents the atmospheric transmission curves at different PWV levels, overlaid with a selection of molecular transitions in the 440--504 GHz window. The transmission curves were estimated using the model of \citet{Pardo2001}, demonstrating the critical importance of favorable weather conditions for observations at these wavelengths. Moreover, several sources listed in Table~\ref{tab:linesurvey} have only been investigated with the Herschel Space Observatory. Even for the same sources, XSMT-15m will significantly reduce beam dilution effects and alleviate line confusion issues thanks to its smaller beams. Therefore, XSMT-15m is expected to deliver a substantial improvement in this frequency range compared to currently available northern-hemisphere facilities.

\begin{table*}[!htbp]
    \centering
    \caption{\textbf{Existing line surveys at the frequency range of 440--504~GHz}
    \label{tab:linesurvey}}
    \begin{tabular}{ccccp{6cm}}
    \hline\hline
\textbf{Source} & \textbf{Class$^a$} & \textbf{frequency range} & \textbf{Telescope} & \textbf{Reference} \\
    \hline
Orion KL        &  HMSF           & 190--900~GHz$^b$ & CSO-10m  & \citet{1995ApJ...451..238S} \\
Orion KL        &  HMSF           & 455--507~GHz & JCMT-15m & \citet{2003AA...407..589W} \\
Orion KL        &  HMSF           & 486--492~GHz & ODIN-1.1m & \citet{2007AA...476..791O} \& \citet{2007AA...476..807P} \\
Orion KL        &  HMSF           & 480--1907~GHz & Herschel-3.5m & \citet{2014ApJ...787..112C} \\
OMC-2 FIR 4     & IMSF            & 480--1907~GHz & Herschel-3.5m & \citet{2013AA...556A..57K} \\
AFGL 2591       & HMSF            & 480--1907~GHz & Herschel-3.5m & \citet{2014AA...567A..53K} \\
Sgr B2(M)       & HMSF            & 480--1907~GHz & Herschel-3.5m & \citet{2021AA...651A...9M} \\
Sgr B2(N)       & HMSF            & 480--1907~GHz & Herschel-3.5m & \citet{2014ApJ...789....8N} \\
NGC6334I        & HMSF           & 480--1907~GHz & Herschel-3.5m & \citet{2012AA...546A..87Z} \\
Orion South     & HMSF           & 480--1900~GHz & Herschel-3.5m & \citet{2016ApJ...832...12T}\\
Orion Bar       & PDR            & 480--1907~GHz & Herschel-3.5m & \citet{2017AA...599A..22N} \\
IRC +10216      & C-rich star    & 488--1907~GHz & Herschel-3.5m & \cite{2014ApJ...787..112C}$^c$ \\
CK Vul          & red nova       & 440--494~GHz$^d$ & APEX-12m & \citet{2017AA...607A..78K} \\
NGC6334I        & HMSF           & 459--462~GHz  & APEX-12m & \citet{2006AA...454L..41S} \\
G327.3$-$0.6    & HMSF           & 459--462~GHz  & APEX-12m & \citet{2006AA...454L..41S} \\ 
\hline\hline
\end{tabular}
    \\
    \noindent
    \textbf{Note.} $^{a}$Source types. SFR=Star formation region; IMSF=Intermediate-mass star formation; HMSF=High-mass star formation; PDR: photodissocation region; $^{b}$A coarse spectral resolution of 0.2~GHz. $^{c}$Only the frequency range of 554.5--636.5GHz is shown in this work; $^{d}$Sparsely covered.
\end{table*}

\subsection{Objectives}
To better investigate molecular evolution across different evolutionary stages, we divide the sample into three categories: low-mass star-forming region, high-mass star-forming region, and evolved stars. Each category is discussed in the following sections.

\textbf{Low-mass star formation.} The molecular complexity of Solar System bodies is most likely intimately related to the earliest stages of star formation \citep[e.g.,][]{2012A&ARv..20...56C}. Understanding molecular evolution throughout the formation of low-mass stars is crucial, as it sheds light on how simple interstellar molecules are transformed and assembled into the more complex chemical species that may eventually seed the ingredients for life. Hence, we selected a sample of prototypical sources listed in Table~\ref{tab:targets_ls}. These sources were selected from the Large Program ``Astrochemical Surveys At IRAM'' \citep{2018MNRAS.477.4792L}. These sources are well-established benchmarks, representing diverse environments at different stages of low-mass star formation. Notably, while previous surveys have covered frequencies up to 272~GHz, the higher-frequency range of 440--504~GHz remains largely unexplored. This makes these sources particularly well-suited for the present study.

\textbf{High-mass star formation.} Massive star formation is a fundamental process that enriches the universe with heavy elements, drives powerful feedback that shapes the structure and evolution of galaxies, and produces luminous beacons that trace cosmic star formation across time \citep[e.g.,][]{2007ARA&A..45..565M}. Crucially, molecular evolution within the parental clouds regulates thermodynamic balance by changing the cooling efficiency, and affects fragmentation pathways that set the initial conditions for the birth of massive stars. Tracking the evolving molecular composition not only provides critical diagnostics of the physical conditions, chemical timescales, and feedback processes at play, but also offers unique insights into environments that differ markedly from those of low-mass star formation regions \citep[e.g.,][]{2020ARA&A..58..727J}. Studying molecular evolution across massive star formation environments is essential to understanding how chemical complexity develops under the specific conditions required for massive star formation, offering a crucial test bed for models of star formation, feedback, and the cycling of matter in galaxies. In this work, we selected a sample of prominent sources (Table~\ref{tab:targets_ls}) widely recognized as prototypical regions, encompassing a diverse range of physical environments and evolutionary stages characteristic of high-mass star formation. Therefore, they serve as prime targets for a comprehensive spectral line survey aimed at understanding molecular evolution during massive star formation.

\textbf{Evolved stars.} The molecular envelopes of evolved stars represent a crucial laboratory for studying the complex interplay between stellar evolution, mass loss, and circumstellar chemistry \citep[e.g.,][]{2006A&A...456.1001C,2018A&ARv..26....1H,2020A&A...637A..59A}. As stars evolve from the AGB phase through the protoplanetary nebula (PPN) stage and into planetary nebulae (PNe), their molecular envelopes undergo dramatic changes in morphology, chemistry, and physical conditions \citep[e.g.,][]{2011IAUS..280..203K}. These transitions mark the recycling of processed stellar material into the ISM, enriching it with dust and complex molecules that play essential roles in galactic chemical evolution and the formation of future stars and planetary systems. Despite their importance, the detailed molecular evolution across these phases remains poorly understood. We selected eight typical sources  to investigate the chemical footprints of C-rich (C/O $>$ 1), O-rich (C/O $<$ 1), and S-type (C/O $\approx$ 1) AGB stars, along with the massive star, protoplanetary nebula, and planetary nebula. These sources have been previously targeted for line surveys at other frequency ranges (see Table~\ref{tab:targets_ls}) or observed at other frequencies, making them well-suited for a spectral line survey of evolved stars.

\begin{table*}[!htbp]
    \centering
    \caption{\textbf{Source sample and their physical properties.}
    \label{tab:targets_ls}}
    \begin{tabular}{ccccc}
    \hline\hline
\textbf{Source} & \textbf{Class$^a$} & \textbf{Distance (pc)} & \textbf{Comments$^a$} & \textbf{Reference$^{b}$} \\
    \hline
    \multicolumn{5}{c}{Low-mass star formation} \\
    \hline
TMC--1            & starless   &  140  & unsaturated COMs & 1,2,3 \\
L1544           & prestellar &  140  & high deuteration & 4 \\
NGC1333~IRAS4A  & Class 0    &  260  & hot corino       & 5 \\
L1527           & Class 0    &  140  & WCCC             & 6 \\
SVS13A          & Class I    &  260  & hot corino       & 7,8 \\
L1448R2         & Jet        &  300  & EHV bullet   & 9  \\
L1157B1         & Bow-shock  & 250   & MHD shock    & 10\\
\hline
\multicolumn{5}{c}{High-mass star formation} \\
\hline
G0.253+0.016    & IRDC  & 8150 & quiescent, the YMC precursor  & 11 \\
G+0.693-0.027   & shock & 8150 & quiescent, Nitriles  & 12 \\
Sgr B2          &  HC/UCHII   & 8150 & COM-rich  & 13 \\
Orion KL        &  HC   & 414  & energetic event & 14, 15 \\
Orion Bar       &  PDR  & 414  & FUV  & 16 \\
G34.26+0.15     & HCHII/UCHII & 1560 & COM-rich  & 17 \\
W3OH            & UCHII & 1950 & COM-rich  &  17, 18 \\
HH 80$-$81      & jet/outflow & 1700 & EHV & 17, 19 \\
\hline
\multicolumn{5}{c}{Evolved stars} \\
\hline
IRC +10216     & AGB          & 130 & C-rich, exotic  &  20, 21\\
IK Tau         & AGB          & 289 & O-rich  & 22  \\
Chi Cyg        & AGB          & 150 & S-type  & 23  \\
W Aql          & AGB          & 395 & S-type  & 24  \\                 
VY CMa         & RSG          & 1200 & O-rich, Ti-bearing  & 25, 26 \\
CRL 618        & PPN          & 900 & transition, methylpolyynes  & 27, 28 \\
CRL 2688       & PPN           & 1000 &  C-rich  & 29 \\
NGC 7027       & PN           & 980 & CO$^{+}$, HeH$^{+}$  & 30, 31 \\
\hline\hline
\end{tabular}
    \\
    \noindent
    \textbf{Note.} $^a$Source types. HC: hot core; IRDC: infrared dark cloud; PDR: photodissocation region; HCHII: hypercompact H{\scriptsize II} region; UCHII: ultracompact H{\scriptsize II} region. AGB: Asymptotic giant branch. RSG: Red supergiant. PPN: Protoplanetary nebula. PN: Planetary nebula. COMs=Complex Organic Molecules; WCCC: Warm carbon chain chemistry; EHV: Extremely high velocity; MHD: magnetohydrodynamics shock; YMC: young massive cluster; MYSO: massive young stellar object. $^{b}$: 1=\citet{2021NatAs...5..188L}; 2=\citet{2021A&A...652L...9C}; 3=\citet{2004PASJ...56...69K}; 4=\citet{2013ApJ...769L..19S}; 5=\citet{2007A&A...463..601B}; 6=\citet{2008ApJ...672..371S}; 7=\citet{2009ApJ...691.1729C}; 8=\citet{2016MNRAS.462L..75C}; 9=\citet{2013A&A...549A..16N}; 10=\citet{1997ApJ...487L..93B}; 11=\citet{2012ApJ...746..117L}; 12=\citet{2020ApJ...899L..28R}; 13=\citet{2013A&A...559A..47B}; 14=\citet{2015A&A...581A..48G}; 15=\citet{2014ApJ...787..112C};
    16=\citet{2017AA...599A..22N}; 17=\citet{2017ApJS..232....3W}, 18=\citet{1997A&AS..124..205H}; 19=\citet{2009ApJ...702L..66Q}; 20=\citet{2015A&A...574A..56G}; 21=\citet{2000A&AS..142..181C};
    22=\citet{2022A&A...667...A74}; 23=\citet{2010A&A...521...L6};
    24=\citet{2021A&A...655...A80}; 25=\citet{2012ApJ...744...23Z}; 26=\citet{2013A&A...551A.113K}; 27=\citet{2001ApJ...546L.123C}; 28=\citet{2007ApJ...661..250P}; 29=\citet{2022ApJS...259...56};
    30=\citet{1993ApJ...419L..97L};31=\citet{2019Natur.568..357G}. 
\end{table*}

In order to study molecules in diverse environments, we propose to carry out the 440--504~GHz spectral line survey toward the selected targets with XSMT-15m. The immediate objectives are the following:

\begin{itemize}
    \item Comprehensive molecular inventory. Establish a complete census of molecular species across diverse astrophysical environments within the 440--504~GHz window, providing a chemical template that can serve as a reference for studies of molecular complexity in the Universe.
    
    \item Chemical differentiation across environments. Reveal how molecular compositions vary across astrophysical environments and uncover the key environmental drivers that shape chemical diversity.

    \item Tracing chemical evolution. Map the chemical evolution along the pathways of low-mass star formation, high-mass star formation, and late stellar evolution, delivering empirical benchmarks that inform and constrain models of chemical evolution.

    \item Constraining astrochemical models. Use the measured molecular abundances to test, challenge, and refine cutting-edge astrochemical models, exploiting the survey's diverse source sample as a natural laboratory spanning a wide range of ages and physical conditions.
     
    \item Expect the unexpected. Open a discovery space for serendipitous findings, including previously unreported molecular transitions or even new species, thereby expanding the cosmic chemical territory and enriching our understanding of the molecular Universe.

\end{itemize}

The complete dataset will serve as an invaluable legacy for the entire community, including astrophysicists, astronomers, chemists, spectroscopists, and theoreticians.

\section{AC: Synergy with Galactic Chemical Evolution}
\label{sec:AC-Galactic-Chemical}

\textbf{Plain summary for non-experts:} \textit{Astrochemistry has made substantial progress in explaining molecular abundances in diverse Galactic environments under solar-like CNO abundance ratios. However, challenges persist in extreme environments such as the outer Galactic disk, nuclear regions, and metal-poor dwarf galaxies, where observed molecular line ratios frequently deviate from standard model predictions. These discrepancies likely arise from variations in CNO abundance ratios ([C/O], [N/O]) influenced by galactic chemical evolution, localized stellar feedback, and environmental factors such as UV radiation fields and cosmic-ray ionization rates (CRIR). This paper outlines a systematic observational and modeling campaign to map CNO-bearing molecules across metallicity gradients in the Milky Way and nearby galaxies. By combining multi-line surveys at 1.3 and 0.6~mm with isotopic (e.g., \(^{13}\mathrm{C}\)) and ionized species diagnostics, we aim to establish empirical constraints on astrochemical networks in non-solar abundance regimes. Key goals include resolving the origin of CO-dark gas, calibrating metallicity-dependent conversion factors, and quantifying ionization states to infer cosmic ray variations.}

\subsection{Motivation}
The ISM's chemical evolution is fundamentally shaped by the nucleosynthetic origins of carbon and oxygen. Oxygen, synthesized predominantly in massive stars ($M \gtrsim 8\,M_\odot$) through the ${}^{12}\text{C}(\alpha,\gamma){}^{16}\text{O}$ reaction during hydrostatic helium burning \citep{Romano2022}, is rapidly dispersed into the ISM via core-collapse supernova explosions. This contrasts sharply with carbon production, which exhibits a bimodal origin: while massive stars contribute through the triple-$\alpha$ process, a significant fraction of its source emerges from low- to intermediate-mass stars (LIMs) during their AGB phase \citep{Berg2019}. The delayed release of carbon from AGB stars, operating on gigayear timescales compared to oxygen's rapid injection, creates characteristic chemical evolution patterns. Nitrogen is believed to form mainly in LIMs, while fast-rotating massive stars can greatly enhance it \citep{Limongi2018}. This temporal decoupling produces a pronounced [C/O] and [N/O] depression at metallicities [O/H] $\sim -$1.5, with outliers likely reflecting exotic evolution history \citep{Romano2022}.  

\begin{figure}[H]
 \centering
\includegraphics[width=0.49\textwidth]{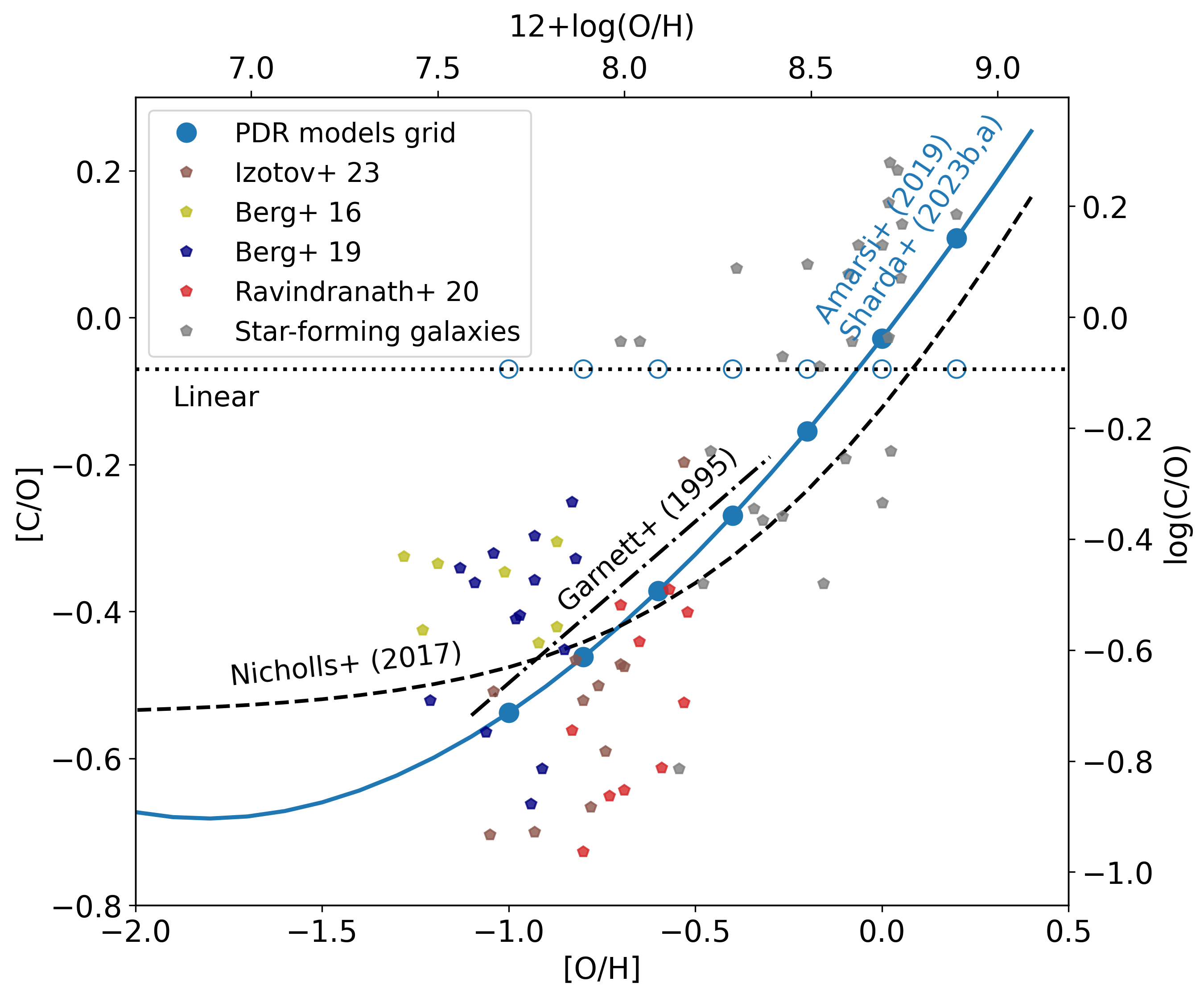}
\caption{Adapted from \citet{Bisbas2025}. 
[C/O]–[O/H] correlation based on \cite{Garnett1995} (dot-dashed), 
\cite{Nicholls2017} (black dashed), and \cite{Amarsi2019} observations, the latter using the best-fit relations of \cite{Sharda2023} (blue solid). The dotted black line shows the case when C/H scales linearly with O/H. Filled blue circles show the [C/O]–[O/H] pairs for which we have performed the PDR modeling. The corresponding values for a linear relationship between C/H and O/H are shown in empty blue circles. The observational data of \citet{Berg2016, Berg2019, Ravindranath2020, Izotov2023} are shown for comparison. The secondary y-axis shows the log(C/O) abundance ratio.}
\label{fig: C-to-O-abundances}
\end{figure}

With advances of modern observational facilities, striking deviations have been revealed from solar abundance patterns in extreme environments. For example, JWST spectroscopy of the high-redshift galaxy GN-z11 ($z \sim 10.6$) measures [C/O] = $-0.57 \pm 0.11$ \citep{Isobe2023}, while ALMA surveys detect elevated [O\,\textsc{iii}]88$\mu$m/[C\,\textsc{ii}]158$\mu$m ratios ($>$10) in $z > 6$ systems \citep{Harikane2020}. These measurements align with chemical evolution models predicting $\alpha$-enhancement in rapidly assembling galaxies in the early epoch of the Universe. Concurrently, sublinear dust-to-gas scaling ($\mathcal{D} \propto Z^{0.7}$, where $\mathcal{D}$ is the dust-to-gas mass ratio relative to Solar) observed in metal-poor dwarfs \citep{Remy2014} exacerbates molecular complexity by reducing UV shielding efficiency and photoelectric heating rates.

However, traditional astrochemical models face three interconnected challenges in this regime. First, the assumption of fixed [C/O] and [N/O] ratios fails to capture metallicity-dependent abundance variations critical for cooling calculations (Figure~\ref{fig: C-to-O-abundances}). Second, nonlinear dust evolution alters H$_2$ formation rates and the penetration depth of radiation fields. Third, conventional molecular tracers become unreliable: CO photodissociation fronts advance at lower column densities while [C\,\textsc{i}] emission weakens disproportionately due to carbon depletion \citep{Michiyama2021}. Hydrodynamic simulations incorporating these effects \citep{Gong2020} predict the emergence of [C\,\textsc{i}]-dark molecular gas phases where H$_2$ mass correlates poorly with either CO or atomic carbon emission. Other, less abundant C-, O- and N- molecules will inevitably be affected. 

\begin{figure*}[!htbp]
 \centering
\includegraphics[width=0.49\textwidth]{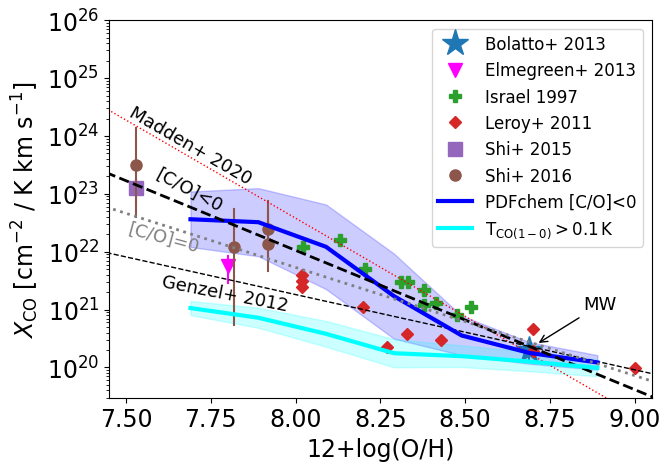}
\includegraphics[width=0.49\textwidth]{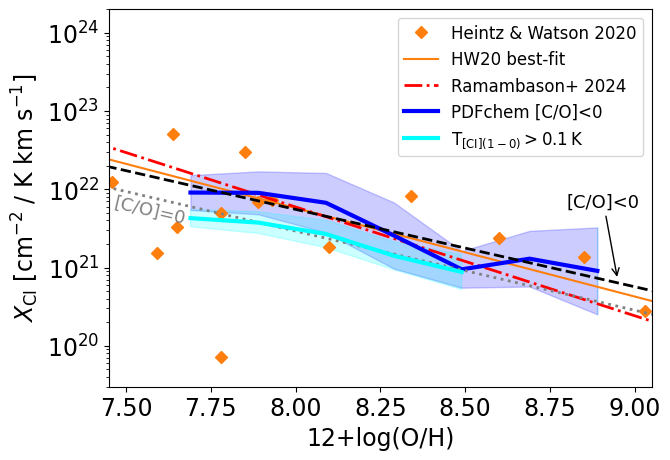}
\caption{Adapted from \citet{Bisbas2025}. The $X_{\rm CO}$-factor (left panel) and the $X_{\rm CI}$-factor (right panel) for all models explored in this work (blue solid lines). The dashed black lines illustrate the best-fit expressions (Eqns 7 and 8 for the $X_{\rm CO}$ and $X_{\rm CI}$, respectively). On the left panel, the observations of \cite{Israel1997, Leroy2011, Elmegreen2013, Shi2015, Shi2016} are shown, including the average MW value marked in large blue star \citep{Bolatto2013} and the best-fit relations of \cite{Genzel2012} and \cite{Madden2020}. On the right panel, the observations of \cite{Heintz2020} and their best-fit relation is shown, as well as the best-fit relation of \cite{Ramambason2024}.} 
\label{fig: Xco Xci}
\end{figure*}

In order to advance our understanding of environments which are not well described by traditional astrochemical models, we aim to address the scientific questions shown below: 
\begin{enumerate}
    \item How do the variations in CNO abundance ratios ([C/O], [N/O]) influence molecular line diagnostics, in the Milky Way and nearby galaxies?
    \item Which physical mechanisms (e.g., CRIR, FUV fields, turbulence) dominate molecular chemistry in non-solar abundance regimes?
    \item Can [C\,\textsc{i}] reliably trace CO-faint gas in \(\alpha\)-enhanced environments?
    \item How can ionization states derived from molecular ions be used to constrain CRIR in extreme environments?
\end{enumerate}

\subsection{Objectives}

To address the questions outlined above, a systematic survey of astrochemical conditions across various extreme conditions is needed. The XSMT-15m offers a great opportunity to perform such a survey at 1.3 and 0.6 mm wavelengths. 
A 1.3~mm and 0.6~mm spectral survey of C-, N-, S-, Si-, and O-bearing molecules (e.g., CO, CN, HCN, HCO$^+$, N$_2$H$^+$, DCO$^+$, CS, SiO, SO, SO$_2$, H$_2$S, C{\sc i}) will be conducted across the Milky Way disk, spanning a wide range of Galactocentric radii and metallicities. Targets include: 

\begin{itemize}
    \item Clouds close to the Galactic Center with supersolar metallicities (\( Z \sim 2\,Z_\odot \)), such as Sgr~B2 and the brick cloud. 
    \item Outer disk regions (\( R_{\mathrm{gal}} \sim 12-25\)~kpc, \citealt{Sun2024}) with subsolar metallicities (\( Z \sim 0.2 - 0.5\,Z_\odot \)) and [C/O] \(< 0\). 
    \item Other environments with extreme UV radiation fields, elevated cosmic ray ionisation rates (\citealt{Yusef-Zadeh2007}), and high turbulence. Classic photon-dominated regions (such as M17), cosmic ray enhanced supernova remnants (such as IC~443, W28), and high turbulence regions (such as RCW~120, NGC~6334, \citealt{Fukui2021}), will be good candidates. 
\end{itemize}

This survey will allow us to address the following scientific objectives:
\begin{itemize}

\item Since most transitions of the main isotopologues are optically thick, it is difficult to measure their abundances. On the other hand, the transitions of their $^{13}$C- or $^{18}$O-bearing isotopologues are optically thin, and can therefore be used to measure molecular abundances, given the $^{13}$C/$^{12}$C ($^{18}$O/$^{16}$O) abundance ratios. Isotopologues (\(^{13}\mathrm{CO}\), C\(^{18}\mathrm{O}\), \(^{13}\mathrm{CN}\)) will enable high-accuracy abundance measurements.

\item Abundance ratios of key species (e.g., CN/HCN, CO/HCO\(^+\), H$_2$CO, CS, HNC, CH$_3$OH, HC$_3$N, N$_{2}$D$^{+}$, etc.) will be compared with chemical models (e.g., \textsc{3D-PDR}, \textsc{UCLCHEM}) under varying [C/O] and [N/O] to identify ``chemical fingerprints" of \(\alpha\)-enhancement and to test predictions of nitrogen deficit under general low-metallicity environments.

\item Both the star-formation rate and the cosmic-ray intensity show systematic variations across the Galactic disk, we plan to observe critical transitions of molecular ions in the 1.3~mm band (211--275~GHz), including HCO$^+$ \(J=3\text{--}2\) (267.56~GHz), N$_2$H$^+$ \(J=3\text{--}2\) (279.51~GHz), and DCO$^+$ \(J=3\text{--}2\) (216.11~GHz). These observations will allow estimates of the cosmic-ray ionization rate \(\zeta_{\text{H}_2}\) by considering the abundances of several ionized species (e.g., HCO$^+$, N$_2$H$^+$, \citealt{Caselli1998}).

\item Combining CO(1--0), [C\,\textsc{i}] (\(^3P_1\)--\(^3P_0\)), and dust continuum maps, we will quantify the spatial correlation between CO-dark gas and [C\,\textsc{i}] emission in low-metallicity regions. Radiative transfer models incorporating \(\mathrm{[C/O]}\)-dependent photodissociation \citep{Bisbas2024, Bisbas2025} will then be used to test whether [C\,\textsc{i}]/CO ratios constrain \(X_{\mathrm{CO}} \) variations (Figure~\ref{fig: Xco Xci}). 

\end{itemize}

\section{Discussion}
\label{sec:data-policy}

To become one of the leading (sub)millimeter telescopes in the Northern hemisphere with unique capability and/or advantages at $>300$~GHz (corresponding to ALMA Band 7 and 8), XSMT-15m will adopt strategies similar to those of other world-class (sub)millimeter observatories. While this white paper does not yet provide details of telescope operations, these will be developed and presented at a later stage. A key component will be the establishment of a science committee or advisory board to guide the selection of first-light observations and to define policies for future observing modes (e.g., regular, large, survey, target-of-opportunity, filler), program selection, and data organization. In parallel, a team dedicated to calibration and data reduction pipeline development will be formed, consisting of members with expertise in (sub)millimeter single-dish observations and the new instruments. First-light observations will be designed to highlight the unique capabilities of XSMT-15m while requiring relatively modest observing times, and will be carried out shortly after the commissioning of the corresponding instruments.

The exact observing overheads and sensitivity/time estimates will ultimately depend on on-telescope instrument testing. We therefore do not include overheads here and emphasize that all time estimates in this white paper are preliminary, although they should not be too far off the final values. For spectral line observations with the heterodyne instruments, our sensitivity calculator agrees with that of the IRAM 30~m telescope\footnote{\url{https://oms.iram.fr/tse/}} to within 10\% when adopting a 30~m diameter, a surface accuracy of 66~$\mu$m, and a total telescope time equal to four times the on-source integration time. This factor arises because the IRAM 30~m time estimator assumes an overall efficiency of 50\% between the total elapsed telescope time and the on+off integration time, in order to account for calibration observations (e.g., pointing, line calibrators), and because the off-position integration time is usually equal to the on-source integration time. For continuum observations with the KID camera, our sensitivity calculator yields an RMS $\sim$3 times lower than that of SCUBA-2\footnote{\url{https://proposals.eaobservatory.org/jcmt/calculator/scuba2/rms}} at 850~$\mu$m in matched-filter mode with the smallest scan pattern (Daisy, $\sim$3~arcmin map) optimized for point-source detection. This performance gain is consistent with expectations, as discussed in the JCMT online documentation\footnote{\url{https://www.eaobservatory.org/jcmt/instrumentation/continuum/}} and in \citet{LiShaoliang2024}, where the authors describe the development of a next-generation KID camera for JCMT. They report that KID detectors have a per-pixel dark noise equivalent power (NEP) of $< 4 \times 10^{-17},\mathrm{W,Hz^{-1/2}}$, corresponding to an RMS improvement by a factor of three relative to the SCUBA-2 TES detector arrays. The XSMT-15m KID camera is expected to reach at least the same  NEP, and therefore should ideally achieve a threefold improvement in RMS compared to SCUBA-2, pending confirmation through on-sky testing. The overheads for continuum mapping depend on the survey area. For example, with 1~h of telescope time using the SCUBA-2 PONG~1800 scan pattern ($\sim$30~arcmin map), the area within $1.1\times$ the central RMS covers $\sim$700~arcmin$^2$ (i.e. $15.5\times$ the instantaneous 45~arcmin$^2$ SCUBA-2 field of view), while the full scanned area is about twice as large. Such factors must be considered when estimating observing overheads, and dedicated modeling of scan patterns and efficiencies will be undertaken by the XSMT-15m Instrument Working Group.

The availability of weather conditions suitable for Band 8 ($\sim$440--510~GHz) is currently being monitored by the XSMT-15m Site-Selection Working Group, and definitive results will be reported in a forthcoming study. Preliminary analyses indicate an average PWV of $<0.87$~mm (XSMT-15m Site-Selection Working Group, priv. comm.), with 27\% of the days in the winter months (October--March) falling below 0.5~mm. Conditions with PWV $\lesssim$0.5~mm are ideal for both Band 8 continuum and spectral line observations (zenith transmission $\sim$0.67), while PWV in the range 0.5--0.9~mm should still provide acceptable performance. Although Table~\ref{tab1} lists the integration times required to reach the continuum confusion limits at each band, we emphasize that super-deep imaging can be achieved by incorporating prior information. Accordingly, Table~\ref{tab1} also includes the integration time needed to reach an RMS of $\sim$0.2~mJy in the continuum. Nevertheless, because the confusion limit and dark NEP increase rapidly with frequency, Band 8 imaging would require prohibitively long integration times to achieve the same RMS as in Bands 6 and 7. This limitation is mitigated by the rise of galaxies’ dust SEDs toward higher frequencies, which partly compensates for the shallower sensitivity of Band 8 observations.

\section{Summary and outlook}

The Xue-shan-mu-chang 15-meter submillimeter telescope (XSMT-15m), to be constructed at a high-altitude site in Qinghai, represents a landmark step for Chinese astronomy. As the first independently developed, world-class submillimeter facility in China mainland, XSMT-15m is designed to address a broad range of scientific questions spanning extragalactic astronomy, Galactic ecology, astrochemistry, and time-domain astrophysics. With state-of-the-art continuum cameras and heterodyne receivers, the telescope will enable ambitious surveys of both nearby and distant objects, with the potential to advance our understanding of galaxy formation and evolution, the processes governing the interstellar medium, astrochemistry linked to the origins of life, and time-domain astrophysics. 

The success of XSMT-15m will establish the Xue-shan-mu-chang site as a premier location for submillimeter astronomy in the northern hemisphere. If the initial phase proves successful, the site’s excellent conditions open the prospect for even more ambitious projects, including the construction of a next-generation 50m class single-dish submillimeter telescope, such as the proposed Atacama Large Aperture Submillimeter Telescope (AtLAST), or the development of an ALMA-like interferometer. These future facilities would greatly enhance the scientific reach of the site, enabling transformative research on the cold Universe. With XSMT-15m as the foundation, Xue-shan-mu-chang has the potential to become a world-leading hub for submillimeter science in the decades to come.


\makeatletter
\Acknowledgements{This work was supported by the Ministry of Science and Technology of China under the National Key R\&D Program (Grant No. 2023YFA1608200), the National Natural Science Foundation of China (Grant Nos. 12427901 and 12333013), and the Chinese Academy of Sciences (Grant No. PTYQ2024BJ0010).}
\ifx\@Acknowledgements\@empty\else
  \section*{Acknowledgements}
  \@Acknowledgements
\fi
\makeatother


\makeatletter
\InterestConflict{The authors declare that they have no conflict of interest.}
\ifx\@InterestConflict\@empty\else
  \section*{Conflict of interest}
  \@InterestConflict
\fi
\makeatother


{\small \setlength{\baselineskip}{-1pt}
\bibliographystyle{raa}
\bibliography{reference}

\begin{thebibliography}{333}
\providecommand\natexlab[1]{#1}
\providecommand\JournalTitle[1]{#1}

\bibitem[{Abazajian} {et~al.}(2019)]{CMB-S4-2019}
{Abazajian}, K., {Addison}, G., {Adshead}, P., {et~al.} 2019, arXiv e-prints,
  arXiv:1907.04473

\bibitem[{Abbott} {et~al.}(2017)]{Abbott2017}
{Abbott}, B.~P., {Abbott}, R., {Abbott}, T.~D., {et~al.} 2017, \apjl, 848, L12

\bibitem[{Adam} {et~al.}(2017)]{Adam2017}
{Adam}, R., {Bartalucci}, I., {Pratt}, G.~W., {et~al.} 2017, \aap, 598, A115

\bibitem[{Ag{\'u}ndez} {et~al.}(2020)]{2020A&A...637A..59A}
{Ag{\'u}ndez}, M., {Mart{\'\i}nez}, J.~I., {de Andres}, P.~L., {Cernicharo},
  J., \& {Mart{\'\i}n-Gago}, J.~A. 2020, \aap, 637, A59

\bibitem[{Allen} {et~al.}(2011)]{Allen2011}
{Allen}, S.~W., {Evrard}, A.~E., \& {Mantz}, A.~B. 2011, \araa, 49, 409

\bibitem[{Alves} {et~al.}(2014)]{2014A&A...569L...1A}
{Alves}, F.~O., {Frau}, P., {Girart}, J.~M., {et~al.} 2014, \aap, 569, L1

\bibitem[{Amarsi} {et~al.}(2019)]{Amarsi2019}
{Amarsi}, A.~M., {Nissen}, P.~E., \& {Sk{\'u}lad{\'o}ttir}, {\'A}. 2019, \aap,
  630, A104

\bibitem[{Andriantsaralaza} {et~al.}(2022)]{2022A&A...667...A74}
{Andriantsaralaza}, M., {Ramstedt}, S., {Vlemmings}, W.~H.~T., \& {De Beck}, E.
  2022, \aap, 667, A74

\bibitem[{Armitage}(2016)]{Armitage2016}
{Armitage}, P.~J. 2016, \apjl, 833, L15

\bibitem[{Audard} {et~al.}(2014)]{Audard2014}
{Audard}, M., {{\'A}brah{\'a}m}, P., {Dunham}, M.~M., {et~al.} 2014, in
  Protostars and Planets VI, ed. H.~{Beuther}, R.~S. {Klessen}, C.~P.
  {Dullemond}, \& T.~{Henning}, 387

\bibitem[{Bachiller} \& {P{\'e}rez Guti{\'e}rrez}(1997)]{1997ApJ...487L..93B}
{Bachiller}, R., \& {P{\'e}rez Guti{\'e}rrez}, M. 1997, \apjl, 487, L93

\bibitem[{Baek} {et~al.}(2020)]{Baek2020}
{Baek}, G., {MacFarlane}, B.~A., {Lee}, J.-E., {et~al.} 2020, \apj, 895, 27

\bibitem[{Baraffe} {et~al.}(2009)]{Baraffe2009}
{Baraffe}, I., {Chabrier}, G., \& {Gallardo}, J. 2009, \apjl, 702, L27

\bibitem[{Barrufet} {et~al.}(2023)]{Barrufet2023}
{Barrufet}, L., {Oesch}, P.~A., {Weibel}, A., {et~al.} 2023, \mnras, 522, 449

\bibitem[{Belloche} {et~al.}(2013)]{2013A&A...559A..47B}
{Belloche}, A., {M{\"u}ller}, H.~S.~P., {Menten}, K.~M., {Schilke}, P., \&
  {Comito}, C. 2013, \aap, 559, A47

\bibitem[{Berg} {et~al.}(2019)]{Berg2019}
{Berg}, D.~A., {Erb}, D.~K., {Henry}, R. B.~C., {Skillman}, E.~D., \&
  {McQuinn}, K. B.~W. 2019, \apj, 874, 93

\bibitem[{Berg} {et~al.}(2016)]{Berg2016}
{Berg}, D.~A., {Skillman}, E.~D., {Henry}, R. B.~C., {Erb}, D.~K., \& {Carigi},
  L. 2016, \apj, 827, 126

\bibitem[{Bertola} {et~al.}(2024)]{Bertola2024}
{Bertola}, E., {Circosta}, C., {Ginolfi}, M., {et~al.} 2024, \aap, 691, A178

\bibitem[{B{\'e}thermin} {et~al.}(2015)]{Bethermin2015}
{B{\'e}thermin}, M., {Daddi}, E., {Magdis}, G., {et~al.} 2015, \aap, 573, A113

\bibitem[{Bietenholz} {et~al.}(2021)]{Bietenholz2021}
{Bietenholz}, M.~F., {Bartel}, N., {Argo}, M., {et~al.} 2021, \apj, 908, 75

\bibitem[{Bietenholz} {et~al.}(2013)]{Bietenholz+13}
{Bietenholz}, M.~F., {Kondratiev}, V., {Ransom}, S., {et~al.} 2013, \mnras,
  431, 2590

\bibitem[{Bietenholz} {et~al.}(2011)]{Bietenholz+11}
{Bietenholz}, M.~F., {Matheson}, H., {Safi-Harb}, S., {Brogan}, C., \&
  {Bartel}, N. 2011, \mnras, 412, 1221

\bibitem[{Bisbas} {et~al.}(2024)]{Bisbas2024}
{Bisbas}, T.~G., {Zhang}, Z.-Y., {Gjergo}, E., {et~al.} 2024, \mnras, 527, 8886

\bibitem[{Bisbas} {et~al.}(2025)]{Bisbas2025}
{Bisbas}, T.~G., {Zhang}, Z.-Y., {Kyrmanidou}, M.-C., {et~al.} 2025, arXiv
  e-prints, arXiv:2503.12073

\bibitem[{Blain} {et~al.}(2002)]{Blain2002}
{Blain}, A.~W., {Smail}, I., {Ivison}, R.~J., {Kneib}, J.~P., \& {Frayer},
  D.~T. 2002, \physrep, 369, 111

\bibitem[{Bocchino} {et~al.}(2010)]{Bocchino+10}
{Bocchino}, F., {Bandiera}, R., \& {Gelfand}, J. 2010, \aap, 520, A71

\bibitem[{Bocchino} {et~al.}(2005)]{Bocchino+05}
{Bocchino}, F., {van der Swaluw}, E., {Chevalier}, R., \& {Bandiera}, R. 2005,
  \aap, 442, 539

\bibitem[{Bolatto} {et~al.}(2008)]{Bolatto2008}
{Bolatto}, A.~D., {Leroy}, A.~K., {Rosolowsky}, E., {Walter}, F., \& {Blitz},
  L. 2008, \apj, 686, 948

\bibitem[{Bolatto} {et~al.}(2013)]{Bolatto2013}
{Bolatto}, A.~D., {Wolfire}, M., \& {Leroy}, A.~K. 2013, \araa, 51, 207

\bibitem[{Borkowski} {et~al.}(2016)]{Borkowski+16}
{Borkowski}, K.~J., {Reynolds}, S.~P., \& {Roberts}, M. S.~E. 2016, \apj, 819,
  160

\bibitem[{Bot} {et~al.}(2010)]{Bot2010}
{Bot}, C., {Ysard}, N., {Paradis}, D., {et~al.} 2010, \aap, 523, A20

\bibitem[{Bottinelli} {et~al.}(2007)]{2007A&A...463..601B}
{Bottinelli}, S., {Ceccarelli}, C., {Williams}, J.~P., \& {Lefloch}, B. 2007,
  \aap, 463, 601

\bibitem[{Brada{\v{c}}} {et~al.}(2008)]{Bradac2008}
{Brada{\v{c}}}, M., {Allen}, S.~W., {Treu}, T., {et~al.} 2008, \apj, 687, 959

\bibitem[Cabral \& Leedom(1993)]{Cabral93}
Cabral, B., \& Leedom, L.~C. 1993, in Proceedings of the 20th Annual Conference
  on Computer Graphics and Interactive Techniques, SIGGRAPH '93 (New York, NY,
  USA: Association for Computing Machinery), 263–270

\bibitem[{Capak} {et~al.}(2011)]{Capak2011}
{Capak}, P.~L., {Riechers}, D., {Scoville}, N.~Z., {et~al.} 2011, \nat, 470,
  233

\bibitem[{Carlstrom} {et~al.}(2002)]{Carlstrom2002}
{Carlstrom}, J.~E., {Holder}, G.~P., \& {Reese}, E.~D. 2002, \araa, 40, 643

\bibitem[{Carniani} {et~al.}(2024)]{Carniani2024}
{Carniani}, S., {Hainline}, K., {D'Eugenio}, F., {et~al.} 2024, \nat, 633, 318

\bibitem[{Caselli} \& {Ceccarelli}(2012)]{2012A&ARv..20...56C}
{Caselli}, P., \& {Ceccarelli}, C. 2012, \aapr, 20, 56

\bibitem[{Caselli} {et~al.}(1998)]{Caselli1998}
{Caselli}, P., {Walmsley}, C.~M., {Terzieva}, R., \& {Herbst}, E. 1998, \apj,
  499, 234

\bibitem[{Casey} {et~al.}(2014)]{Casey2014}
{Casey}, C.~M., {Narayanan}, D., \& {Cooray}, A. 2014, \physrep, 541, 45

\bibitem[{CCAT-Prime Collaboration} {et~al.}(2023)]{CCAT-prime-2023}
{CCAT-Prime Collaboration}, {Aravena}, M., {Austermann}, J.~E., {et~al.} 2023,
  \apjs, 264, 7

\bibitem[{CDMS}(2025)]{websiteOfCDMS}
{CDMS}. 2025, {The Cologne Database for Molecular Spectroscopy -- Molecules in
  Space}, \url{https://cdms.astro.uni-koeln.de/classic/molecules}, accessed:
  2025-03-31

\bibitem[{Cernicharo} {et~al.}(2021)]{2021A&A...652L...9C}
{Cernicharo}, J., {Ag{\'u}ndez}, M., {Kaiser}, R.~I., {et~al.} 2021, \aap, 652,
  L9

\bibitem[{Cernicharo} {et~al.}(2000)]{2000A&AS..142..181C}
{Cernicharo}, J., {Gu{\'e}lin}, M., \& {Kahane}, C. 2000, \aaps, 142, 181

\bibitem[{Cernicharo} {et~al.}(2001)]{2001ApJ...546L.123C}
{Cernicharo}, J., {Heras}, A.~M., {Tielens}, A.~G.~G.~M., {et~al.} 2001, \apjl,
  546, L123

\bibitem[{Chastenet} {et~al.}(2022)]{chastenet+22}
{Chastenet}, J., {De Looze}, I., {Hensley}, B.~S., {et~al.} 2022, \mnras, 516,
  4229

\bibitem[{Chawner} {et~al.}(2020)]{chawner+20}
{Chawner}, H., {Gomez}, H.~L., {Matsuura}, M., {et~al.} 2020, \mnras, 493, 2706

\bibitem[{Chen} {et~al.}(2009)]{2009ApJ...691.1729C}
{Chen}, X., {Launhardt}, R., \& {Henning}, T. 2009, \apj, 691, 1729

\bibitem[{Chen} {et~al.}(2025)]{ChenZhiwei2025}
{Chen}, Z., {Johnstone}, D., {Contreras Pe{\~n}a}, C., {et~al.} 2025, arXiv
  e-prints, arXiv:2506.08389

\bibitem[{Cherchneff}(2006)]{2006A&A...456.1001C}
{Cherchneff}, I. 2006, \aap, 456, 1001

\bibitem[{Chiappini}(2009)]{Chiappini2009}
{Chiappini}, C. 2009, in IAU Symposium, Vol. 254, The Galaxy Disk in
  Cosmological Context, ed. J.~{Andersen}, {Nordstr{\"o}ara}, B.~{m}, \&
  J.~{Bland-Hawthorn}, 191

\bibitem[{CHIME/FRB Collaboration} {et~al.}(2020)]{CHIME2020}
{CHIME/FRB Collaboration}, {Andersen}, B.~C., {Bandura}, K.~M., {et~al.} 2020,
  \nat, 587, 54

\bibitem[{Ching} {et~al.}(2022)]{2022ApJ...941..122C}
{Ching}, T.-C., {Qiu}, K., {Li}, D., {et~al.} 2022, \apj, 941, 122

\bibitem[{Cicone} {et~al.}(2014)]{Cicone2014}
{Cicone}, C., {Maiolino}, R., {Sturm}, E., {et~al.} 2014, \aap, 562, A21

\bibitem[{Cicone} {et~al.}(2015)]{Cicone2015}
{Cicone}, C., {Maiolino}, R., {Gallerani}, S., {et~al.} 2015, \aap, 574, A14

\bibitem[{Cicone} {et~al.}(2019)]{Cicone2019}
{Cicone}, C., {De Breuck}, C., {Chen}, C.-C., {et~al.} 2019, \baas, 51, 82

\bibitem[{Circosta} {et~al.}(2021)]{Circosta2021}
{Circosta}, C., {Mainieri}, V., {Lamperti}, I., {et~al.} 2021, \aap, 646, A96

\bibitem[{Codella} {et~al.}(2016)]{2016MNRAS.462L..75C}
{Codella}, C., {Ceccarelli}, C., {Bianchi}, E., {et~al.} 2016, \mnras, 462, L75

\bibitem[{Cody} {et~al.}(2017)]{Cody2017}
{Cody}, A.~M., {Hillenbrand}, L.~A., {David}, T.~J., {et~al.} 2017, \apj, 836,
  41

\bibitem[{Contreras Pe{\~n}a} {et~al.}(2020)]{ContrerasPena2020}
{Contreras Pe{\~n}a}, C., {Johnstone}, D., {Baek}, G., {et~al.} 2020, \mnras,
  495, 3614

\bibitem[{Crockett} {et~al.}(2014)]{2014ApJ...787..112C}
{Crockett}, N.~R., {Bergin}, E.~A., {Neill}, J.~L., {et~al.} 2014, \apj, 787,
  112

\bibitem[{Crutcher}(2012)]{2012ARA&A..50...29C}
{Crutcher}, R.~M. 2012, \araa, 50, 29

\bibitem[{Crutcher} \& {Kemball}(2019)]{2019FrASS...6...66C}
{Crutcher}, R.~M., \& {Kemball}, A.~J. 2019, Frontiers in Astronomy and Space
  Sciences, 6, 66

\bibitem[{Dale} {et~al.}(2014)]{Dale2014}
{Dale}, D.~A., {Helou}, G., {Magdis}, G.~E., {et~al.} 2014, \apj, 784, 83

\bibitem[{Dale} {et~al.}(2012)]{Dale2012}
{Dale}, D.~A., {Aniano}, G., {Engelbracht}, C.~W., {et~al.} 2012, \apj, 745, 95

\bibitem[{Danilovich} {et~al.}(2021)]{2021A&A...655...A80}
{Danilovich}, T., {Van de Sande}, M., {Plane}, J.~M.~C., {et~al.} 2021, \aap,
  655, A80

\bibitem[{De Looze} {et~al.}(2019)]{deLooze+19}
{De Looze}, I., {Barlow}, M.~J., {Bandiera}, R., {et~al.} 2019, \mnras, 488,
  164

\bibitem[{Dessauges-Zavadsky} {et~al.}(2019)]{Dessauges2019}
{Dessauges-Zavadsky}, M., {Richard}, J., {Combes}, F., {et~al.} 2019, Nature
  Astronomy, 3, 1115

\bibitem[{Dole} {et~al.}(2006)]{Dole2006}
{Dole}, H., {Lagache}, G., {Puget}, J.~L., {et~al.} 2006, \aap, 451, 417

\bibitem[{Draine}(1978)]{Draine1978}
{Draine}, B.~T. 1978, \apjs, 36, 595

\bibitem[{Draine}(2003)]{Draine2003}
{Draine}, B.~T. 2003, \araa, 41, 241

\bibitem[{Draine} {et~al.}(2014)]{Draine2014}
{Draine}, B.~T., {Aniano}, G., {Krause}, O., {et~al.} 2014, \apj, 780, 172

\bibitem[{Du} {et~al.}(2025)]{Du2025}
{Du}, K., {Shi}, Y., {Wang}, T., \& {Shu}, C. 2025, \aap, 695, A46

\bibitem[{Dunham} {et~al.}(2010)]{Dunham2010}
{Dunham}, M.~M., {Evans}, II, N.~J., {Terebey}, S., {Dullemond}, C.~P., \&
  {Young}, C.~H. 2010, \apj, 710, 470

\bibitem[{Dunne} {et~al.}(2003)]{dunne+03}
{Dunne}, L., {Eales}, S., {Ivison}, R., {Morgan}, H., \& {Edmunds}, M. 2003,
  Nat, 424, 285

\bibitem[{Dunne} {et~al.}(2009)]{dunne+09}
{Dunne}, L., {Maddox}, S.~J., {Ivison}, R.~J., {et~al.} 2009, \mnras, 394, 1307

\bibitem[{Dutta} {et~al.}(2024)]{Dutta2024}
{Dutta}, S., {Lee}, C.-F., {Johnstone}, D., {et~al.} 2024, \aj, 167, 72

\bibitem[{Eales} {et~al.}(2012)]{Eales2012}
{Eales}, S., {Smith}, M. W.~L., {Auld}, R., {et~al.} 2012, \apj, 761, 168

\bibitem[{Eisenhardt} {et~al.}(2012)]{Eisenhardt2012}
{Eisenhardt}, P. R.~M., {Wu}, J., {Tsai}, C.-W., {et~al.} 2012, \apj, 755, 173

\bibitem[{Ejlali} {et~al.}(2024)]{Ejlali2024}
{Ejlali}, G., {Adam}, R., {Ade}, P., {et~al.} 2024, in European Physical
  Journal Web of Conferences, Vol. 293, mm Universe 2023 - Observing the
  Universe at mm Wavelengths, 00016

\bibitem[{Ejlali} {et~al.}(2025)]{Ejlali2025}
{Ejlali}, G., {Tabatabaei}, F.~S., {Roussel}, H., {et~al.} 2025, \aap, 693, A88

\bibitem[{Elbaz} {et~al.}(2011)]{Elbaz2011}
{Elbaz}, D., {Dickinson}, M., {Hwang}, H.~S., {et~al.} 2011, \aap, 533, A119

\bibitem[{Elmegreen}(2000)]{Elmegreen2000}
{Elmegreen}, B.~G. 2000, \apj, 530, 277

\bibitem[{Elmegreen} {et~al.}(2013)]{Elmegreen2013}
{Elmegreen}, B.~G., {Rubio}, M., {Hunter}, D.~A., {et~al.} 2013, \nat, 495, 487

\bibitem[{Emonts} {et~al.}(2016)]{Emonts2016}
{Emonts}, B.~H.~C., {Lehnert}, M.~D., {Villar-Mart{\'\i}n}, M., {et~al.} 2016,
  Science, 354, 1128

\bibitem[{Evans}(2008)]{2008ASPC..390...52E}
{Evans}, II, N.~J. 2008, in Astronomical Society of the Pacific Conference
  Series, Vol. 390, Pathways Through an Eclectic Universe, ed. J.~H. {Knapen},
  T.~J. {Mahoney}, \& A.~{Vazdekis}, 52

\bibitem[{Event Horizon Telescope Collaboration}
  {et~al.}(2019)]{EHT-Collaboration-2019}
{Event Horizon Telescope Collaboration}, {Akiyama}, K., {Alberdi}, A., {et~al.}
  2019, \apjl, 875, L1

\bibitem[{Fiore} {et~al.}(2017)]{Fiore2017}
{Fiore}, F., {Feruglio}, C., {Shankar}, F., {et~al.} 2017, \aap, 601, A143

\bibitem[{Fischer} {et~al.}(2023)]{Fischer2023}
{Fischer}, W.~J., {Hillenbrand}, L.~A., {Herczeg}, G.~J., {et~al.} 2023, in
  Astronomical Society of the Pacific Conference Series, Vol. 534, Protostars
  and Planets VII, ed. S.~{Inutsuka}, Y.~{Aikawa}, T.~{Muto}, K.~{Tomida}, \&
  M.~{Tamura}, 355

\bibitem[{Fischer} {et~al.}(2019)]{Fischer2019}
{Fischer}, W.~J., {Safron}, E., \& {Megeath}, S.~T. 2019, \apj, 872, 183

\bibitem[{Fischer} {et~al.}(2017)]{Fischer2017}
{Fischer}, W.~J., {Megeath}, S.~T., {Furlan}, E., {et~al.} 2017, \apj, 840, 69

\bibitem[{Francis} {et~al.}(2022)]{Francis2022}
{Francis}, L., {Johnstone}, D., {Lee}, J.-E., {et~al.} 2022, \apj, 937, 29

\bibitem[{Franco} {et~al.}(2018)]{Franco2018}
{Franco}, M., {Elbaz}, D., {B{\'e}thermin}, M., {et~al.} 2018, \aap, 620, A152

\bibitem[{Friberg} {et~al.}(2016)]{2016SPIE.9914E..03F}
{Friberg}, P., {Bastien}, P., {Berry}, D., {et~al.} 2016, in Society of
  Photo-Optical Instrumentation Engineers (SPIE) Conference Series, Vol. 9914,
  Millimeter, Submillimeter, and Far-Infrared Detectors and Instrumentation for
  Astronomy VIII, ed. W.~S. {Holland} \& J.~{Zmuidzinas}, 991403

\bibitem[{Fukui} {et~al.}(2021)]{Fukui2021}
{Fukui}, Y., {Habe}, A., {Inoue}, T., {Enokiya}, R., \& {Tachihara}, K. 2021,
  \pasj, 73, S1

\bibitem[{Galametz} {et~al.}(2011)]{Galametz2011}
{Galametz}, M., {Madden}, S.~C., {Galliano}, F., {et~al.} 2011, \aap, 532, A56

\bibitem[{Galametz} {et~al.}(2009)]{Galametz2009}
{Galametz}, M., {Madden}, S., {Galliano}, F., {et~al.} 2009, \aap, 508, 645

\bibitem[{Galametz} {et~al.}(2012)]{Galametz2012}
{Galametz}, M., {Kennicutt}, R.~C., {Albrecht}, M., {et~al.} 2012, \mnras, 425,
  763

\bibitem[{Galliano} {et~al.}(2018)]{Galliano2018}
{Galliano}, F., {Galametz}, M., \& {Jones}, A.~P. 2018, \araa, 56, 673

\bibitem[{Galliano} {et~al.}(2005)]{Galliano2005}
{Galliano}, F., {Madden}, S.~C., {Jones}, A.~P., {Wilson}, C.~D., \& {Bernard},
  J.~P. 2005, \aap, 434, 867

\bibitem[{Galliano} {et~al.}(2003)]{Galliano2003}
{Galliano}, F., {Madden}, S.~C., {Jones}, A.~P., {et~al.} 2003, \aap, 407, 159

\bibitem[{Galliano} {et~al.}(2011)]{Galliano2011}
{Galliano}, F., {Hony}, S., {Bernard}, J.~P., {et~al.} 2011, \aap, 536, A88

\bibitem[{Gao} {et~al.}(2013)]{GaoH2013}
{Gao}, H., {Ding}, X., {Wu}, X.-F., {Zhang}, B., \& {Dai}, Z.-G. 2013, \apj,
  771, 86

\bibitem[{Gao} {et~al.}(2007)]{2007ApJ...660L..93G}
{Gao}, Y., {Carilli}, C.~L., {Solomon}, P.~M., \& {Vanden Bout}, P.~A. 2007,
  \apjl, 660, L93

\bibitem[{Gao} \& {Solomon}(2004)]{2004ApJ...606..271G}
{Gao}, Y., \& {Solomon}, P.~M. 2004, \apj, 606, 271

\bibitem[{Garnett} {et~al.}(1995)]{Garnett1995}
{Garnett}, D.~R., {Skillman}, E.~D., {Dufour}, R.~J., {et~al.} 1995, \apj, 443,
  64

\bibitem[{Genzel} {et~al.}(2012)]{Genzel2012}
{Genzel}, R., {Tacconi}, L.~J., {Combes}, F., {et~al.} 2012, \apj, 746, 69

\bibitem[{Gerin} {et~al.}(2016)]{2016ARA&A..54..181G}
{Gerin}, M., {Neufeld}, D.~A., \& {Goicoechea}, J.~R. 2016, \araa, 54, 181

\bibitem[{Gomez} {et~al.}(2012{\natexlab{a}})]{gomez+12b}
{Gomez}, H.~L., {Krause}, O., {Barlow}, M.~J., {et~al.} 2012{\natexlab{a}},
  ApJ, 760, 96

\bibitem[{Gomez} {et~al.}(2012{\natexlab{b}})]{gomez+12a}
{Gomez}, H.~L., {Clark}, C.~J.~R., {Nozawa}, T., {et~al.} 2012{\natexlab{b}},
  \mnras, 420, 3557

\bibitem[{Gong} {et~al.}(2020)]{Gong2020}
{Gong}, M., {Ostriker}, E.~C., {Kim}, C.-G., \& {Kim}, J.-G. 2020, \apj, 903,
  142

\bibitem[{Gong} {et~al.}(2015{\natexlab{a}})]{2015A&A...574A..56G}
{Gong}, Y., {Henkel}, C., {Spezzano}, S., {et~al.} 2015{\natexlab{a}}, \aap,
  574, A56

\bibitem[{Gong} {et~al.}(2015{\natexlab{b}})]{2015A&A...581A..48G}
{Gong}, Y., {Henkel}, C., {Thorwirth}, S., {et~al.} 2015{\natexlab{b}}, \aap,
  581, A48

\bibitem[{Gould} \& {Salpeter}(1963)]{Gould1963}
{Gould}, R.~J., \& {Salpeter}, E.~E. 1963, \apj, 138, 393

\bibitem[{Green} {et~al.}(2004)]{Green+04}
{Green}, D.~A., {Tuffs}, R.~J., \& {Popescu}, C.~C. 2004, \mnras, 355, 1315

\bibitem[{G{\"u}sten} {et~al.}(2019)]{2019Natur.568..357G}
{G{\"u}sten}, R., {Wiesemeyer}, H., {Neufeld}, D., {et~al.} 2019, \nat, 568,
  357

\bibitem[{Hacar} {et~al.}(2016)]{Hacar2016}
{Hacar}, A., {Alves}, J., {Burkert}, A., \& {Goldsmith}, P. 2016, \aap, 591,
  A104

\bibitem[{Han}(2017)]{2017ARA&A..55..111H}
{Han}, J.~L. 2017, \araa, 55, 111

\bibitem[{Hand} {et~al.}(2012)]{Hand2012}
{Hand}, N., {Addison}, G.~E., {Aubourg}, E., {et~al.} 2012, \prl, 109, 041101

\bibitem[{Harikane} {et~al.}(2020)]{Harikane2020}
{Harikane}, Y., {Ouchi}, M., {Inoue}, A.~K., {et~al.} 2020, \apj, 896, 93

\bibitem[{Harsono} {et~al.}(2015)]{Harsono2015}
{Harsono}, D., {van Dishoeck}, E.~F., {Bruderer}, S., {Li}, Z.~Y., \&
  {J{\o}rgensen}, J.~K. 2015, \aap, 577, A22

\bibitem[{Hasselfield} {et~al.}(2013)]{Hasselfield2013}
{Hasselfield}, M., {Hilton}, M., {Marriage}, T.~A., {et~al.} 2013, \jcap, 2013,
  008

\bibitem[{Heiles}(2000)]{2000AJ....119..923H}
{Heiles}, C. 2000, \aj, 119, 923

\bibitem[{Heintz} \& {Watson}(2020)]{Heintz2020}
{Heintz}, K.~E., \& {Watson}, D. 2020, \apjl, 889, L7

\bibitem[{Helmich} \& {van Dishoeck}(1997)]{1997A&AS..124..205H}
{Helmich}, F.~P., \& {van Dishoeck}, E.~F. 1997, \aaps, 124, 205

\bibitem[{Hennebelle} \& {Inutsuka}(2019)]{2019FrASS...6....5H}
{Hennebelle}, P., \& {Inutsuka}, S.-i. 2019, Frontiers in Astronomy and Space
  Sciences, 6, 5

\bibitem[{Hensley} \& {Draine}(2023)]{hensley+23}
{Hensley}, B.~S., \& {Draine}, B.~T. 2023, \apj, 948, 55

\bibitem[{Herbst} \& {van Dishoeck}(2009)]{2009ARA&A..47..427H}
{Herbst}, E., \& {van Dishoeck}, E.~F. 2009, \araa, 47, 427

\bibitem[{Herbst} \& {Yates}(2013)]{2013ChRv..113.8707H}
{Herbst}, E., \& {Yates}, Jr., J.~T. 2013, Chemical Reviews, 113, 8707

\bibitem[{Herczeg} {et~al.}(2017)]{Herczeg2017}
{Herczeg}, G.~J., {Johnstone}, D., {Mairs}, S., {et~al.} 2017, \apj, 849, 43

\bibitem[{Heyer} {et~al.}(2009)]{Heyer2009}
{Heyer}, M., {Krawczyk}, C., {Duval}, J., \& {Jackson}, J.~M. 2009, \apj, 699,
  1092

\bibitem[{Hillenbrand} {et~al.}(2013)]{Hillenbrand2013}
{Hillenbrand}, L.~A., {Miller}, A.~A., {Covey}, K.~R., {et~al.} 2013, \aj, 145,
  59

\bibitem[{H{\"o}fner} \& {Olofsson}(2018)]{2018A&ARv..26....1H}
{H{\"o}fner}, S., \& {Olofsson}, H. 2018, \aapr, 26, 1

\bibitem[{Hsieh} {et~al.}(2019)]{Hsieh2019}
{Hsieh}, T.-H., {Murillo}, N.~M., {Belloche}, A., {et~al.} 2019, \apj, 884, 149

\bibitem[{Hu} {et~al.}(2019)]{2019NatAs...3..776H}
{Hu}, Y., {Yuen}, K.~H., {Lazarian}, V., {et~al.} 2019, Nature Astronomy, 3,
  776

\bibitem[{Huang} {et~al.}(2011)]{Huang2011}
{Huang}, J.~S., {Zheng}, X.~Z., {Rigopoulou}, D., {et~al.} 2011, \apjl, 742,
  L13

\bibitem[{Hughes} {et~al.}(1998)]{Hughes1998}
{Hughes}, D.~H., {Serjeant}, S., {Dunlop}, J., {et~al.} 1998, \nat, 394, 241

\bibitem[{Hull} \& {Zhang}(2019)]{2019FrASS...6....3H}
{Hull}, C. L.~H., \& {Zhang}, Q. 2019, Frontiers in Astronomy and Space
  Sciences, 6, 3

\bibitem[{Hunt} {et~al.}(2015)]{Hunt2015}
{Hunt}, L.~K., {Draine}, B.~T., {Bianchi}, S., {et~al.} 2015, \aap, 576, A33

\bibitem[{Isobe} {et~al.}(2023)]{Isobe2023}
{Isobe}, Y., {Ouchi}, M., {Tominaga}, N., {et~al.} 2023, \apj, 959, 100

\bibitem[{Israel}(1997)]{Israel1997}
{Israel}, F.~P. 1997, \aap, 328, 471

\bibitem[{Izotov} {et~al.}(2023)]{Izotov2023}
{Izotov}, Y.~I., {Schaerer}, D., {Worseck}, G., {et~al.} 2023, \mnras, 522,
  1228

\bibitem[{Izumi} {et~al.}(2023)]{Izumi2023}
{Izumi}, T., {Wada}, K., {Imanishi}, M., {et~al.} 2023, Science, 382, 554

\bibitem[{Jiao} {et~al.}(2022)]{jiao2022SCPMA}
{Jiao}, S., {Lin}, Y., {Shui}, X., {et~al.} 2022, Science China Physics,
  Mechanics, and Astronomy, 65, 299511

\bibitem[{Johnstone} {et~al.}(2013)]{Johnstone2013}
{Johnstone}, D., {Hendricks}, B., {Herczeg}, G.~J., \& {Bruderer}, S. 2013,
  \apj, 765, 133

\bibitem[{Johnstone} {et~al.}(2022)]{Johnstone2022}
{Johnstone}, D., {Lalchand}, B., {Mairs}, S., {et~al.} 2022, \apj, 937, 6

\bibitem[{J{\o}rgensen} {et~al.}(2020)]{2020ARA&A..58..727J}
{J{\o}rgensen}, J.~K., {Belloche}, A., \& {Garrod}, R.~T. 2020, \araa, 58, 727

\bibitem[{J{\o}rgensen} {et~al.}(2008)]{Jorgensen2008}
{J{\o}rgensen}, J.~K., {Johnstone}, D., {Kirk}, H., {et~al.} 2008, \apj, 683,
  822

\bibitem[{Justtanont} {et~al.}(2010)]{2010A&A...521...L6}
{Justtanont}, K., {Decin}, L., {Sch{\"o}ier}, F.~L., {et~al.} 2010, \aap, 521,
  L6

\bibitem[{Kaifu} {et~al.}(2004)]{2004PASJ...56...69K}
{Kaifu}, N., {Ohishi}, M., {Kawaguchi}, K., {et~al.} 2004, \pasj, 56, 69

\bibitem[{Kama} {et~al.}(2013)]{2013AA...556A..57K}
{Kama}, M., {L{\'o}pez-Sepulcre}, A., {Dominik}, C., {et~al.} 2013, \aap, 556,
  A57

\bibitem[{Kami{\'n}ski} {et~al.}(2017)]{2017AA...607A..78K}
{Kami{\'n}ski}, T., {Menten}, K.~M., {Tylenda}, R., {et~al.} 2017, \aap, 607,
  A78

\bibitem[{Kami{\'n}ski} {et~al.}(2013)]{2013A&A...551A.113K}
{Kami{\'n}ski}, T., {Gottlieb}, C.~A., {Menten}, K.~M., {et~al.} 2013, \aap,
  551, A113

\bibitem[{Katsioli} {et~al.}(2024)]{Katsioli2024}
{Katsioli}, S., {Adam}, R., {Ade}, P., {et~al.} 2024, in European Physical
  Journal Web of Conferences, Vol. 293, mm Universe 2023 - Observing the
  Universe at mm Wavelengths, 00026

\bibitem[{Ka{\'z}mierczak-Barthel} {et~al.}(2014)]{2014AA...567A..53K}
{Ka{\'z}mierczak-Barthel}, M., {van der Tak}, F.~F.~S., {Helmich}, F.~P.,
  {et~al.} 2014, \aap, 567, A53

\bibitem[{Kennicutt} {et~al.}(2011)]{Kennicutt2011}
{Kennicutt}, R.~C., {Calzetti}, D., {Aniano}, G., {et~al.} 2011, \pasp, 123,
  1347

\bibitem[{Kenyon} {et~al.}(1990)]{Kenyon1990}
{Kenyon}, S.~J., {Hartmann}, L.~W., {Strom}, K.~M., \& {Strom}, S.~E. 1990,
  \aj, 99, 869

\bibitem[{Kim} {et~al.}(2017)]{Kim2017}
{Kim}, S., {Schulze}, S., {Resmi}, L., {et~al.} 2017, \apjl, 850, L21

\bibitem[{Klochkov} {et~al.}(2016)]{Klochkov+16}
{Klochkov}, D., {Suleimanov}, V., {Sasaki}, M., \& {Santangelo}, A. 2016, \aap,
  592, L12

\bibitem[{Knapen} {et~al.}(2017)]{Knapen2017}
{Knapen}, J.~H., {Lee}, J.~C., \& {Gil de Paz}, A., eds. 2017, Astrophysics and
  Space Science Library, Vol. 434, {Outskirts of Galaxies}

\bibitem[{Koch} {et~al.}(2012)]{2012ApJ...747...79K}
{Koch}, P.~M., {Tang}, Y.-W., \& {Ho}, P. T.~P. 2012, \apj, 747, 79

\bibitem[{Koda} {et~al.}(2014)]{Koda2014}
{Koda}, J., {Blake}, C., {Davis}, T., {et~al.} 2014, \mnras, 445, 4267

\bibitem[{Kravtsov} \& {Borgani}(2012)]{Kravtsov2012}
{Kravtsov}, A.~V., \& {Borgani}, S. 2012, \araa, 50, 353

\bibitem[{Kryukova} {et~al.}(2014)]{2014AJ....148...11K}
{Kryukova}, E., {Megeath}, S.~T., {Hora}, J.~L., {et~al.} 2014, \aj, 148, 11

\bibitem[{Kunitomo} {et~al.}(2017)]{Kunitomo2017}
{Kunitomo}, M., {Guillot}, T., {Takeuchi}, T., \& {Ida}, S. 2017, \aap, 599,
  A49

\bibitem[{Kwok}(2011)]{2011IAUS..280..203K}
{Kwok}, S. 2011, in IAU Symposium, Vol. 280, The Molecular Universe, ed.
  J.~{Cernicharo} \& R.~{Bachiller}, 203

\bibitem[{Lada} \& {Lada}(2003)]{Lada2003}
{Lada}, C.~J., \& {Lada}, E.~A. 2003, \araa, 41, 57

\bibitem[{Larson}(1981)]{Larson1981}
{Larson}, R.~B. 1981, \mnras, 194, 809

\bibitem[{Laskar} {et~al.}(2019)]{Laskar2019}
{Laskar}, T., {Alexander}, K.~D., {Gill}, R., {et~al.} 2019, \apjl, 878, L26

\bibitem[{Latter} {et~al.}(1993)]{1993ApJ...419L..97L}
{Latter}, W.~B., {Walker}, C.~K., \& {Maloney}, P.~R. 1993, \apjl, 419, L97

\bibitem[{Lee} {et~al.}(2018)]{Lee2018}
{Lee}, B.~E., {Le Brun}, A.~M.~C., {Haq}, M.~E., {et~al.} 2018, \mnras, 479,
  890

\bibitem[{Lee}(2007)]{LeeJE2007}
{Lee}, J.-E. 2007, Journal of Korean Astronomical Society, 40, 83

\bibitem[{Lee} {et~al.}(2019)]{LeeJE2019}
{Lee}, J.-E., {Lee}, S., {Baek}, G., {et~al.} 2019, Nature Astronomy, 3, 314

\bibitem[{Lee} {et~al.}(2020)]{LeeYH2020}
{Lee}, Y.-H., {Johnstone}, D., {Lee}, J.-E., {et~al.} 2020, \apj, 903, 5

\bibitem[{Lee} {et~al.}(2021)]{LeeYH2021}
{Lee}, Y.-H., {Johnstone}, D., {Lee}, J.-E., {et~al.} 2021, \apj, 920, 119

\bibitem[{Lefloch} {et~al.}(2018)]{2018MNRAS.477.4792L}
{Lefloch}, B., {Bachiller}, R., {Ceccarelli}, C., {et~al.} 2018, \mnras, 477,
  4792

\bibitem[{Leroy} {et~al.}(2011)]{Leroy2011}
{Leroy}, A.~K., {Bolatto}, A., {Gordon}, K., {et~al.} 2011, \apj, 737, 12

\bibitem[{Leroy} {et~al.}(2021)]{Leroy2021}
{Leroy}, A.~K., {Schinnerer}, E., {Hughes}, A., {et~al.} 2021, \apjs, 257, 43

\bibitem[{Li} {et~al.}(2021)]{2021MNRAS.503.4508L}
{Li}, F., {Wang}, J., {Gao}, F., {et~al.} 2021, \mnras, 503, 4508

\bibitem[{Li} \& {Henning}(2011)]{2011Natur.479..499L}
{Li}, H.-B., \& {Henning}, T. 2011, \nat, 479, 499

\bibitem[{Li} {et~al.}(2015)]{2015Natur.520..518L}
{Li}, H.-B., {Yuen}, K.~H., {Otto}, F., {et~al.} 2015, \nat, 520, 518

\bibitem[Li {et~al.}(2025)]{research.0586}
Li, J., Deng, X., Li, Y., {et~al.} 2025, Research, 8, 0586

\bibitem[{Li} {et~al.}(2024)]{LiShaoliang2024}
{Li}, S., {Bintley}, D., {Ho}, P. T.~P., {et~al.} 2024, in Society of
  Photo-Optical Instrumentation Engineers (SPIE) Conference Series, Vol. 13096,
  Ground-based and Airborne Instrumentation for Astronomy X, ed. J.~J.
  {Bryant}, K.~{Motohara}, \& J.~R.~D. {Vernet}, 130967K

\bibitem[{Limongi} \& {Chieffi}(2018)]{Limongi2018}
{Limongi}, M., \& {Chieffi}, A. 2018, \apjs, 237, 13

\bibitem[{Lin} {et~al.}(2025)]{Lin2025}
{Lin}, L., {Zhang}, Z.-Y., {Wang}, J., {et~al.} 2025, Nature Astronomy, 9, 406

\bibitem[{Liu} {et~al.}(2018)]{Liu2018}
{Liu}, D., {Daddi}, E., {Dickinson}, M., {et~al.} 2018, \apj, 853, 172

\bibitem[{Liu} {et~al.}(2019)]{Liudz2019b}
{Liu}, D., {Schinnerer}, E., {Groves}, B., {et~al.} 2019, \apj, 887, 235

\bibitem[{Liu} {et~al.}(2022)]{2022FrASS...9.3556L}
{Liu}, J., {Zhang}, Q., \& {Qiu}, K. 2022, Frontiers in Astronomy and Space
  Sciences, 9, 943556

\bibitem[{Liu} {et~al.}(2023)]{2023ApJ...945..160L}
{Liu}, J., {Zhang}, Q., {Koch}, P.~M., {et~al.} 2023, \apj, 945, 160

\bibitem[{Longmore} {et~al.}(2012)]{2012ApJ...746..117L}
{Longmore}, S.~N., {Rathborne}, J., {Bastian}, N., {et~al.} 2012, \apj, 746,
  117

\bibitem[{Loomis} {et~al.}(2021)]{2021NatAs...5..188L}
{Loomis}, R.~A., {Burkhardt}, A.~M., {Shingledecker}, C.~N., {et~al.} 2021,
  Nature Astronomy, 5, 188

\bibitem[{Ludlow} {et~al.}(2012)]{Ludlow2012}
{Ludlow}, A.~D., {Navarro}, J.~F., {Li}, M., {et~al.} 2012, \mnras, 427, 1322

\bibitem[{Mac Low} \& {Klessen}(2004)]{2004RvMP...76..125M}
{Mac Low}, M.-M., \& {Klessen}, R.~S. 2004, Reviews of Modern Physics, 76, 125

\bibitem[{Macci{\`o}} {et~al.}(2007)]{Maccio2007}
{Macci{\`o}}, A.~V., {Dutton}, A.~A., {van den Bosch}, F.~C., {et~al.} 2007,
  \mnras, 378, 55

\bibitem[{Madau} \& {Dickinson}(2014)]{Madau2014}
{Madau}, P., \& {Dickinson}, M. 2014, \araa, 52, 415

\bibitem[{Madden} {et~al.}(2020)]{Madden2020}
{Madden}, S.~C., {Cormier}, D., {Hony}, S., {et~al.} 2020, \aap, 643, A141

\bibitem[{Magdis} {et~al.}(2012)]{Magdis2012}
{Magdis}, G.~E., {Daddi}, E., {B{\'e}thermin}, M., {et~al.} 2012, \apj, 760, 6

\bibitem[{Mairs} {et~al.}(2019)]{Mairs2019}
{Mairs}, S., {Lalchand}, B., {Bower}, G.~C., {et~al.} 2019, \apj, 871, 72

\bibitem[{Mairs} {et~al.}(2024)]{Mairs2024}
{Mairs}, S., {Lee}, S., {Johnstone}, D., {et~al.} 2024, \apj, 966, 215

\bibitem[{Matthews} {et~al.}(2009)]{2009ApJS..182..143M}
{Matthews}, B.~C., {McPhee}, C.~A., {Fissel}, L.~M., \& {Curran}, R.~L. 2009,
  \apjs, 182, 143

\bibitem[{Maud} {et~al.}(2019)]{Maud2019}
{Maud}, L.~T., {Cesaroni}, R., {Kumar}, M.~S.~N., {et~al.} 2019, \aap, 627, L6

\bibitem[{McGuire}(2022)]{2022ApJS..259...30M}
{McGuire}, B.~A. 2022, \apjs, 259, 30

\bibitem[{McKee} \& {Ostriker}(2007{\natexlab{a}})]{McKee2007}
{McKee}, C.~F., \& {Ostriker}, E.~C. 2007{\natexlab{a}}, \araa, 45, 565

\bibitem[{McKee} \& {Ostriker}(2007{\natexlab{b}})]{2007ARA&A..45..565M}
{McKee}, C.~F., \& {Ostriker}, E.~C. 2007{\natexlab{b}}, \araa, 45, 565

\bibitem[{McKee} \& {Tan}(2003)]{McKee2003}
{McKee}, C.~F., \& {Tan}, J.~C. 2003, \apj, 585, 850

\bibitem[{McKellar}(1940)]{1940PASP...52..187M}
{McKellar}, A. 1940, \pasp, 52, 187

\bibitem[{M{\'e}ndez-Delgado} {et~al.}(2022)]{Mendez-Delgado2022}
{M{\'e}ndez-Delgado}, J.~E., {Amayo}, A., {Arellano-C{\'o}rdova}, K.~Z.,
  {et~al.} 2022, \mnras, 510, 4436

\bibitem[{Meny} {et~al.}(2007)]{Meny2007}
{Meny}, C., {Gromov}, V., {Boudet}, N., {et~al.} 2007, \aap, 468, 171

\bibitem[{Micelotta} {et~al.}(2018)]{micelotta+18}
{Micelotta}, E.~R., {Matsuura}, M., \& {Sarangi}, A. 2018, \ssr, 214, 53

\bibitem[{Michiyama} {et~al.}(2021)]{Michiyama2021}
{Michiyama}, T., {Saito}, T., {Tadaki}, K.-i., {et~al.} 2021, \apjs, 257, 28

\bibitem[{Miley} {et~al.}(2004)]{Miley2004}
{Miley}, G.~K., {Overzier}, R.~A., {Tsvetanov}, Z.~I., {et~al.} 2004, \nat,
  427, 47

\bibitem[{Miller} {et~al.}(2018)]{Miller2018}
{Miller}, T.~B., {Chapman}, S.~C., {Aravena}, M., {et~al.} 2018, \nat, 556, 469

\bibitem[{Miville-Desch{\^e}nes} {et~al.}(2017)]{Miville-Deschenes2017}
{Miville-Desch{\^e}nes}, M.-A., {Murray}, N., \& {Lee}, E.~J. 2017, \apj, 834,
  57

\bibitem[{Molinari} {et~al.}(2010)]{Molinari+10}
{Molinari}, S., {Swinyard}, B., {Bally}, J., {et~al.} 2010, \pasp, 122, 314

\bibitem[{M{\"o}ller} {et~al.}(2021)]{2021AA...651A...9M}
{M{\"o}ller}, T., {Schilke}, P., {Schmiedeke}, A., {et~al.} 2021, \aap, 651, A9

\bibitem[{Morgan} {et~al.}(2003)]{morgan+03}
{Morgan}, H.~L., {Dunne}, L., {Eales}, S.~A., {Ivison}, R.~J., \& {Edmunds},
  M.~G. 2003, \apjl, 597, L33

\bibitem[{Morton} {et~al.}(2007)]{Morton+07}
{Morton}, T.~D., {Slane}, P., {Borkowski}, K.~J., {et~al.} 2007, \apj, 667, 219

\bibitem[{Nagy} {et~al.}(2017)]{2017AA...599A..22N}
{Nagy}, Z., {Choi}, Y., {Ossenkopf-Okada}, V., {et~al.} 2017, \aap, 599, A22

\bibitem[{Neill} {et~al.}(2014)]{2014ApJ...789....8N}
{Neill}, J.~L., {Bergin}, E.~A., {Lis}, D.~C., {et~al.} 2014, \apj, 789, 8

\bibitem[{Nicholls} {et~al.}(2017)]{Nicholls2017}
{Nicholls}, D.~C., {Sutherland}, R.~S., {Dopita}, M.~A., {Kewley}, L.~J., \&
  {Groves}, B.~A. 2017, \mnras, 466, 4403

\bibitem[{Nisini} {et~al.}(2013)]{2013A&A...549A..16N}
{Nisini}, B., {Santangelo}, G., {Antoniucci}, S., {et~al.} 2013, \aap, 549, A16

\bibitem[{Nozawa} {et~al.}(2011)]{nozawa+11}
{Nozawa}, T., {Maeda}, K., {Kozasa}, T., {et~al.} 2011, \apj, 736, 45

\bibitem[{{\"O}berg} {et~al.}(2011)]{2011ApJ...740...14O}
{{\"O}berg}, K.~I., {van der Marel}, N., {Kristensen}, L.~E., \& {van
  Dishoeck}, E.~F. 2011, \apj, 740, 14

\bibitem[{Olofsson} {et~al.}(2007)]{2007AA...476..791O}
{Olofsson}, A.~O.~H., {Persson}, C.~M., {Koning}, N., {et~al.} 2007, \aap, 476,
  791

\bibitem[{Onishi} {et~al.}(1999)]{1999PASJ...51..871O}
{Onishi}, T., {Kawamura}, A., {Abe}, R., {et~al.} 1999, \pasj, 51, 871

\bibitem[Pardo {et~al.}(2001)]{Pardo2001}
Pardo, J., Cernicharo, J., \& Serabyn, E. 2001, IEEE Transactions on Antennas
  and Propagation, 49, 1683

\bibitem[{Pardo} {et~al.}(2007)]{2007ApJ...661..250P}
{Pardo}, J.~R., {Cernicharo}, J., {Goicoechea}, J.~R., {Gu{\'e}lin}, M., \&
  {Asensio Ramos}, A. 2007, \apj, 661, 250

\bibitem[{Park} {et~al.}(2024)]{Park2024}
{Park}, G., {Johnstone}, D., {Pe{\~n}a}, C.~C., {et~al.} 2024, \aj, 168, 122

\bibitem[{Pattle} {et~al.}(2017)]{2017ApJ...846..122P}
{Pattle}, K., {Ward-Thompson}, D., {Berry}, D., {et~al.} 2017, \apj, 846, 122

\bibitem[{Peebles}(1980)]{Peebles1980}
{Peebles}, P.~J.~E. 1980, {The large-scale structure of the universe}
  ({Princeton University Press})

\bibitem[{Persson} {et~al.}(2007)]{2007AA...476..807P}
{Persson}, C.~M., {Olofsson}, A.~O.~H., {Koning}, N., {et~al.} 2007, \aap, 476,
  807

\bibitem[{Pillai} {et~al.}(2020)]{2020NatAs...4.1195P}
{Pillai}, T. G.~S., {Clemens}, D.~P., {Reissl}, S., {et~al.} 2020, Nature
  Astronomy, 4, 1195

\bibitem[{Planck Collaboration} {et~al.}(2011)]{Planck2011}
{Planck Collaboration}, {Ade}, P.~A.~R., {Aghanim}, N., {et~al.} 2011, \aap,
  536, A17

\bibitem[{Planck Collaboration} {et~al.}(2014)]{Planck2014a}
{Planck Collaboration}, {Ade}, P.~A.~R., {Aghanim}, N., {et~al.} 2014, \aap,
  571, A1

\bibitem[{Planck Collaboration} {et~al.}(2015)]{planck2015}
{Planck Collaboration}, {Ade}, P.~A.~R., {Aghanim}, N., {et~al.} 2015, \aap,
  576, A104

\bibitem[{Planck Collaboration} {et~al.}(2016{\natexlab{a}})]{Planck2016}
{Planck Collaboration}, {Ade}, P.~A.~R., {Aghanim}, N., {et~al.}
  2016{\natexlab{a}}, \aap, 594, A24

\bibitem[{Planck Collaboration} {et~al.}(2016{\natexlab{b}})]{planck+16}
{Planck Collaboration}, {Arnaud}, M., {Ashdown}, M., {et~al.}
  2016{\natexlab{b}}, \aap, 586, A134

\bibitem[{Planck Collaboration} {et~al.}(2016{\natexlab{c}})]{Planck2016a}
{Planck Collaboration}, {Ade}, P.~A.~R., {Aghanim}, N., {et~al.}
  2016{\natexlab{c}}, \aap, 586, A140

\bibitem[{Priestley} {et~al.}(2022)]{Priestley+22}
{Priestley}, F.~D., {Chawner}, H., {Barlow}, M.~J., {et~al.} 2022, \mnras, 516,
  2314

\bibitem[{Qiu} {et~al.}(2022)]{2022ApJS...259...56}
{Qiu}, J.-J., {Zhang}, Y., {Zhang}, J.-S., \& {Nakashima}, J.-i. 2022, \apjs,
  259, 56

\bibitem[{Qiu} \& {Zhang}(2009)]{2009ApJ...702L..66Q}
{Qiu}, K., \& {Zhang}, Q. 2009, \apjl, 702, L66

\bibitem[{Ramambason} {et~al.}(2024)]{Ramambason2024}
{Ramambason}, L., {Lebouteiller}, V., {Madden}, S.~C., {et~al.} 2024, \aap,
  681, A14

\bibitem[{Ravindranath} {et~al.}(2020)]{Ravindranath2020}
{Ravindranath}, S., {Monroe}, T., {Jaskot}, A., {Ferguson}, H.~C., \&
  {Tumlinson}, J. 2020, \apj, 896, 170

\bibitem[{Reipurth} \& {Schneider}(2008)]{2008hsf1.book...36R}
{Reipurth}, B., \& {Schneider}, N. 2008, {Star Formation and Young Clusters in
  Cygnus}, Vol.~4, Handbook of Star Forming Regions, Volume I, Vol.~4
  (Astronomical Society of the Pacific), 36

\bibitem[{R{\'e}my-Ruyer} {et~al.}(2014)]{Remy2014}
{R{\'e}my-Ruyer}, A., {Madden}, S.~C., {Galliano}, F., {et~al.} 2014, \aap,
  563, A31

\bibitem[{Reynolds} {et~al.}(2006)]{Reynolds+06}
{Reynolds}, S.~P., {Borkowski}, K.~J., {Hwang}, U., {et~al.} 2006, \apjl, 652,
  L45

\bibitem[{Rho} {et~al.}(2023)]{Rho+23}
{Rho}, J., {Ravi}, A.~P., {Tram}, L.~N., {et~al.} 2023, \mnras, 522, 2279

\bibitem[{Riechers} {et~al.}(2013)]{Riechers2013}
{Riechers}, D.~A., {Bradford}, C.~M., {Clements}, D.~L., {et~al.} 2013, \nat,
  496, 329

\bibitem[{Riechers} {et~al.}(2014)]{Riechers2014}
{Riechers}, D.~A., {Carilli}, C.~L., {Capak}, P.~L., {et~al.} 2014, \apj, 796,
  84

\bibitem[{Ritacco} {et~al.}(2017)]{2017A&A...599A..34R}
{Ritacco}, A., {Ponthieu}, N., {Catalano}, A., {et~al.} 2017, \aap, 599, A34

\bibitem[{Ritacco} {et~al.}(2024)]{2024PASP..136k5001R}
{Ritacco}, A., {Bizzarri}, L., {Savorgnano}, S., {et~al.} 2024, \pasp, 136,
  115001

\bibitem[{Rivilla} {et~al.}(2020)]{2020ApJ...899L..28R}
{Rivilla}, V.~M., {Mart{\'i}n-Pintado}, J., {Jim{\'e}nez-Serra}, I., {et~al.}
  2020, \apjl, 899, L28

\bibitem[{Romano}(2022)]{Romano2022}
{Romano}, D. 2022, \aapr, 30, 7

\bibitem[{Roseboom} {et~al.}(2010)]{Roseboom2010}
{Roseboom}, I.~G., {Oliver}, S.~J., {Kunz}, M., {et~al.} 2010, \mnras, 409, 48

\bibitem[{Rowan-Robinson} {et~al.}(2016)]{Rowan2016}
{Rowan-Robinson}, M., {Oliver}, S., {Wang}, L., {et~al.} 2016, \mnras, 461,
  1100

\bibitem[{Saintonge} {et~al.}(2018)]{Saintonge2018}
{Saintonge}, A., {Wilson}, C.~D., {Xiao}, T., {et~al.} 2018, \mnras, 481, 3497

\bibitem[{Sakai} {et~al.}(2008)]{2008ApJ...672..371S}
{Sakai}, N., {Sakai}, T., {Hirota}, T., \& {Yamamoto}, S. 2008, \apj, 672, 371

\bibitem[{Sarangi} {et~al.}(2018)]{sarangi+18}
{Sarangi}, A., {Matsuura}, M., \& {Micelotta}, E.~R. 2018, Space Science
  Review, 214, 63

\bibitem[{Sayers} {et~al.}(2021)]{Sayers2021}
{Sayers}, J., {Sereno}, M., {Ettori}, S., {et~al.} 2021, \mnras, 505, 4338

\bibitem[{Sayers} {et~al.}(2013)]{Sayers2013}
{Sayers}, J., {Mroczkowski}, T., {Zemcov}, M., {et~al.} 2013, \apj, 778, 52

\bibitem[{Sayers} {et~al.}(2019)]{Sayers2019}
{Sayers}, J., {Monta{\~n}a}, A., {Mroczkowski}, T., {et~al.} 2019, \apj, 880,
  45

\bibitem[{Schilke} {et~al.}(2006)]{2006AA...454L..41S}
{Schilke}, P., {Comito}, C., {Thorwirth}, S., {et~al.} 2006, \aap, 454, L41

\bibitem[{Schinnerer} {et~al.}(2007)]{Schinnerer2007}
{Schinnerer}, E., {Smol{\v{c}}i{\'c}}, V., {Carilli}, C.~L., {et~al.} 2007,
  \apjs, 172, 46

\bibitem[{Schneider} {et~al.}(2013)]{2013ApJ...766L..17S}
{Schneider}, N., {Andr{\'e}}, P., {K{\"o}nyves}, V., {et~al.} 2013, \apjl, 766,
  L17

\bibitem[{Schreiber} {et~al.}(2015)]{Schreiber2015}
{Schreiber}, C., {Pannella}, M., {Elbaz}, D., {et~al.} 2015, \aap, 575, A74

\bibitem[{Scoville} \& {Wilson}(2004)]{Scoville2004}
{Scoville}, N.~Z., \& {Wilson}, C.~D. 2004, in Astronomical Society of the
  Pacific Conference Series, Vol. 322, The Formation and Evolution of Massive
  Young Star Clusters, ed. H.~J.~G.~L.~M. {Lamers}, L.~J. {Smith}, \&
  A.~{Nota}, 245

\bibitem[{Scoville} {et~al.}(2014)]{Scoville2014}
{Scoville}, N., {Aussel}, H., {Sheth}, K., {et~al.} 2014, \apj, 783, 84

\bibitem[{Scoville} {et~al.}(2016)]{Scoville2016}
{Scoville}, N., {Sheth}, K., {Aussel}, H., {et~al.} 2016, \apj, 820, 83

\bibitem[{Serabyn} \& {Weisstein}(1995)]{1995ApJ...451..238S}
{Serabyn}, E., \& {Weisstein}, E.~W. 1995, \apj, 451, 238

\bibitem[{Sharda} {et~al.}(2023)]{Sharda2023}
{Sharda}, P., {Amarsi}, A.~M., {Grasha}, K., {et~al.} 2023, \mnras, 518, 3985

\bibitem[{Sheehan} {et~al.}(2025)]{Sheehan2025}
{Sheehan}, P.~D., {Johnstone}, D., {Contreras Pe{\~n}a}, C., {et~al.} 2025,
  \apj, 982, 176

\bibitem[{Shi} {et~al.}(2014)]{Shi2014}
{Shi}, Y., {Armus}, L., {Helou}, G., {et~al.} 2014, \nat, 514, 335

\bibitem[{Shi} {et~al.}(2015)]{Shi2015}
{Shi}, Y., {Wang}, J., {Zhang}, Z.-Y., {et~al.} 2015, \apjl, 804, L11

\bibitem[{Shi} {et~al.}(2016)]{Shi2016}
{Shi}, Y., {Wang}, J., {Zhang}, Z.-Y., {et~al.} 2016, Nature Communications, 7,
  13789

\bibitem[{Shi} {et~al.}(2024)]{Shi2024}
{Shi}, Y., {Zhang}, P., {Mao}, S., \& {Gu}, Q. 2024, \mnras, 528, 4922

\bibitem[{Sillassen} {et~al.}(2024)]{Sillassen2024}
{Sillassen}, N.~B., {Jin}, S., {Magdis}, G.~E., {et~al.} 2024, \aap, 690, A55

\bibitem[{Simpson} {et~al.}(2019)]{Simpson2019}
{Simpson}, J.~M., {Smail}, I., {Swinbank}, A.~M., {et~al.} 2019, \apj, 880, 43

\bibitem[{Smail} {et~al.}(1997)]{Smail1997}
{Smail}, I., {Ivison}, R.~J., \& {Blain}, A.~W. 1997, \apjl, 490, L5

\bibitem[{Smith} {et~al.}(2012)]{Smith2012}
{Smith}, M.~W.~L., {Eales}, S.~A., {Gomez}, H.~L., {et~al.} 2012, \apj, 756, 40

\bibitem[{Smith} {et~al.}(2021)]{Smith2021}
{Smith}, M. W.~L., {Eales}, S.~A., {Williams}, T.~G., {et~al.} 2021, \apjs,
  257, 52

\bibitem[Soler(2019)]{soler2019}
Soler, J.~D. 2019, Astronomy and Astrophysics, 629, A96, aDS Bibcode:
  2019A\&A...629A..96S

\bibitem[{Spezzano} {et~al.}(2013)]{2013ApJ...769L..19S}
{Spezzano}, S., {Br{\"u}nken}, S., {Schilke}, P., {et~al.} 2013, \apjl, 769,
  L19

\bibitem[{Spilker} {et~al.}(2023)]{Spilker2023}
{Spilker}, J.~S., {Phadke}, K.~A., {Aravena}, M., {et~al.} 2023, \nat, 618, 708

\bibitem[{Springel} \& {Farrar}(2007)]{Springel2007}
{Springel}, V., \& {Farrar}, G.~R. 2007, \mnras, 380, 911

\bibitem[{Stevens} {et~al.}(2005)]{Stevens2005}
{Stevens}, J.~A., {Amure}, M., \& {Gear}, W.~K. 2005, \mnras, 357, 361

\bibitem[{Strandet} {et~al.}(2017)]{Strandet2017}
{Strandet}, M.~L., {Weiss}, A., {De Breuck}, C., {et~al.} 2017, \apjl, 842, L15

\bibitem[{Su} {et~al.}(2019)]{2019ApJS..240....9S}
{Su}, Y., {Yang}, J., {Zhang}, S., {et~al.} 2019, \apjs, 240, 9

\bibitem[{Sun} {et~al.}(2015)]{Sun2015}
{Sun}, Y., {Xu}, Y., {Yang}, J., {et~al.} 2015, \apjl, 798, L27

\bibitem[{Sun} {et~al.}(2024)]{Sun2024}
{Sun}, Y., {Yang}, J., {Yan}, Q.-Z., {et~al.} 2024, \apjs, 275, 35

\bibitem[{Sunyaev} \& {Zeldovich}(1972)]{Sunyaev1972}
{Sunyaev}, R.~A., \& {Zeldovich}, Y.~B. 1972, Comments on Astrophysics and
  Space Physics, 4, 173

\bibitem[{Sunyaev} \& {Zeldovich}(1980)]{Sunyaev1980}
{Sunyaev}, R.~A., \& {Zeldovich}, Y.~B. 1980, \mnras, 190, 413

\bibitem[{Swings} \& {Rosenfeld}(1937)]{1937ApJ....86..483S}
{Swings}, P., \& {Rosenfeld}, L. 1937, \apj, 86, 483

\bibitem[{Tabatabaei} {et~al.}(2014)]{Tabatabaei2014}
{Tabatabaei}, F.~S., {Braine}, J., {Xilouris}, E.~M., {et~al.} 2014, \aap, 561,
  A95

\bibitem[{Tahani} {et~al.}(2016)]{2016ApJ...832...12T}
{Tahani}, K., {Plume}, R., {Bergin}, E.~A., {et~al.} 2016, \apj, 832, 12

\bibitem[{Tahani}(2022)]{2022FrASS...9.0027T}
{Tahani}, M. 2022, Frontiers in Astronomy and Space Sciences, 9, 940027

\bibitem[{Takahashi} {et~al.}(2024)]{Takahashi2024}
{Takahashi}, S., {Machida}, M.~N., {Omura}, M., {et~al.} 2024, \apj, 964, 48

\bibitem[{Tassis} \& {Mouschovias}(2004)]{2004ApJ...616..283T}
{Tassis}, K., \& {Mouschovias}, T.~C. 2004, \apj, 616, 283

\bibitem[{Terry} {et~al.}(2019)]{CCAT-prime-2019}
{Terry}, H., {Battaglia}, N., {Basu}, K., {et~al.} 2019, in Bulletin of the
  American Astronomical Society, Vol.~51, 213

\bibitem[{The Simons Observatory Collaboration}
  {et~al.}(2025)]{SimonsObservatory2025}
{The Simons Observatory Collaboration}, {Abitbol}, M., {Abril-Cabezas}, I.,
  {et~al.} 2025, arXiv e-prints, arXiv:2503.00636

\bibitem[{Torne} {et~al.}(2017)]{Torne2017}
{Torne}, P., {Desvignes}, G., {Eatough}, R.~P., {et~al.} 2017, \mnras, 465, 242

\bibitem[{Tram} {et~al.}(2025)]{2025arXiv250116079T}
{Tram}, L.~N., {Hoang}, T., {Lazarian}, A., {et~al.} 2025, arXiv e-prints,
  arXiv:2501.16079

\bibitem[{Tsai} {et~al.}(2015)]{Tsai2015}
{Tsai}, C.-W., {Eisenhardt}, P. R.~M., {Wu}, J., {et~al.} 2015, \apj, 805, 90

\bibitem[{Tumlinson} {et~al.}(2017)]{Tumlinson2017}
{Tumlinson}, J., {Peeples}, M.~S., \& {Werk}, J.~K. 2017, \araa, 55, 389

\bibitem[{van der Giessen} {et~al.}(2024)]{vanderGiessen2024}
{van der Giessen}, S.~A., {Matsumoto}, K., {Relano}, M., {et~al.} 2024, \aap,
  692, A39

\bibitem[{van Dishoeck} \& {Blake}(1998)]{1998ARA&A..36..317V}
{van Dishoeck}, E.~F., \& {Blake}, G.~A. 1998, \araa, 36, 317

\bibitem[{Vayner} {et~al.}(2024)]{Vayner2024}
{Vayner}, A., {Zakamska}, N.~L., {Ishikawa}, Y., {et~al.} 2024, \apj, 960, 126

\bibitem[{Vieira} {et~al.}(2013)]{Vieira2013}
{Vieira}, J.~D., {Marrone}, D.~P., {Chapman}, S.~C., {et~al.} 2013, \nat, 495,
  344

\bibitem[{Walter} {et~al.}(2012)]{Walter2012}
{Walter}, F., {Decarli}, R., {Carilli}, C., {et~al.} 2012, \nat, 486, 233

\bibitem[{Wang} {et~al.}(2011)]{2011MNRAS.416L..21W}
{Wang}, J., {Zhang}, Z., \& {Shi}, Y. 2011, \mnras, 416, L21

\bibitem[{Wang} {et~al.}(2016{\natexlab{a}})]{Wang2016b}
{Wang}, T., {Elbaz}, D., {Daddi}, E., {et~al.} 2016{\natexlab{a}}, \apj, 828,
  56

\bibitem[{Wang} {et~al.}(2016{\natexlab{b}})]{Wang2016a}
{Wang}, T., {Elbaz}, D., {Schreiber}, C., {et~al.} 2016{\natexlab{b}}, \apj,
  816, 84

\bibitem[{Wang} {et~al.}(2019)]{Wang2019}
{Wang}, T., {Schreiber}, C., {Elbaz}, D., {et~al.} 2019, \nat, 572, 211

\bibitem[{White} {et~al.}(2003)]{2003AA...407..589W}
{White}, G.~J., {Araki}, M., {Greaves}, J.~S., {Ohishi}, M., \& {Higginbottom},
  N.~S. 2003, \aap, 407, 589

\bibitem[{Widicus Weaver} {et~al.}(2017)]{2017ApJS..232....3W}
{Widicus Weaver}, S.~L., {Laas}, J.~C., {Zou}, L., {et~al.} 2017, \apjs, 232, 3

\bibitem[{Wilson} \& {Rood}(1994)]{Wilson1994}
{Wilson}, T.~L., \& {Rood}, R. 1994, \araa, 32, 191

\bibitem[{Wolfire} {et~al.}(2003)]{Wolfire2003}
{Wolfire}, M.~G., {McKee}, C.~F., {Hollenbach}, D., \& {Tielens}, A.~G.~G.~M.
  2003, \apj, 587, 278

\bibitem[{Wolszczan} {et~al.}(1991)]{Wolszczan+91}
{Wolszczan}, A., {Cordes}, J.~M., \& {Dewey}, R.~J. 1991, \apjl, 372, L99

\bibitem[{Wu} {et~al.}(2005)]{2005ApJ...635L.173W}
{Wu}, J., {Evans}, II, N.~J., {Gao}, Y., {et~al.} 2005, \apjl, 635, L173

\bibitem[{Wu} {et~al.}(2012)]{Wu2012}
{Wu}, J., {Tsai}, C.-W., {Sayers}, J., {et~al.} 2012, \apj, 756, 96

\bibitem[{Yi} {et~al.}(2014)]{Yi2014}
{Yi}, S.-X., {Gao}, H., \& {Zhang}, B. 2014, \apjl, 792, L21

\bibitem[{Yoon} {et~al.}(2022)]{Yoon2022}
{Yoon}, S.-Y., {Herczeg}, G.~J., {Lee}, J.-E., {et~al.} 2022, \apj, 929, 60

\bibitem[{Yuan} {et~al.}(2016)]{Yuan2016}
{Yuan}, Q., {Wang}, Q.~D., {Lei}, W.-H., {Gao}, H., \& {Zhang}, B. 2016,
  \mnras, 461, 3375

\bibitem[{Yuan} \& {Han}(2020)]{Yuan2020}
{Yuan}, Z.~S., \& {Han}, J.~L. 2020, \mnras, 497, 5485

\bibitem[{Yusef-Zadeh} {et~al.}(2007)]{Yusef-Zadeh2007}
{Yusef-Zadeh}, F., {Muno}, M., {Wardle}, M., \& {Lis}, D.~C. 2007, \apj, 656,
  847

\bibitem[{Zavala} {et~al.}(2018)]{Zavala2018}
{Zavala}, J.~A., {Monta{\~n}a}, A., {Hughes}, D.~H., {et~al.} 2018, Nature
  Astronomy, 2, 56

\bibitem[{Zernickel} {et~al.}(2012)]{2012AA...546A..87Z}
{Zernickel}, A., {Schilke}, P., {Schmiedeke}, A., {et~al.} 2012, \aap, 546, A87

\bibitem[{Zhang} {et~al.}(2012)]{2012ApJ...744...23Z}
{Zhang}, B., {Reid}, M.~J., {Menten}, K.~M., \& {Zheng}, X.~W. 2012, \apj, 744,
  23

\bibitem[{Zhang} {et~al.}(2014{\natexlab{a}})]{2014ApJ...792..116Z}
{Zhang}, Q., {Qiu}, K., {Girart}, J.~M., {et~al.} 2014{\natexlab{a}}, \apj,
  792, 116

\bibitem[{Zhang} {et~al.}(2014{\natexlab{b}})]{2014ApJ...784L..31Z}
{Zhang}, Z.-Y., {Gao}, Y., {Henkel}, C., {et~al.} 2014{\natexlab{b}}, \apjl,
  784, L31

\bibitem[{Zhou} {et~al.}(2024{\natexlab{a}})]{Zhou2024}
{Zhou}, L., {Wang}, T., {Daddi}, E., {et~al.} 2024{\natexlab{a}}, \aap, 684,
  A196

\bibitem[{Zhou} {et~al.}(2024{\natexlab{b}})]{ZhouWei2024}
{Zhou}, W., {Chen}, Z., {Jiang}, Z., {Feng}, H., \& {Jiang}, Y.
  2024{\natexlab{b}}, \apjl, 969, L6

\bibitem[{Zhu} {et~al.}(2014)]{Zhu+14}
{Zhu}, H., {Tian}, W.~W., \& {Zuo}, P. 2014, \apj, 793, 95

\bibitem[{Zhu} {et~al.}(2010)]{ZhuZhaohuan2010}
{Zhu}, Z., {Hartmann}, L., {Gammie}, C.~F., {et~al.} 2010, \apj, 713, 1134

\end{thebibliography}
}

\begin{appendix}





\section{Comparison with Astro2020 Science Themes}
\label{appendix:A}
Table~\ref{tab:astro2020} compares the XSMT-15m science cases with the three highest-priority science themes for the coming decade as identified in the US Decadal Survey Astro2020: ``Worlds and Suns in Context'', ``New Messengers and New Physics'', and ``Cosmic Ecosystems''.

\begin{table*}[!htbp]
    \centering
    \caption{Astro2020 science coverage.}
    \label{tab:astro2020}
    \begin{tabular}{|l|c|c|c|}
    \hline
         \multicolumn{1}{|c|}{\textbf{Astro2020}} & \textbf{New Physics} & \textbf{Cosmic Ecosystems} & \textbf{Worlds and Suns} \\
    \hline
         EG: GMC & & $\checkmark$ & \\
         EG: Dust & & $\checkmark$ & \\
         EG: CGM & & $\checkmark$ & \\
         EG: SZ & & $\checkmark$ & \\
         EG: Galaxy Evolution & & $\checkmark$ & \\
         EG: ngEHT & $\checkmark$ & & \\
         MW: Magnetic Fields & $\checkmark$ & $\checkmark$ & \\
         MW: Molecular Clouds & & $\checkmark$ & \\
         MW: Dense Gas & & $\checkmark$ & \\
         MW: Supernova Remnants & & $\checkmark$ & \\
         TD: Protostellar & $\checkmark$ & & \\
         TD: High-energy & $\checkmark$ & & \\
         AC: Molecules in Environments & & $\checkmark$ & $\checkmark$ \\
         AC: Galactic Chemical Evolution & & $\checkmark$ & $\checkmark$ \\
    \hline
    \end{tabular}
\end{table*}

\section{Comparison with CCAT-prime, CMB-S4 and Simons Observatory Science Cases}
\label{appendix:B}

CCAT-prime (aka. Fred Young Submillimeter Telescope, FYST; \citealt{CCAT-prime-2019,CCAT-prime-2023}, CMB-S4 \citep{CMB-S4-2019}, and Simons Observatory (SO LAT; \citealt{SimonsObservatory2025}) are smaller submillimeter telescopes (0.5--6~m) optimized for high-frequency, large-area surveys of galaxies and the CMB. 

With the capability to probe a wide frequency range from 200 to 850~GHz, CCAT-prime 6m has listed seven main driving science cases in \cite{CCAT-prime-2023}: (i) the epoch of reionization, (ii) tracing galaxy evolution over cosmic time, (iii) measuring CMB foregrounds, (iv) galactic polarization, (v) galaxy-cluster evolution, (vi) Rayleigh scattering, and (vii) time domain phenomena. Many of these overlap with the science goals of the XSMT-15m. However, CCAT-prime offers advantages in its much larger field of view and its unique ability to pursue the challenging detection of CMB photon Rayleigh scattering. In contrast, XSMT-15m provides higher angular resolution, important northern-sky time-domain coverage, and new baselines for ngEHT.

The CMB-S4 project focuses on four major science themes: (i) primordial gravitational waves and inflation (i.e., the imprint of primordial gravitational waves on the CMB polarization anisotropy), (ii) the dark Universe, including searching for any departures from the prediction of the Standard Model, and the massive neutrinos' effect on matter power spectrum, the kSZ effect, cosmic birefringence, cold axion energy density, etc., (iii) mapping matter in the cosmos, and (iv) the time-variable millimeter sky. CMB-S4 has a strong focus on testing fundamental physics with the deepest statistics on large-scale CMB and matter properties thanks to its extremely powerful mapping capability --- it can cover 70\% of the sky every day, with $\sim$~130,000 superconducting detectors in total in its cameras. Since CMB-S4 is dedicated to large-scale mapping up to about 270~GHz, XSMT-15m will play a distinct and complementary role, with unique strengths in high-frequency spectral line surveys and rapid mapping at moderately high angular resolution.

The SO LAT 6m telescopes operate at frequencies from 27~GHz to 280~GHz, and the observatory has a list of key science goals on: (i) primordial perturbations, (ii) the earliest snapshot of the Universe, (iii) cosmic birefringence in the linear polarization of CMB photons (for parity-breaking BSM physics), (iv) new large-scale view of dark matter, baryons, and galaxy clusters,(v) extragalactic sources: time-variable blazars and dusty galaxies, (vi) polarized Galactic ISM, (vii) interstellar dust, (viii) solar system bodies and exo-Oort clouds, (ix) asteroid regoliths, and (x) the millimeter transient sky. They have additional science cases on the fundamental physics with the CMB, solar system bodies and asteroids, given their advantages in field of view and multiple 6m LAT telescopes. In contrast, XSMT-15m will provide higher frequency coverage, finer angular resolution, and unique northern time-domain access, making it a powerful and complementary facility to SO LAT.

\end{appendix}

\end{multicols}
\end{document}